\documentclass[format=acmsmall, review=false, screen=true]{acmart}

\usepackage{booktabs} 

\usepackage[ruled]{algorithm2e} 

\SetAlFnt{\small}
\SetAlCapFnt{\small}
\SetAlCapNameFnt{\small}
\SetAlCapHSkip{0pt}
\IncMargin{-\parindent}

\usepackage
	{todonotes}

\usepackage[utf8x]{inputenc} 
\usepackage[greek,english]{babel}
\usepackage[T1]{fontenc}
\usepackage{multirow}
\usepackage{pdflscape}
\usepackage{nth}
\usepackage{longtable}
\usepackage{array}
\usepackage{footmisc}
\usepackage{afterpage}

\acmJournal{CSUR}
\acmVolume{52}
\acmNumber{5}
\acmArticle{89}
\acmYear{2019}
\copyrightyear{2019}

\setcopyright{acmlicensed}

\acmDOI{10.1145/3344548}

\begin{document}

\title[Extraction and Analysis of Fictional Character Networks: A Survey]{Extraction and Analysis of Fictional Character Networks:\\A Survey}

\author{Vincent Labatut}
\orcid{0000-0002-2619-2835}
\affiliation{%
  \institution{Laboratoire Informatique d'Avignon -- LIA EA 4128}
  \streetaddress{339 chemin des Meinajaries, Agroparc BP 91228}
  \city{Avignon cedex 9}
  \postcode{84911}
  \country{France}}
\email{vincent.labatut@univ-avignon.fr}

\author{Xavier Bost}
\orcid{0000-0002-2619-2835}
\affiliation{%
  \institution{Orkis}
  \streetaddress{610 rue Georges Claude, P\^ole d'activit\'e d'Aix-en-Provence}
  \city{Aix-En-Provence}
  \postcode{13290}
  \country{France}
}
\affiliation{%
  \institution{Laboratoire Informatique d'Avignon -- LIA EA 4128}
  \streetaddress{339 chemin des Meinajaries, Agroparc BP 91228}
  \city{Avignon cedex 9}
  \postcode{84911}
  \country{France}}
\email{xbost@orkis.com}


\begin{abstract}
A \textit{character network} is a graph extracted from a narrative, in which vertices represent characters and edges correspond to interactions between them. A number of narrative-related problems can be addressed automatically through the analysis of character networks, such as summarization, classification, or role detection. Character networks are particularly relevant when considering \textit{works of fictions} (e.g. novels, plays, movies, TV series), as their exploitation allows developing information retrieval and recommendation systems. However, works of fiction possess specific properties making these tasks harder.

This survey aims at presenting and organizing the scientific literature related to the extraction of character networks from works of fiction, as well as their analysis. We first describe the extraction process in a generic way, and explain how its constituting steps are implemented in practice, depending on the medium of the narrative, the goal of the network analysis, and other factors. We then review the descriptive tools used to characterize character networks, with a focus on the way they are interpreted in this context. We illustrate the relevance of character networks by also providing a review of applications derived from their analysis. Finally, we identify the limitations of the existing approaches, and the most promising perspectives.
\end{abstract}

%


\maketitle

\vspace{-0.3cm}
\textbf{Note :} This is a longer and slightly updated version of the official ACM CS article, including the supplementary material. In particular, it contains additional figures extracted from the surveyed articles, and Table~\ref{tab:ArticleList} has been completed with a number of bibliographic references published after the original article.

\setcounter{tocdepth}{1}
\tableofcontents

\section{Introduction}
\label{sec:Introduction}
The first works of fiction possibly date back to as far as the Paleolithic, and have constituted a major part of human culture since then~\cite{Robson2018}. Nowadays, it is estimated that, in average, adults are in contact with fictional stories for $6\%$ of their time awake~\cite{Robson2018}. Besides their artistic and entertainment aspects, fictions are assumed to fulfill various social and psychological purposes, e.g. improvement of communication~\cite{Kruger2003}, development of empathy and collaboration skills~\cite{Robson2016, Smith2017, Smith2017a}, elaboration of social norms~\cite{Robson2016}, proxy to understand the real world~\cite{TandiweMyambo2016}, assessment of social strategies~\cite{Robson2018}, constitution of a collective memory~\cite{Robson2018}. It is therefore natural that they are abundantly studied by academia, and that fiction-related business is a significant part of the economy~\cite{Ryan2017, Batchelor2018, Augereau2018, McNary2019}.

A work of fiction takes the form of a \textit{narrative}, i.e. a report of events telling a story. This report can be conducted through a variety of communication means: text, speech, image, music, gesture, and others, under a variety of forms: fables, tales, novels, plays, but also movies, TV series, video games, cartoons, and comics. The collection of events explicitly reported by the narrative constitutes its \textit{plot}. These events are often ordered to form a chronological and/or causal chain~\cite{Bordwell1993}. By comparison, the \textit{story} contains all the plot events, plus those imagined or inferred by the audience, based on both the plot and a number of contextual factors~\cite{Bordwell1993}. As an illustration, an \textit{ellipsis} consists in removing events from the plot without affecting the story, as the audience will interpolate the missing parts. Put differently, the plot is \textit{what} is told, whereas the narrative is \textit{how} it is told, and the story is what the audience \textit{perceives} of the plot through the narrative.

Historically, narratives have been studied from the Aristotelian perspective, which argues that the most important part of a narrative is its \textit{plot}. However, more modern approaches focus on \textit{characters} instead~\cite{Bamman2014}, and consider that they are the agents that advance the plot through their actions~\cite{Min2016}. This is exemplified by Woloch in the field of \textit{literary analysis}~\cite{Woloch2003}. He defines the notion of \textit{character-space} as the narrative environment of characters in a novel, i.e. their position relative to the other elements of the plot (place, time, other characters). In other words, this is how characters are described in the narrative. The concept of \textit{character-system} extends this notion to the narrative as a whole, and corresponds to the union of all character-spaces. This approach has been noticeably used to study and understand how writers and directors build a narrative.

In addition to the characters themselves, researchers have then started to take into account the way characters \textit{interact}, which is considered as the backbone of the narrative~\cite{Cipresso2016, Prado2016}. In such a context, graphs are a natural modeling paradigm, as they allow representing and studying a system through the interactions of its constituting elements. A \textit{character network} is a graph describing a narrative by representing the characters through its vertices, and the interactions between them through its edges. As we will see later, there are many methods to extract this type of network from some raw data representing the considered work, depending not only on the nature of these data, but also on the information that one wants to encode in the produced network, and on what one wants to do with it eventually. Moretti has shown that such an approach allows to handle more formally Woloch's concepts~\cite{Moretti2011a}. In a graph, the subgraph induced by a vertex and its neighborhood can be seen as a projection of the social aspects of the notion of character-space, whereas the whole graph, which contains all characters and their relations, represents the character-system~\cite{Rochat2014}. Woloch emphasizes the fact that character-spaces must be considered jointly, and this is precisely what graphs, a naturally relational modeling framework, allow.

\subsection{Value of Character Networks}
\label{sec:IntroCharNetLit}
This relevance of graphs for modeling works of fiction is illustrated by the number of articles dealing with character networks in the literature, and the variety of purposes for which they are used. We distinguish three categories of such articles.

First, in the context of \textit{Narrative Analysis}, character networks are generally extracted manually, for a very small number of narratives (typically, a single one). Authors use them in a ``distant reading'' fashion to obtain a simplification of the plot~\cite{Moretti2011a}, characterize the plot structure at various levels~\cite{Rochat2017}, detect relevant patterns and narrative events, identify character roles (e.g. protagonist vs. antagonist) or particularly important characters~\cite{Rochat2014}, assess the validity of literary theories~\cite{Elson2010, Falk2016}, and produce graphical representations~\cite{Oelke2012, Venturini2016, Xanthos2016}. In addition to the description of individual plots, they are also used to compare them, for example among episodes of a given series~\cite{Grandjean2015}, or works belonging to the same genre~\cite{Rochat2017}, period~\cite{Jayannavar2015} or author~\cite{Rieck2016}. In other social science domains, character networks are also used for educational purposes~\cite{Bolioli2013}, and to study certain psychological mechanisms~\cite{Bossaert2013}. 

Second, another category of works also adopts a descriptive and comparative approach, but relying on a \textit{Complex Systems} paradigm. These authors consider that character networks are a type of \textit{Complex Network}, and as such they apply the standard tools developed to analyze them~\cite{Prado2016}, and/or propose new ones~\cite{Choi2007}. The network itself is the object of the study. Like for Narrative Analysis, these works generally consider a few narratives, as the networks are often extracted manually. Many articles of this type compare the topological properties of character networks with those of other kinds of complex networks, e.g. real-world social networks~\cite{MacCarron2012, Alberich2002}, random models~\cite{Choi2007}, or other fictions~\cite{Tan2014a}.

Third, a large number of works originate from the \textit{Artificial Intelligence} domain. They focus more on automating the network extraction process, which requires solving various text, speech, image, and/or video processing problems, depending on the media used. Compared to both other categories, this allows using much larger corpora. These works also consider character networks as models of the plot, and take advantage of this to solve higher-level problems: role detection~\cite{Jung2013a}, genre classification~\cite{Suen2013, Ardanuy2014}, storyline detection~\cite{Weng2007}, story segmentation~\cite{Weng2009}, movie scene segmentation~\cite{Liang2009a}, video abstraction~\cite{Tsai2013}, recommendation systems~\cite{Lee2018}, and others. The results obtained by solving some of these problems can be used to treat higher-level tasks, e.g. the detected roles can help summarize a plot. Certain authors directly relate character networks to novel fields such as \textit{movie information retrieval}~\cite{Park2011}, which consists in obtaining and exploiting valuable information from collections of movies. 

Character networks even reach the mainstream audience, mainly for their relevance as a visualization tool. Numerous non-academic or educational Web pages display character graphs extracted from popular culture works, such as \textit{Star Wars}~\cite{Gabasova2015, Benzi2016, Gabasova2016, Ruths2016}, \textit{Harry Potter}~\cite{Rinehart2017, Kunegis2019, Bratanic2021}, The Witcher~\cite{Janosov2021b}, \textit{Marvel} movies~\cite{Ker2018}, \textit{Love Actually}~\cite{Robinson2015}, \textit{Game of Thrones}~\cite{Cukier2013, Hickman2016, Glander2017}, \textit{Star Trek}~\cite{Pettit2016}, \textit{The Simpsons}~\cite{Pudney2014, Gaerber2021}, \textit{South Park}~\cite{Gaerber2021a}, \textit{The Office}~\cite{SortedHat2018, NotRudeDude2018, Duong2021}, \textit{Seinfeld}~\cite{Stoltzman2016}, \textit{Curb your Enthusiasm}~\cite{Fischer2018}, \textit{Grey's Annatomy}~\cite{Weissman2011, Lind2012}, and \textit{Friends}~\cite{Blatt2014, Albright2015, Schoch2015, Simchoni2017}; as well as from classics like \textit{Sherlock Holmes}~\cite{Boudourides2015}, European drama~\cite{Fischer2018a, Young2019} and Shakespeare's plays~\cite{Pierson2014, Boudourides2016, Grandjean2015a, Iuhfd2018}.

\subsection{Specific Features of Fiction Works}
\label{sec:IntroSpecifFiction}
The extraction and use of character networks concern all types of works, including non-fictional ones. For instance, certain authors focus on biographies~\cite{Camp2011}, professional meetings~\cite{Garg2008}, journal articles~\cite{Sudhahar2015}, and broadcast news~\cite{Vinciarelli2007}. So in theory, it is possible to apply these methods developed for non-fiction to deal with fiction. However, in practice this does not necessary leads to good results, because works of fiction possess some specific features, absent from non-fiction. These can result in specific issues, whose resolution requires suitable processing, but they can also correspond to additional information one can leverage through appropriate methods to improve performance. We give examples of both aspects in the following.


First, there can be differences in the \textit{structure} of the narrative. For instance, plays and TV or movie scripts are \textit{semi-structured}, in the sense that scenes are explicitly bounded and speakers are explicitly named. This feature can be harnessed during network extraction~\cite{Nalisnick2013}, and does not appear in non-fictional narratives (or most other types of fictional ones, for that matters). In video-based narratives, the set of camera and editing rules, conventions and guidelines, sometimes metaphorically called \textit{film grammar}~\cite{Bost2018a}, is quite different in fiction and non-fiction works. For instance, the so-called \textit{180 degree} rule states that during a scene, the relative positions of the characters on the screen must not change. The \textit{shot alternation} (or \textit{shot/countershot}) rule is particularly used during conversations: it specifies that consecutive shots alternatively show the involved characters. Yeh \textit{et al}.~\cite{Yeh2012} leverage both of them to improve character detection in movies. Comics and animated films are apart, as in their cases, the medium itself is unlike anything related to real-life. Characters can be highly deformed human, non-anthropomorphic beings, or even inanimate objects, which makes ineffective the methods designed to detect faces or persons in photographs~\cite{Augereau2018} or live action movies~\cite{Valls-Vargas2014a}. Moreover, the structure of comics narratives is unique, in the sense that they include information under a variety of forms encoded in both text (captions, speech balloons, onomatopoeia) and drawings (pose, graphical conventions) not found in other media, and whose extraction requires specific methods~\cite{Augereau2018}.

Second, there are generally significant \textit{stylistic} differences. In texts, literary prose is considered as more complex than journalistic prose~\cite{Elson2012}, and even more so when the work is older~\cite{Grayson2016}. One of the effects of style is actually to give a unique identity to the work, and to distinguish it from both non-fictional works and other fictions~\cite{Bornet2017}. Stylistic differences are so marked that it is possible to assign works automatically to their creators~\cite{Ardanuy2014}. They significantly affect the performance of generic methods on a variety of NLP tasks: plot modeling, character detection and story generation~\cite{Elsner2012}, text summarization~\cite{Kazantseva2010}, named entity recognition (NER)~\cite{Ardanuy2015, Vala2015}, co-reference resolution~\cite{Krug2015, Vala2015}. There are a number of reasons for this drop in performance. For instance, for character detection in novels~\cite{Elsner2012}: many characters are relatives and share the same last name; they bear nicknames; some fictional characters are inanimate objects in real life; writers use specific honorifics corresponding to complex, possibly outdated and even imaginary social conventions; and they craft names in order to convey certain meaning or function. In fact, this task is difficult even for humans, enough to requires a specific annotation process~\cite{Vala2016}. For co-reference resolution, the problem comes from longer sentences, more frequent use of pronouns and direct speech, more numerous and shorter co-reference chains~\cite{Krug2015}.
Similarly to text, certain characteristics of fiction works make generic audiovisual processing tools inefficient~\cite{Yeh2012}. For instance, movie directors use a variety of complex, possibly genre-related, editing techniques~\cite{Li2004a}. At a lower level, the same face can appear under a variety of lights, colors, angles, expression, and other deformations, which do not correspond at all to the very controlled conditions under which non-fictional works are recorded (e.g. news or talk show). Speech-wise, conversations are subject to background noise or music, involve more participants, and a way of speaking that is unlike that found in other forms of audiovisual productions~\cite{Bredin2016}.

Third, fictions often are \textit{closed-worlds}, in the sense that they are self-contained and involve recurring entities, possibly with made up names. Generic tools ignore this characteristic, which sometimes can help handling certain tasks~\cite{Jovanovic2014} such as alias resolution (finding the different variants of a character's name). On the contrary, most generic tools rely either on a training corpus or on external databases: in both cases, the described entities are likely to be completely different from those occurring in a fiction. For example, a standard approach when performing face-matching in news is to leverage pictures from press articles and their captions: this cannot be done for movies containing fictional characters~\cite{Zhang2009e}. If anything, this method is more likely to return the actor's rather than the character's name. Similarly, many NER systems rely on gazetteers or services such as DBPedia, Wikidata, or YAGO~\cite{Bornet2017}, which are likely to include only the main characters (if any) of the considered fiction. For instance, none of the proper nouns used in Tolkien's \textit{The Lord of the Rings} would be present in a standard list of first names or places.

Fourth, there is also a difference in the way characters interact in works of fiction, compared to real life~\cite{Rochat2014}. A real-world social network represents an auto-organized system, whose structure emerges from the interactions between some agents acting according to their own agenda. By comparison, the writer or director controls all actions of fictional characters, and arrange them according to a plot. Put differently, real-world networks are the result of microscopic processes, whereas fictional ones are caused by a macroscopic process~\cite{Rochat2014}. There is no reason to suppose that the writer tries to mimic actual social relationships when producing the work of fiction. As we will see later, studies show that this is generally not the case, as numerous character networks extracted from fictions do not exhibit realistic topological properties. This is because other constraints come into play, such as the intelligibility and appeal of the plot. Analyzing a different structure is likely to require specific tools, compared to real-world networks (including non-fictional character networks).

\subsection{Perimeter and Organization of the Survey}
\label{sec:IntroPeriOrga}
The first publications related to fictional character networks date back to the early 2000s, e.g.~\cite{Alberich2002, Stiller2003}. As explained before, both extracting and leveraging these networks involve solving specific problems. However, there is no synthetic review describing the solutions proposed in the literature. With this survey, we want to fill this gap. Not only do we consider articles directly related to fictional character network extraction and/or usage, but also articles focusing on certain specific steps of this process (without necessarily trying to deal with such networks). Note that certain authors extract other types of graphs (non-character-based) from works of fiction, such as scene transition graphs~\cite{Yeung1996}, or narrative structure graphs~\cite{Jung2004}, but we do not include them in this review.

Our contributions include the description of the tools currently available and the approaches currently adopted to detect characters and their interactions from all forms of narratives, as well as the methods leveraging them to build character networks. We also contribute by identifying open problems at all levels of the extraction and analysis processes, and proposing perspectives to solve them.

Terminology-wise, we need to distinguish scientific work from work of fiction. For this purpose, we will use the words \textit{author} and \textit{article} to refer to scientific authors and their work, whereas \textit{writer} (or \textit{director}, \textit{playwright}, or any medium-specific term) and simply \textit{work} will refer to artistic authors and their works of fiction.

The rest of the survey is organized in two parts. We first focus on the process of extracting a character network from a work of fiction. We introduce it in a generic way (Section~\ref{sec:Overview}), before describing its three main steps: the identification of characters and their occurrences (Section~\ref{sec:CharacterIdentification}), the detection of their interactions over the narrative (Section~\ref{sec:InteractionDetection}), and the extraction of the graph itself (Section~\ref{sec:GraphExtraction}). In the second part, we focus on how to leverage character networks. We first discuss the descriptive tools used in the literature to characterize them, and then examine a selection of more elaborate tools developed to solve specific problems (Section~\ref{sec:AnalApps}). Finally, we identify the current main issues of the field, and conclude with some some perspectives (Section~\ref{sec:Perspectives}).

\section{Overview of the Extraction Process}
\label{sec:Overview}
The process of extracting character networks from works of fiction depends a lot on the form of the considered narrative, e.g. novels are not treated like movies. In order to give the reader a general overview, in this section we make abstraction of these differences and present this process in a very generic way. In the rest of the survey, on the contrary, we focus on their differences.

\begin{figure}[htb!]
	\centering
	\includegraphics[width=\textwidth]{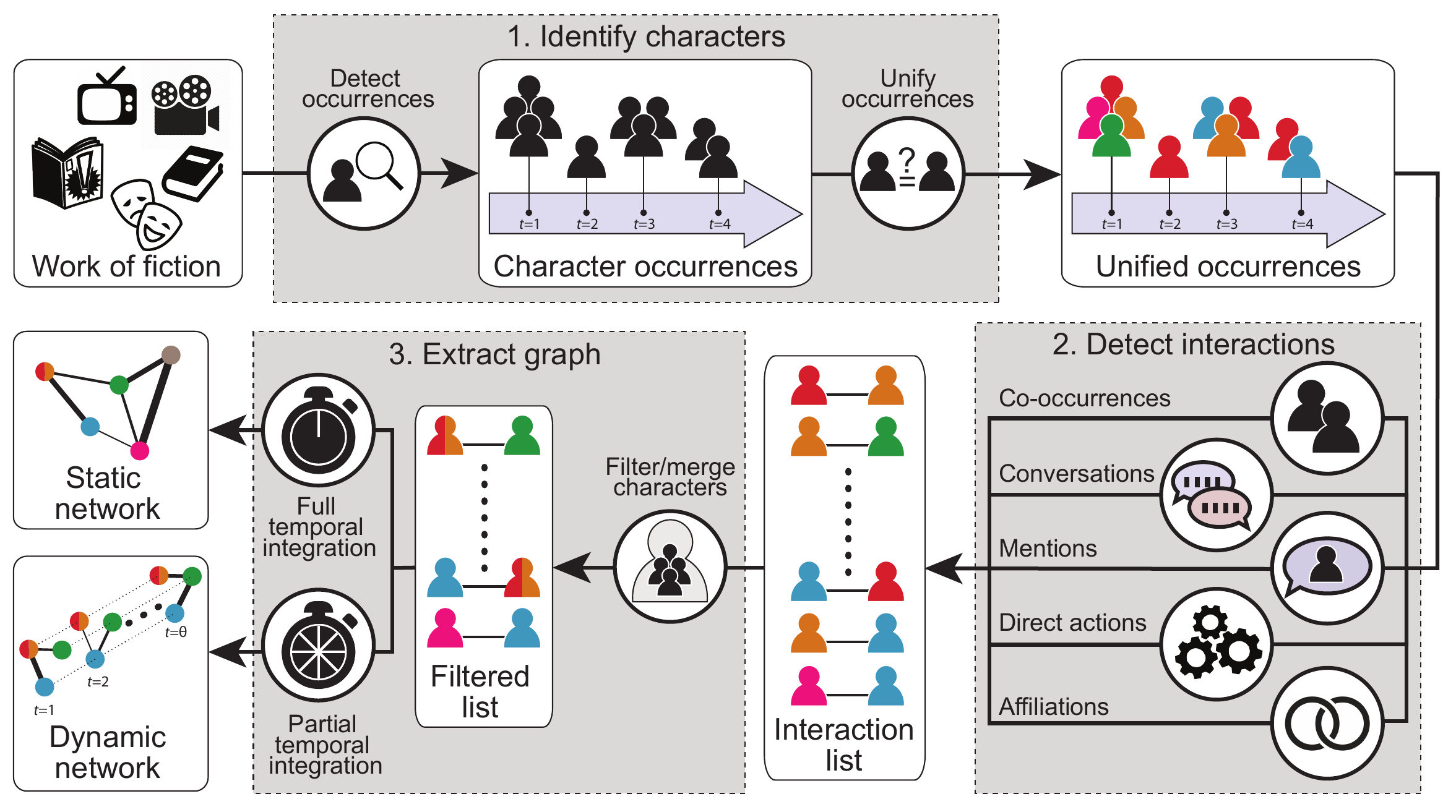}
	\caption{Overview of the generic character network extraction process. Figure available at \href{https://doi.org/10.6084/m9.figshare.7993040}{10.6084/m9.figshare.7993040} under CC-BY license.}
	\label{fig:process}
\end{figure}

We consider that this process consists of three main steps, represented in Figure~\ref{fig:process}: 1) the identification of characters; 2) of their interactions; and 3) the extraction of the proper graph. Each of them can be conducted in a number of ways, depending not only on the nature of the considered narrative, but also on the planned usage of the character network, and on certain methodological choices.

The first step is the most dependent on the form of the narrative, as it starts with the raw material, i.e. the work of fiction itself. We distinguish two substeps. The first is to detect occurrences of characters in the narrative, for instance looking for people names in a novel, or looking for faces in a movie. The second is to unify these occurrences, i.e. to determine which ones correspond to the same character. In a text, the same character can appear under different names, whereas in a movie, the same face can be shown under a variety of scales, colors, lights, and angles. The output of this step takes the form of a chronological sequence of unified character occurrences.

The second step consists in detecting interactions between characters. Note that it is sometimes more efficient or convenient to conduct parts of this process during the first step, but this is generally not the case. We identify five different definitions for the notion of interaction. Many authors consider that a simple \textit{co-occurrence} between two characters is enough to infer an interaction between them. Others prefer to identify \textit{explicit} interactions, which is generally a more difficult process. One way of doing this is to take into account \textit{conversations}, and to consider that two characters interact when one talks to the other. With certain forms of narrative such as plays, in which speakers are given, this task is relatively straightforward. An alternative is to focus on the content of the conversations, and to leverage \textit{mentions}, i.e. situations where one character talks about the other. Some authors consider all sorts of actions one character can perform on the other (besides conversing). This is particularly the case with novels, a form of narrative in which such actions are explicitly described. Finally, certain authors do not focus on actions and prefer to use \textit{affiliations}, i.e. explicit or inferred social relationships such as being married, being relatives, or working together. Note that it is possible to combine these definitions of the notion of interaction, for instance by looking for both co-occurrences and conversations.

The output of the second step is a chronological sequence of interactions between characters. The third step is therefore relatively generic, as it relies only on this list and is thus independent of the nature of the original narrative. We distinguish two substeps. The first, which is optional, consists in simplifying this sequence by \textit{filtering and/or merging} some of the characters under certain conditions. For example, when considering co-occurrences, some authors merge characters that always appear together: this allows simplifying the network. The second substep defines how the graph is extracted through \textit{temporal integration}, i.e. the aggregation of the previously identified interactions. There are a number of approaches for this purpose, which we separate into two groups: those performing a \textit{full} integration and therefore leading to a \textit{static} network, and those performing only a \textit{partial} integration, and producing a \textit{dynamic} network.

\section{Character Identification}
\label{sec:CharacterIdentification}
\textit{Character identification} consists in detecting \textit{which} characters appear in the considered narrative, and \textit{when} exactly they appear in this narrative. As mentioned before, the form under which characters appear in the narrative varies much depending on the medium. In the case of text, they can be represented in three ways~\cite{Elson2012, Trovati2014}: \textit{proper nouns} (e.g. ``Sherlock Holmes''), \textit{pronouns} (e.g. ``He''), and \textit{nominals}, i.e. anaphoric noun phrases referring to characters (e.g. ``The consulting detective''). For videos, they either can appear onscreen, or be mentioned in the audio stream (again as a proper noun, pronoun, or nominal). In comics, characters can either appear as drawings or be mentioned in the text (again, under the same three forms).

Automating character identification is quite challenging, which explains why many non-specialists prefer to perform this task manually. We describe this manual approach separately, as there are various ways of proceeding (Section~\ref{sec:CharIdManual}). But our focus is rather on automatic approaches, for which we distinguish two subproblems: first, finding character occurrences in the narrative (Section~\ref{sec:CharIdOccurrences}); and second, determining which of these occurrences represent the same character (Section~\ref{sec:CharIdUnification}). For both subproblems, there are two very different categories of approaches, which depend on whether characters are represented in a textual vs. audiovisual way. Note that this dichotomy does not necessarily matches the type of narrative, for instance a movie can be treated as a video or as a text (through its transcript or script).

Besides strict character identification, certain authors perform some additional processing in order to extract individual attributes to describe characters (e.g. age, gender...), and/or to filter them. We discuss this in Section~\ref{sec:CharIdAddProc}.

\subsection{Manual Approaches}
\label{sec:CharIdManual}
Some authors adopt a fully manual approach to detect character occurrences, in which case there is no need to distinguish occurrence detection from occurrence unification, as both tasks are conducted at once.

\subsubsection{Direct Annotation}
The most widespread method is direct annotation, which consists for the authors in annotating by themselves the narrative they want to study, e.g.~\cite{Bossaert2013, Kydros2015a} for novels,~\cite{MacCarron2012, Miranda2013} for myths,~\cite{Moretti2011a, Sparavigna2013} for plays,~\cite{Tran2015, Cipresso2016} for movies, and~\cite{Weng2007, Bazzan2018} for TV series. Opting for such a manual approach can be due to technical limitations, e.g. the authors do not have access to efficient automatic methods~\cite{Moretti2011a}. However, it can also be a methodological choice, e.g. to better focus on the assessment of the other steps of the network extraction and/or analysis process~\cite{Agarwal2012, Agarwal2013, Bazzan2018}. 

Though rarely mentioned explicitly~\cite{Grandjean2015}, the authors that adopt manual approaches for later extracting \textit{co-occurrence} networks (cf. Section~\ref{sec:InterDetCooc}) often ignore mentions of characters which are just named by others, but do not physically participate in the action, as in~\cite{Prado2016}. It is also not clear exactly how annotators deal with occurrence unification. However, context generally suggests they perform this task, and do so manually (e.g.~\cite{Moretti2011a}), as the extra cost is marginal (cognitively speaking).

\subsubsection{Character Index}
Instead of doing the annotation work themselves, certain authors take advantage of predefined resources, which are also manually constituted. For certain classic novels, literary experts have constituted so-called character indexes, indicating at which point of the plot each character appears. This is for instance the case for Rousseau's \textit{Les confessions} in~\cite{Rochat2014, Rochat2014a, Rochat2015}, and Park's \textit{Toji} in~\cite{Park2013, Park2013a}. Several authors proceed similarly for comics~\cite{Alberich2002, Gleiser2007}, as they study the \textit{Marvel universe} by taking advantage of the \textit{Marvel Chronology Project}\footnote{\url{http://www.chronologyproject.com/}}, an online database listing the occurrences of all significant Marvel characters. 

Even if such indices are elaborated by experts of the considered work of fiction, it is difficult, or even impossible to assess their reliability. Moreover, it is important to notice that they impose a predefined level of precision on the rest of the extraction process. For instance, character occurrences are expressed in terms of pages for \textit{Les confessions}, and comic issues for the \textit{Marvel universe}. This lack of control can be considered as a limitation, since the level of precision affects certain subsequent extraction steps (e.g. it constrains the selection of a narrative unit when extracting co-occurrence networks, cf. Section~\ref{sec:InterDetCooc}). 

Like for direct annotation, the elaboration of indices is likely to include some form of character occurrence unification. However, it is difficult to determine whether it is the case for a given index, and according to which procedure exactly. Indeed, this task is conducted by the writers of the index, not those of the study that take advantage of this index for network extraction. Moreover, the index is often not properly documented regarding this aspect. Only a very few articles mention occurrence unification, but they nevertheless reveal some differences in the way they handle this task. For instance, in~\cite{Rochat2014, Rochat2014a}, the character index considers all variants of the character names, but not pronominal references, whereas the index used in~\cite{Rochat2015} includes both.

\subsubsection{Crowdsourcing}
In practice, it is hard to handle more than a few works of fiction when using either of the previous approaches (direct annotation vs. predefined character index). A workaround is to turn to \textit{crowdsourcing}, as Rochat \& Kaplan do in~\cite{Rochat2017} to constitute their own indices. Interestingly, this study is also characterized by its multimedia nature, as the authors consider a corpus of science-fiction works including novels, comics, movies, TV series, and video games. The manual approach has the advantage of allowing a more uniform network extraction process over the variety of considered media, and therefore makes it possible to compare them. They can also select their own level of precision during the elaboration of the indices: pages for novels and comics, and intervals of one minute for movies and TV series. For video games, they experiment with three different base materials: walk-throughs (i.e. texts explaining how to finish the game), which are treated like novels; cinematic scenes, which are treated like other videos (movies and TV series), and transcriptions of these scenes, which are treated like scripts.

\subsection{Detection of Character Occurrences}
\label{sec:CharIdOccurrences}
We now switch to approaches that are at least partially automated. As mentioned before, the process of character identification largely differs depending on whether the narrative is visual (Section~\ref{sec:CharIdOccVideo}) or textual, which is why we separate them in our description. Moreover, certain texts such as plays and scripts possess a structure which can be leveraged for character occurrence detection, so we distinguish such semi-structured text (Section~\ref{sec:CharIdOccSemiStruc}) from free text (Section~\ref{sec:CharIdOccFreeText}).

\subsubsection{Free Text}
\label{sec:CharIdOccFreeText}
As mentioned before, a character can appear under three forms in text: proper noun, nominal, and pronoun. The methods used in the literature all handle the first form, but not necessarily the two others, as detecting them is generally a much harder problem, and they are often not considered as informative. A simple way to detect character names is to use a predefined list of these names and proceed through exact matching~\cite{Hutchinson2012, Beveridge2016, Bonato2016, Lee2016d, Grener2017}. Such a list is generally constituted manually, either by the authors themselves or through an external source such as the Wikipedia page of the considered novel. Constituting it is not a trivial task, as characters can be referred to through a variety of \textit{aliases}, i.e. variations of their name. For instance, Sherlock Holmes can also be called simply ``Sherlock'', ``Holmes'', or ``Mr. Holmes''. Some authors perform a manual verification after the exact matching step~\cite{Grener2017}.

Detecting character names can be viewed as a specific version of the Named Entity Recognition (NER) problem. NER consists in finding expressions in the text corresponding to proper nouns~\cite{Nadeau2007}, and to identify their category (e.g. \textit{Location}, \textit{Person}, \textit{Organization}). A number of authors apply off-the-shelf NER tools to novels, e.g.~\cite{Elson2010, Agarwal2013a, Hettinger2015, Srivastava2016, Chaturvedi2017}, and then only retain the \textit{Person} entities. Incidentally, those are generally much more frequent than other proper nouns in literary texts~\cite{Dalen-Oskam2014}. Dekker \textit{et al}. perform an empirical comparison of four such tools in the context of character extraction based on novels~\cite{Dekker2019}. It is possible for a NER tool to assign different categories to distinct instances of the \textit{same} string, because of contextual differences. For instance, ``France'' is a country, but also a first name. However, in the context of novels, such a situation is likely an error: it is generally agreed upon that a novel is a small, self-contained world~\cite{Ardanuy2014}, and that the writer would not confuse the reader by using the same name to denote entities of different types (such as a person and a place)~\cite{Ardanuy2015}. A straightforward solution to the multiple category issue is then to keep the majority category, as in~\cite{Ardanuy2014, Ardanuy2015}. Valls-Vargas \textit{et al}. specifically train a classifier to distinguish characters from other types of mentions~\cite{Valls-Vargas2014a, Valls-Vargas2014a}. A very few authors consider other categories in addition to \textit{Person}, such as \textit{Location}~\cite{Lee2012f, Marazzato2014} and \textit{Organization}~\cite{Amancio2015b}, which eventually results in a network with multiple types of vertices.

Fiction texts have certain characteristics which are leveraged by some authors, either to perform some post-processing after having applied an off-the-shelf NER tool, in order to find missed mentions and/or discard incorrectly detected ones, or to design new fiction-specific NER tools. A simple method is to remove infrequent names, as they are likely to be errors. For instance, Elsner~\cite{Elsner2012} and Sack~\cite{Sack2012} remove names appearing fewer than five times. Some authors also perform a manual verification to fix the errors of the automatic tools~\cite{Sack2012}. Some automatic approaches use honorifics (titles such as ``Sir'' or ``Madam''), generally by relying on manually predefined resources. One can take advantage of a list of honorifics to detect them in the text and check the surrounding text for character names~\cite{Ardanuy2014, Ardanuy2015}, or look for a set of patterns describing the various possible combinations of honorifics, first names and last names~\cite{Trovati2014}. Some approaches proceed similarly with action verbs, as only characters are likely to be their subjects. For instance, Ardanuy \& Sporleder~\cite{Ardanuy2014} use a manually constituted list of speech verbs (e.g. to say, to discuss), while Goh \textit{et al}. leverage WordNet to focus on \textit{human} action verbs only~\cite{Goh2012}. Zhang \textit{et al}. also use the grammatical structure of the sentence through part-of-speech (PoS) tagging~\cite{Zhang2003a}. Finally, some approaches consist in looking for relations of possession (through genitive marks, such as ``'s'' in English)~\cite{Trovati2014}, as only characters are supposed to own things. These approaches are more robust than generic NER tools, in the sense that they allow detecting non-human characters behaving as humans~\cite{Karsdorp2012}. In~\cite{Bornet2017}, Bornet \& Kaplan propose an ensemble-based method associating most of the aspects listed above, for French texts. It relies on the vote-based combination of the outputs of six basic classifiers. Each one focuses on the detection of a specific type of clues: presence of honorifics, position in the sentence, semantics of neighboring words, grammatical structure, occurrence in external resources, and presence of nearby explicit quotes.

These last methods are likely to return not only proper nouns, but also nominals (anaphoric noun phrases referring to characters). Some authors propose methods specifically designed to detect these nominals, generally through regular expression matching. Elson \textit{et al}.~\cite{Elson2010a} look for structures of the form: a \textit{determiner} (article, possessive, number...), an optional \textit{modifier} (e.g. an adjective), and a \textit{head noun} (not necessarily a proper noun). They manually compile lists of determiners and head nouns based on their corpus and external linguistic resources such as WordNet. They use them to detect the determiner and head noun first, and consider the text located in between as the modifier. The task of detecting pronouns is more or less difficult depending on the considered language. For English, exact matching based on a manually defined list is a simple and efficient approach~\cite{Elson2012}.

\subsubsection{Semi-Structured Text}
\label{sec:CharIdOccSemiStruc}
A number of narratives can take the form of a script: theatric plays, movies, TV series. A script is essentially a conversation-based text, with specific structure and formatting described as \textit{semi-regular} by certain authors~\cite{Agarwal2014b}: scene boundaries are clearly indicated, the characters involved in a scene are explicitly listed at its beginning in uppercase, and the name of the character speaking a line is indicated right before it, also in uppercase. When the script is properly formatted and structured, it is relatively straightforward to extract this information automatically. Authors have proposed methods based on exact string matching~\cite{Mutton2004, Ding2010, Jung2013a}, regular expression~\cite{Xanthos2016, Suen2013}, or a custom parser~\cite{Rieck2016}. 

\begin{figure}[htb!]
	\centering
	\begin{tabular}{c@{} c@{} c}
    	\includegraphics[height=2.9cm]{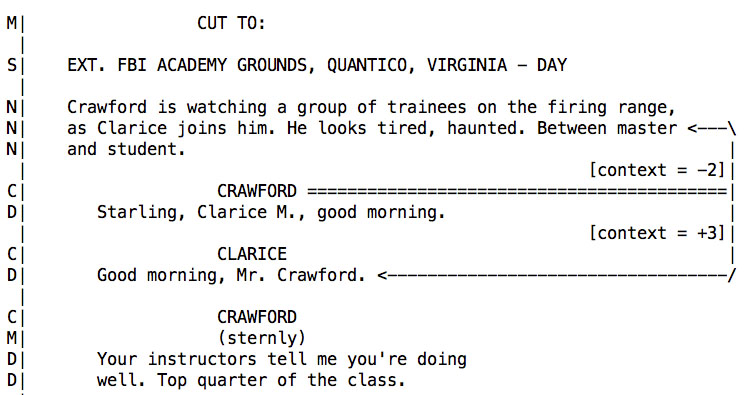} & 
    	\includegraphics[height=2.9cm]{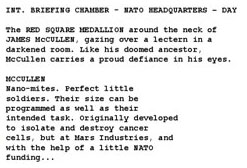} &
    	\includegraphics[height=2.9cm]{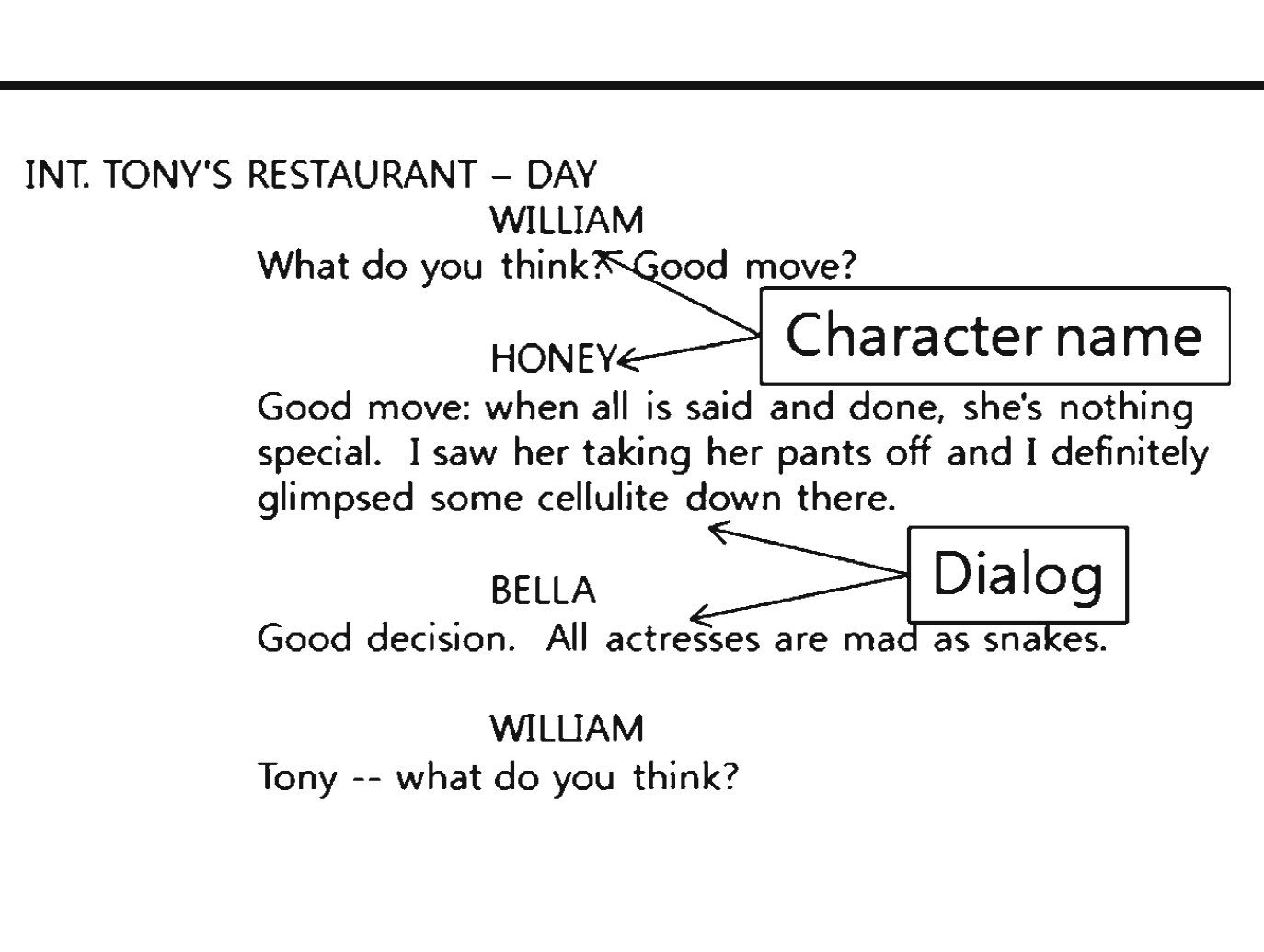} \\
    	a) & b) & c)\\
	\end{tabular}
	\caption{Examples of semi-structured text: a) from~\cite{Agarwal2014b}, b) from~\cite{Ding2010} c) from~\cite{Jung2013a}.}
	\label{fig:semistruct}
\end{figure}

However, this structure and formatting is not a proper standard, and can vary from one script to the other~\cite{Makris2016}, as illustrated in Figure~\ref{fig:semistruct}. It is even possible to find inconsistencies in the same script. Machine learning can help solving this issue. Agarwal \textit{et al}.~\cite{Agarwal2014b} propose a method to identify which parts of the script correspond to character lists, dialogues, speaker names, scene boundaries, and scene instructions. Tan \textit{et al}.~\cite{Tan2014a} take advantage of this type of decomposition to only focus on character mentions associated to utterances, in order to ignore passive characters which are present in a scene but do not intervene. 

Discrepancies can also appear in the speaker names. In this case, a simple approach consists in using an \textit{a priori} list of the characters involved in the script~\cite{Xanthos2016, Tan2014a, Krishnan2015}, with their associated aliases. This list is generally constituted manually, or by taking advantage of publicly available resources (generally also constituted manually), such as the Wikipedia page of the considered work of fiction. Again, machine learning based method can be more robust than such simple matching-based approaches. Makris \& Vikatos~\cite{Makris2016} take advantage of the Wikipedia pages of the movies they study to train a classifier into identifying which character speaks which line. Certain authors directly apply off-the-shelf NER tools~\cite{Zhang2009e, Gorinski2015} to detect speaker names.

Identifying the characters involved or speaking during a scene is enough when extracting co-occurrence (cf. Section~\ref{sec:InterDetCooc}) or conversational networks (Section~\ref{sec:InterDetConv}), respectively. Dealing with other types of interactions between characters requires identifying character mentions in the rest of the text~\cite{Mutton2004, Kwon2017}: not only explicitly identified speakers, but also scene metadata, spoken lines, and/or stage directions. In this case, one can apply similar approaches to those already described for free text. For instance, Krishnan \& Eisenstein~\cite{Krishnan2015} train a classifier to detect addressees mentioned in utterances, in order to determine who speaks to whom exactly. They detect not only proper nouns, but also nominals, including titles and placeholder names (e.g. ``bro'', ``dude'', ``sir'').

Choi \textit{et al}.~\cite{Choi2007} apply an approach relatively similar to those designed for scripts, but rather to a biographic dictionary of fictional characters: the classic \textit{Dictionary of Greek and Roman mythology} by Grant \& Hazel. This work takes the form of a series of entries, each one summarizing the biography of a given character. Choi \textit{et al}. constitute the character list by parsing the entry keys, and identify their occurrences in the entry bodies through exact matching.

\subsubsection{Visual Narratives}
\label{sec:CharIdOccVideo}
In audiovisual narratives, detecting character occurrences amounts to solving several distinct but related problems, depending on whether one focuses on the video or the audio stream. In videos, these are face detection and face tracking~\cite{Yeh2012, Yeh2014}. 

\begin{figure}[htb!]
	\centering
	\begin{tabular}{c c c}
    	\includegraphics[height=4cm]{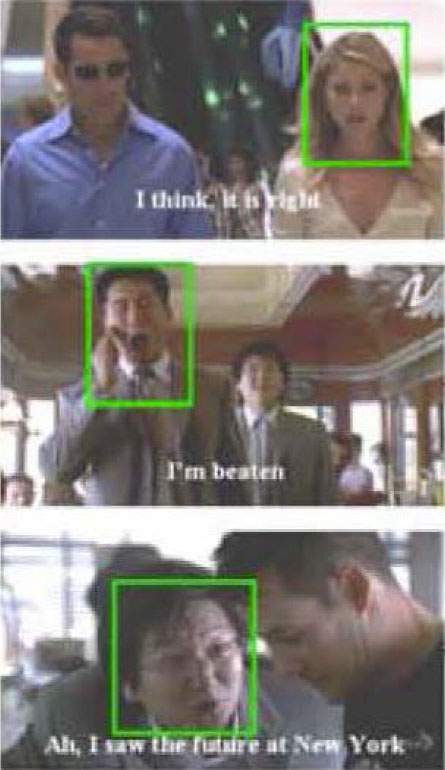} & 
    	\includegraphics[height=4cm]{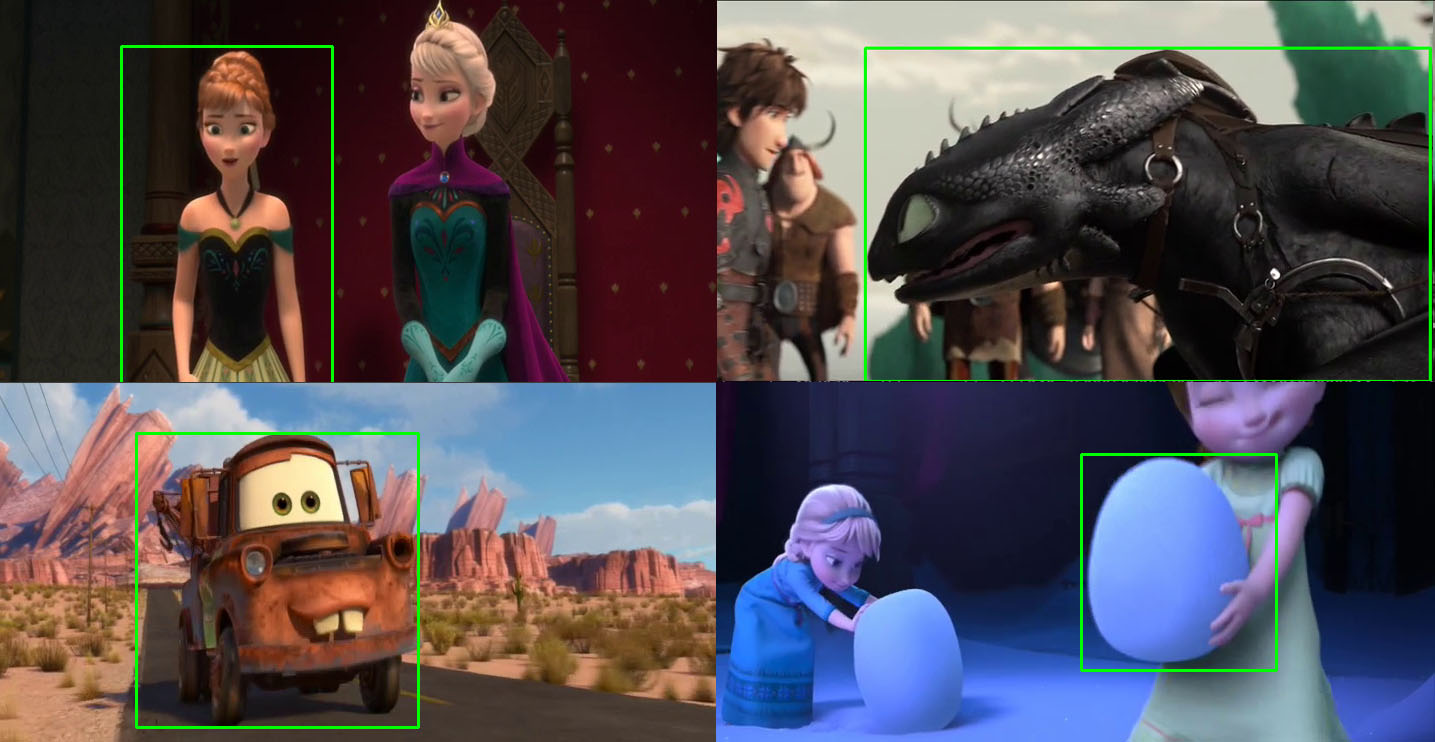} &
    	\includegraphics[height=4cm]{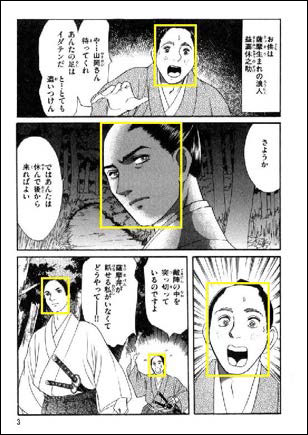} \\
    	a) & b) & c)\\
	\end{tabular}
	\caption{Examples of face and body detection in: a) live-action movies~\cite{Park2009}; b) animated movies~\cite{Somandepalli2018}; c) comics~\cite{Chu2017a}.}
	\label{fig:faces}
\end{figure}

\paragraph{Video Streams.}
\textit{Face detection} consists in identifying which parts of a still image correspond to faces, as illustrated in Figure~\ref{fig:faces}. See~\cite{Kumar2018} for a recent review of the field. Jung \textit{et al}. note that current face detection methods are efficient mainly on front views of the faces~\cite{Jung2013a}: this is a strong limitation in our context, as a character can be filmed under a variety of angles. Weng \textit{et al}. also observe that current automatic methods do not reach satisfying enough performances, which is why they first proceed manually~\cite{Weng2007}. However, they later train their own model to obtain acceptable performances on their dataset~\cite{Weng2007a, Weng2009}. They experimentally find the community structure of the extracted networks to be relatively robust to face detection errors. A number of authors proceed automatically using off-the-shelf tools~\cite{Liang2009a, Park2009, Yeh2012, Tsai2013, Nan2015}.

The face detection problem is relatively similar when dealing with comics, except that the images are drawings (cf. Figure~\ref{fig:faces}). This implies a number of additional difficulties: the characters can be very deformed, non-human, or even non-anthropomorphic. Moreover, the structural lines defining the characters and objects composing the panels are mixed with textures, screentones, and stylistic elements. For these reasons, methods designed to handle photographs generally perform poorly on comics, which require specific approaches~\cite{Chu2017a, Stricker2018}. One such approach consists in adapting features or models originally developed for photographs, e.g. facial landmarks (points corresponding to specific parts of the face such as eyes or mouth) in~\cite{Stricker2018}. Takayama \textit{et al}.~\cite{Takayama2012} handcraft features to fit the specific case of mangas (skin and hair colors, jaw line shape, symmetry). The fact that a pattern appears frequently in the narrative is also used as a hint to distinguish characters from other objects~\cite{Ho2013}. More recent articles focus on training Deep Neural Networks~\cite{Chu2017a}, but there is not enough publicly annotated data yet to reach the full potential of such approaches~\cite{Augereau2018}. Finally, it is worth noticing that before being applied, many face detection approaches require some preprocessing, in particular detecting panel bounds and speech bubbles~\cite{Chu2017a}, which in turn constitute specific problems~\cite{Stommel2012, Rigaud2015a}.

\textit{Face tracking} builds upon face detection, and aims at identifying chronological sequences of faces corresponding to the same person in a video. These sequences are called \textit{face tracks}, and can be considered as character occurrences in videos. Performing face tracking requires accounting for changes in pose, scale, rotation, expression, color, light, angle, and blur. Most authors use off-the-shelf tools~\cite{Zhang2009e, Liang2009a}, usually based on some form of similarity-based classification of the detected faces.

Somandepalli \textit{et al}.~\cite{Somandepalli2018} detect characters in \textit{animated} movies. This task proves to be much more difficult than with live-action videos, as the design of the characters can vary widely, including non-human, and even non-anthropomorphic shapes (cf. Figure~\ref{fig:faces}). The authors first list character candidates by detecting salient objects in a generic way, before taking advantage of graphical and saliency features to discard irrelevant ones. They then use an off-the-shelf tool to track deformable objects.

\paragraph{Audio Streams.}
When using the audio stream, detecting character occurrences amounts to solving the \textit{speaker segmentation} (or \textit{speaker change detection}) problem~\cite{Moattar2012}. It consists in partitioning the audio stream into segments associated to unique speakers. Put differently, one wants to find the moments corresponding to switches between speakers. This task is sometimes performed simultaneously with that of \textit{segment clustering}, which consists in grouping the segments spoken by the same person. Performing both these tasks sequentially or simultaneously is called \textit{speaker diarization}. However, we treat this later in Section~\ref{sec:CharIdUnifVideo}, and focus here only on speaker segmentation. 

Certain existing systems work well in controlled environments, but this performance strongly drops when applied to fiction works, e.g. movie trailers and cartoons~\cite{Clement2011}, and TV series~\cite{Ercolessi2013}. This is mainly due to the presence of background music and sound effects, the higher number of speakers~\cite{Bost2016}, the spontaneous (though acted) nature of the exchanges, and the shorter speech turns~\cite{Bredin2016}. Results improve when using methods specifically designed or trained on fictional audiovisual narratives, e.g.~\cite{Bassiou2010} for movies,~\cite{Bost2014} for TV series. It is worth noting that compared to video-based methods, audio-only tools do not allow identifying characters that appear in a scene \textit{without speaking}~\cite{Jung2013a}.

\paragraph{Multimodal Approaches.}
Certain approaches try to combine several types of information, be it video-, audio-, or language-based. A few multimodal methods able to perform speaker segmentation using both audio and video have been proposed, but we describe them later as they all additionally solve speaker clustering (and therefore speaker diarization). Scripts can be used to distinguish speakers or on-screen characters as in~\cite{Park2009, Park2011, Jung2013a}. A script is not time-stamped, so this approach requires first solving an additional problem called \textit{script alignment}, which consists in determining the exact time at which each line contained in the script occurs in the video. In~\cite{Chen2016h}, Chen \textit{et al}. use transcripts extracted from TV series. They apply off-the-shelf tools to detect all three forms of textual character mentions (proper nouns, nominals, pronouns).

\subsection{Unification of Character Occurrences}
\label{sec:CharIdUnification}
The second step of character identification is occurrence unification, which consists in determining, for each detected occurrence, to which character it corresponds. Like for occurrence detection, the methods proposed for this purpose vary much depending on the medium of the narrative: textual (Section~\ref{sec:CharIdUnifText}) vs. visual (Section~\ref{sec:CharIdUnifVideo}). However, this time there is no distinction to make between free and semi-structured text, as all the additional information of the latter has already been used during the detection step.

As the literature shows, character unification is often not performed at all. There are mainly two reasons for this: this task is generally harder than occurrence detection (especially in text); and in certain situations it is simply unnecessary. For instance, when extracting a purely conversational network (Section~\ref{sec:InterDetConv}) from a clean script, the speakers explicitly named in the script are enough, e.g.~\cite{Moretti2011a, Ding2010, Park2009}. Or when extracting a network from a novel by considering chapter co-occurrences (Section~\ref{sec:InterDetCooc})~\cite{Knuth1993}: it is likely that all characters participating in the chapter will be explicitly named. Some authors show this empirically, e.g. Seo \textit{et al}.~\cite{Seo2014} argue that restricting their analysis to explicit proper nouns (and thus, ignoring pronouns and noun phrases) is enough to perform their targeted tasks (character ranking and edge prediction) without significant performance loss.

\subsubsection{Textual Narratives}
\label{sec:CharIdUnifText}
As mentioned before, characters occurrences appear under three forms in text: proper nouns, nominals, and pronouns. Unifying these occurrences can be considered as a specific version of the \textit{coreference resolution} problem, which consists in identifying sequences of expressions, called \textit{coreference chains}, that represent the same concept (see~\cite{Sukthanker2018} for a recent review). Generic tools exist to solve this problem, but their performance does not necessarily translate to fiction works~\cite{Vala2015}. In particular, they tend to overlook minor characters such as those mentioned only through nominals (e.g. ``the detective'')~\cite{Vala2015}. Moreover, in our case the referents are necessarily persons (characters), a category of entities possessing certain characteristics (e.g. gender) which can be leveraged to improve performance. Two variants of the problem appear in the literature: certain authors focus only on \textit{alias resolution} (e.g.~\cite{Elson2010}), which consists in grouping proper nouns referring to the same character, while others additionally solve pronominal and/or nominal anaphoras.

The task of alias resolution arises because of the variability of proper nouns appearing in fiction works. On the one hand, in addition to their full name, characters are generally called by a variety of aliases depending on context, style, and other factors. For instance, Sherlock Holmes can also be called ``Mr. Holmes'' or ``Sherlock''. On the other hand, some aliases cannot be unequivocally associated to a character, e.g. ``Mr. Holmes'' can refer to both Sherlock Holmes and his brother Mycroft Holmes. Most authors use some form of \textit{name clustering} to perform alias resolution, each cluster corresponding to all the names encountered for a specific character. Roughly, They use two factors to determine that two aliases point at the same character: \textit{string similarity} and \textit{gender compatibility}~\cite{Elson2010a, Elsner2012, Oelke2012, Ardanuy2014, Vala2015}. The gender of a character mention can be detected using gendered honorifics (e.g. ``Mr.'' vs. ``Mrs.'') and gendered first names (e.g. ``Stephen'' vs. ``Stephanie''), matched to a manually constituted list or some external resource such as WordNet~\cite{Elson2010a, Oelke2012}. 

A straightforward approach to compare strings is to use an appropriate distance function~\cite{Jannidis2016}. However, by doing so, one ignores the structure of the names: potential presence of honorifics, initials, multiple first or last names, distinction between first and last names. Moreover, a number of conventions are culture-specific (e.g. the use of patronyms in Russian names). Also, the relative proportions of first and last name occurrences is likely to vary considerably from one work to the other, as it is tied to stylistic aspects: it is assumed to reflect the level of intimacy in the narrative~\cite{Dalen-Oskam2014}. Certain authors propose to perform direct comparisons through predefined patterns~\cite{Oelke2012} or rules~\cite{Ardanuy2014, Ardanuy2015}. Elsner~\cite{Elsner2012} first compares only multiword names, in order to deal with the ambiguity of isolated first or last names. He constitutes clusters of similar and compatible multiword names, and only then assigns single word names to these clusters whenever possible. The remaining names are assigned based on spatial proximity in the text and lexical frequency. Certain authors use a generative approach~\cite{Elson2010, Vala2015}: based on multiword names found in the text, they produce potential variants thanks to predefined recombination rules (e.g. addition of honorifics, omission of first names) and resources such as gazetteers. These artificial names are then matched to those found in the text. Vala \textit{et al}.~\cite{Vala2015} use additional constraints to prevent certain names from being grouped together: co-occurring names, names with the same last name but different first names, names containing different honorifics.

The other types of anaphoras are more difficult to handle, as they convey additional issues. Some pronouns or nominals may not be connected to any proper noun (and therefore character), if their referent is missing. They can also have split referents, e.g. ``They'' and ``the Holmes brothers'' can both refer to ``Sherlock and Mycroft Holmes''. Certain anaphoric expressions can also refer to non-character entities. Many authors use off-the-shelf tools to solve automatically pronominal anaphoras, e.g.~\cite{Sudhahar2013, Trovati2014, Srivastava2016}. Lee \& Yeung additionally define a distance limit between the reference and the referent, in order to discard relations deemed too remote~\cite{Lee2012f}. Vala \textit{et al}.~\cite{Vala2015} extend to pronouns and nominals the cluster-based approach used for alias resolution. As for proper nouns, gender compatibility can be leveraged for certain pronouns (ex. ``she'' vs. ``he'') and nominals (ex. ``uncle'' vs. ``aunt''). In order to identify anaphoras referring to characters (by opposition to other types of entities), they constitute a list of verb-noun co-occurrences considered as frequent in novels, and perform a grammatical dependency parsing: only the expressions involved in such situations are considered as character mentions. In~\cite{Jannidis2016}, Jannidis \textit{et al}. use a co-reference resolution tool that they previously developed specifically for novels in German~\cite{Krug2015}. In particular, they use linguistic resources to associate close synonym nominals to the same character.

\subsubsection{Visual Narratives}
\label{sec:CharIdUnifVideo}
Like before, handling audiovisual narratives amounts to solving very different problems, which depend on whether one uses video or audio data. When dealing with videos, the problems at hand are face track clustering, and possibly face-name matching.

\textit{Face track clustering} consists in identifying groups of face tracks (output from the occurrence detection step) corresponding to the same face, hence character. Certain authors use off-the-shelf tools~\cite{Liang2009a, Yeh2012, Tsai2013}, but others consider that these generic methods are not sufficiently efficient when applied to fiction works~\cite{Weng2007}. Their results can be improved by training them on a corpus of such works~\cite{Weng2007a, Weng2009}, but this requires additional work and resources. Zhang \textit{et al}.~\cite{Zhang2009e} propose a new method combining the Earth-mover's distance with constrained $k$-means clustering, followed by an additional pruning step.

As mentioned before, when using the audio stream, the problem is to perform \textit{speaker clustering} based on the output of the speaker segmentation step, the whole process being called \textit{speaker diarization}~\cite{Moattar2012}. Speaker clustering consists in grouping audio segments spoken by the same character (akin to track face clustering). Generic method are subject to the same limitations as observed for speaker segmentation~\cite{Clement2011, Ercolessi2013, Bredin2016}. Methods specifically developed for fictions obtain better results. In~\cite{Bost2014}, Bost \& Linarès treat TV series through a two-step method first solving the problem locally at the scene level, then combining these partial results at the global level in order to deal with the whole character set. In~\cite{Bost2015}, they turn to a multimodal approach to enhance their method: in addition to their audio-based tool, they independently perform speaker diarization based on low-level video and audio features, before performing optimal matching to combine both resulting outputs. However, after further experimentation, Bost \textit{et al}. consider the obtained performance is not sufficient, and eventually turn to manual annotation~\cite{Bost2016, Bost2016b}.

Like for character occurrence detection, certain authors adopt multimodal approaches. Some prefer to extract the transcript of the work and apply text-based methods instead of directly using the video or audio stream. This is the case for Chen \textit{et al}.~\cite{Chen2016h} who, after having detected character occurrences in this text, use existing off-the-shelf co-reference tools to detect chains of mentions to the same character. They identify the concerned character by using predefined rules, exploiting the presence of a proper noun in the chain, or connections to utterances whose speaker could be identified. They propose an automatic method based on agglomerative convolutional networks to take advantage of the latter type of information~\cite{Chen2017t} when solving co-references and identifying characters associated to co-reference chains. In~\cite{Bredin2016}, Bredin \& Gelly combine face track clustering and speaker diarization: they first detect speakers through standard speech activity detection tools, before using face embeddings to cluster the face tracks corresponding to the resulting speech segments.

An additional issue specific to audiovisual narratives is to determine the name of the characters detected by grouping faces or speech segments. In the former case, this is called \textit{Face-name matching}, and in the latter, \textit{speaker identification}. In both cases, solving this problem requires using linguistic information: speech content (via transcripts, scripts, or subtitles), text overlaid in the video, predefined list of characters, other external resources. For instance, based on the assumption that the title of the fiction work is known, Tran \textit{et al}.~\cite{Tran2017c} retrieve its list of characters from IMDb, look for their picture using Google Image, and leverage this information to infer character names in the movie through matching. In the context of fiction works though, the favored approach is to leverage scripts~\cite{Liang2009a, Zhang2009e, Park2009, Jung2013a}. This again raises the issue of \textit{script alignment} (with the corresponding video, transcript, or subtitle), as scripts are not time-stamped. After alignment, the script directly allows recovering speaker names, and inferring addressee names (for instance by crosschecking names mentioned in the conversation and on-screen faces).

For comics, a variety of methods have been proposed for \textit{character detection} (or \textit{face recognition}~\cite{Takayama2012}), i.e. to match the multiple occurrences of a character's face. The general approach consists in defining some form of similarity measure, which is then leveraged to group occurrences corresponding to the same character. This is what Takayama \textit{et al}.~\cite{Takayama2012} do, based on the features they use for face detection (cf. Section~\ref{sec:CharIdOccVideo}). Stricker \textit{et al}.~\cite{Stricker2018} adopt a similar approach, but to compare their sets of facial landmarks. The problem is difficult and still open, its resolution will likely require large annotated corpora~\cite{Augereau2018}. Ho \textit{et al}.~\cite{Ho2013} represent character occurrences by graphs of adjacent graphical subregions, and use approximate graph matching to group occurrences corresponding to the same objects. Sun \textit{et al}.~\cite{Sun2013d}  propose a method based on local features.

\subsection{Additional Processing}
\label{sec:CharIdAddProc}
In addition to character occurrence detection and unification, certain authors perform some additional processing related to character identification. We identify two such operations (which are not mutually exclusive): filtering characters (or character occurrences) considered to amount to noise (Section~\ref{sec:CharIdAddProcCharFilter}), and get a more detailed individual description of the characters (Section~\ref{sec:CharIdAddProcCharAttr}).

\subsubsection{Character Filtering}
\label{sec:CharIdAddProcCharFilter}
Certain authors remove characters deemed not frequent enough~\cite{Elson2010, Elsner2012, Hutchinson2012, Sudhahar2013}. For instance, Elson \textit{et al}.~\cite{Elson2010} remove a character if it appears three times or fewer in a novel, or if he amounts to $1\%$ or less of all occurrences. They consider these as noise, possibly generated by their alias resolution method: the occurrences are likely to refer to another existing character, and not a separate infrequent one. When extracting conversational networks, Suen \textit{et al}.~\cite{Suen2013} ignore a character if he speaks fewer than five times over the plot. Character filtering can concern a large part of the novel, e.g. Hutchinson \textit{et al}.~\cite{Hutchinson2012} remove approximately $80\%$ of the characters to study the \textit{Harry Potter} novels.

Bossaert \& Meidert~\cite{Bossaert2013} remove characters depending on their social category. Indeed, they want to use the character network of the \textit{Harry Potter} series of novels as a proxy to study peer support among adolescents. For this reason, they remove all characters that are not Harry Potter's schoolmates.

\subsubsection{Attribute Extraction}
\label{sec:CharIdAddProcCharAttr}
In addition to simply identifying the characters, certain authors extract some additional information allowing to describe them. It typically takes the form of nodal attributes in the final character network. These are retrieved directly from the work of fiction itself, but also from external sources.

The most popular character trait used in the literature is probably their gender, which can be assessed manually~\cite{Rydberg2011, Bossaert2013}, through some external resource~\cite{Gleiser2007}, or automatically~\cite{Elson2010a, Ardanuy2014, Ardanuy2015}. Some authors define a categorical attribute representing the various sides present in the considered narrative; in the articles reviewed in this survey, this is always performed manually. For instance, when dealing with the \textit{Marvel Universe}, Gleiser extracts the characters' \textit{alignment} (\textit{hero} vs. \textit{villain}). In~\cite{Yose2017}, Yose \textit{et al}. study the medieval Irish epic text \textit{Cogadh Gaedhel re Gallaibh}, which narrates a war between Irishmen and Vikings. For their analysis, they distinguish three categories of vertices: \textit{Irishmen}, \textit{Vikings}, and \textit{Others}. When studying Homer's \textit{Iliad}, Kydros \textit{et al}. identify four mutually exclusive categories of characters: \textit{Greek}, \textit{Trojans}, \textit{Gods} and \textit{Others}.

Certain authors focus on more sociological traits. In~\cite{Rochat2017}, Rochat \& Triclot are interested in the representation and relative position of science and politics in science-fiction works. Through crowdsourcing, they manually define an attribute representing whether the main occupation of each character is mainly related to \textit{Politics}, \textit{Technology}, or \textit{Science}, but also \textit{Family}, \textit{Religion}, or \textit{Art}. They use the same attribute to identify \textit{Animals}, and the remaining characters are noted as \textit{Undefined}. When studying the \textit{Harry Potter} series of novels, Bossaert \& Meidert manually retrieve the school house and school year of each student (in addition to his or her gender)~\cite{Bossaert2013}. In~\cite{Rydberg2011}, Rydberg studies Greek tragedies. He manually extracts the so-called social class of attributes: a mix between race, social status and narrative role (\textit{Gods}, \textit{Upper/lower class mortals}, \textit{Chorus}, and \textit{Non-speaking characters}. 

The resulting nodal attributes have mainly two usages. Certain authors use them directly to study the considered narrative. For instance, Bossaert \& Meidert want to assess the effect of certain character traits on the psychological mechanism of peer support, and therefore on the character network structure~\cite{Bossaert2013}. Other authors take advantage of these attributes to solve some higher-level problem. For instance, Ardanuy \& Sporleder~\cite{Ardanuy2014, Ardanuy2015} use character gender to build features later leveraged to classify novels automatically.

\section{Interaction Detection}
\label{sec:InteractionDetection}
Based on the character occurrences, the next step of the extraction process consists in detecting all interactions happening in the narrative between each pair of characters. Such an interaction can be explicitly described, but also inferred from the narrative, depending on what one considers to be an interaction. We identify five distinct approaches in the literature.

The first (Section~\ref{sec:InterDetCooc}) is co-occurrence-based, and relies on a decomposition of the narrative into smaller narrative units. Two characters are considered to interact when they \textit{jointly appear} in the same such unit. The second approach (Section~\ref{sec:InterDetConv}) considers only \textit{direct verbal interactions} between the characters. This is particularly appropriate for dialogue-oriented narratives such as plays. The third (Section~\ref{sec:InterDetMent}) requires one character to \textit{explicitly mention} another one to infer an interaction between them. The fourth (Section~\ref{sec:InterDetDirAct}) takes into account other types of \textit{direct interaction} than conversation (e.g. fighting, kissing). The fifth (Section~\ref{sec:InterDetAffil}) focuses on explicitly expressed \textit{affiliations}, such as family relationships or being coworkers. Finally, it is also possible to combine several of these approaches, in various ways (Section~\ref{sec:InterDetMult}).

\subsection{Co-occurrences}
\label{sec:InterDetCooc}
The co-occurrence-based approach is the most widespread in the literature, probably because it is the easiest to apply: detecting interactions in a more precise way can be a difficult problem, even for humans~\cite{Rochat2017}. This approach consists in breaking down the considered work into smaller \textit{narrative units}, and in assuming that two characters interact when they occur together within the same unit. A few authors use additional constraints, to ensure that co-occurrences actually capture interactions. Some want the narrative unit to contain \textit{only} the two characters of interest, and no one else~\cite{Elsner2012, Iyyer2016}. Others take into account only \textit{consecutive} occurrences, i.e. not separated by another character~\cite{Grayson2016}.

Using co-occurrences presents several limitations mainly caused by their imprecise nature. Indeed, co-occurrence is only a proxy for actual interaction, as it is possible for two characters to appear together without interacting at all (e.g. they both are spectators of some event~\cite{Min2016}, or one mentions the other in his absence~\cite{Prado2016}). The first limitation is that this imprecision propagates to the network itself: the set of co-occurrence-based interactions theoretically contains the conversation-, mention-, action- and affiliation-based ones, plus some false positives\footnote{False negatives are also a possibility, but they require the characters to interact without co-occurrence or mention, which is highly unlikely.}. In practice though, Ardanuy \& Sporleder~\cite{Ardanuy2015} argue that false positives are rare in the sense that two co-occurring characters are almost always related in one way or the other. But this holds only for their experimental results, obtained by integrating co-occurrences over whole narratives. On the contrary, Edwards \textit{et al}. note that co-occurrence networks are denser~\cite{Edwards2018}.

The second limitation also directly comes from the imprecise nature of co-occurrences. As they encompass a number of different types of interactions, it is not possible to assign them a direction, and they are therefore regarded as some form of bilateral interaction. Furthermore, for Kwon \& Shim~\cite{Kwon2017}, due to their imprecise nature, co-occurrences ignore intimate aspects of interactions, such as opinions and emotions. 
The third limitation, according to Prado \textit{et al}.~\cite{Prado2016}, is that using co-occurrences results in more importance being given to otherwise minor characters, when later analyzing the obtained character network. As discussed in our introduction, all these arguments must be balanced accordingly to the possibly very specific nature of the considered narrative.

We discuss the choice of the narrative unit in Section~\ref{sec:InterDetCoocNarrUnit}, as it depends on the type of narrative and can affect the end result. Besides the detection of interactions under the form of co-occurrences, certain authors additionally assign a numerical score to each interaction in order to include more information in the character network eventually extracted: we review such approaches in Section~\ref{sec:InterDetCoocScore}.

\subsubsection{Narrative Unit}
\label{sec:InterDetCoocNarrUnit}

\paragraph{Novels}
In novels, Rochat \& Kaplan use the page as a narrative unit~\cite{Rochat2014}, as imposed by the predefined character index they leverage during character identification (see Subsection~\ref{sec:CharIdManual}). Such a partitioning of the text, based on purely \textit{physical} (and therefore arbitrary) aspects, results in the possible split of chapters, paragraphs, or even sentences. It is therefore very likely to miss co-occurrences. Rochat \& Kaplan try to overcome this problem through a two-page unit with a one-page overlap (instead of consecutive pairs of pages). 

Other authors use smaller narrative units that can be considered as more natural, in the sense that they at least avoid such arbitrary splitting: one sentence~\cite{Lee2012f}, ten sentences~\cite{Park2013}, one paragraph~\cite{Elsner2012}, ten paragraphs~\cite{Elson2010}. However, the length of sentences and paragraphs can vary considerably from one writer to the other, to the point where one writer's paragraph can be shorter than an other's sentence~\cite{Iyyer2016}. Using word spans instead solves this problem: the literature contains examples ranging from five~\cite{Hutchinson2012} to $1{,}000$ words~\cite{Grener2017}. Certain authors take chapter breaks into account, so the span can be smaller when reaching the end of a chapter~\cite{Grener2017}. However, word spans are as arbitrary a narrative unit as the page, and suffer from the same limitation.

Certain authors prefer to use a larger unit, especially the chapter~\cite{Knuth1993, Ardanuy2015, Min2016}, or the pericope~\cite{Grandjean2015} (its counterpart in the context of biblical writings). Like the sentence and the paragraph, however, it does not lead to split sentences but its size can vary significantly from one writer to the other. Moreover, it can be considered as too long a narrative unit for this usage, as many events can take place in the same chapter. Bolioli \textit{et al}.~\cite{Bolioli2013} use what they call a \textit{narrative sequence}: a small portion of text characterized by its unity of location and involved characters (basically, a scene). Prado \textit{et al}.~\cite{Prado2016} proceed similarly when studying \textit{La chanson de Roland}, as this song of heroic deeds is originally divided into stanzas of very irregular size.

\paragraph{Scripts}
For scripts, certain authors use the line~\cite{Xanthos2016}, but the most widespread narrative unit is simply the scene, e.g.~\cite{Ding2010, Suen2013, Gorinski2015, Rieck2016}. However, as noted by Suen \textit{et al}.~\cite{Suen2013}, if a script contains long scenes, this can lead to the connection of characters involved in completely different parts of the story. To solve this issue, Stiller \textit{et al}.~\cite{Stiller2003, Stiller2005} prefer to use \textit{subscenes}, i.e. parts of a scene identified by the fixed set of involved characters. Put differently, as soon as one character leaves or enters the stage, the current subscene ends and a new one begins.

\paragraph{Comics}
In comics, Rochat uses a two-page narrative unit~\cite{Rochat2017}, like he does for novels (see above). In~\cite{Alberich2002, Gleiser2007}, the authors use the comic book issue, but this narrative unit is dictated by the predefined database they use as raw material. There seems to be a lot of room for exploration here, as comic book formats are very diverse, and there are very few articles dealing with this question. For instance, a seemingly natural choice would be to transpose the concept of scene, and consider a sequence of panels involving the same characters as a narrative unit. However, this requires efficiently solving certain lower-level problems, in particular panel identification and panel ordering. Existing methods to detect the boundaries of panels take advantage of the black lines generally outlining them, or of the white space called \textit{gutter} separating them~\cite{Stommel2012, Augereau2018}. But a number of artists use complex page layouts, which makes both panel detection and ordering much harder: overlapping panels, panels joined by other objects (speech bubbles, caption) partly open panels, or even panels with no explicit boundary~\cite{Stommel2012}.

\paragraph{Videos}
In videos, certain authors use a fixed time interval as a narrative unit, e.g. ten seconds~\cite{Lee2018}, one minute~\cite{Rochat2017}. Similarly to using the page as a narrative unit for novels, this approach seems quite arbitrary. Indeed, it is likely to split the plot at any moment, including for instance the middle of an action. Several more \textit{natural} units are traditionally used when dealing with videos~\cite{Rui1999, Park2011}: The \textit{frame} is the smallest, it is a single image. The \textit{shot} is a sequence of frames continuously recorded by a single camera. The \textit{scene} is a collection of consecutive and semantically related shots. The \textit{group} is a logical subdivision of a scene (e.g. a conversation, in a scene containing several of them~\cite{Park2009, Park2011}), and the \textit{sequence} is a series of consecutive scenes constituting a short story.

A number of authors prefer to work with scenes, but this requires solving the difficult problem of scene boundary detection. Certain authors perform this task manually, e.g. Weng \textit{et al}. in~\cite{Weng2007}. They later adopt a semi-automatic approach, consisting in applying first an existing automatic tool and then manually detecting and correcting its errors~\cite{Weng2007a, Weng2009}. Others proceed automatically, using an off-the-shelf tool~\cite{Zhang2009e, Tsai2013}. In~\cite{Tran2015a}, Tran \textit{et al}. segment movies by detecting stable periods in terms of character on-screen presence, and merging them according to criteria related to character similarity and temporal proximity.

\subsubsection{Interaction Score}
\label{sec:InterDetCoocScore}

\paragraph{Positive Scores}
While most authors adopt a boolean approach when detecting interactions (i.e. presence vs. absence), certain authors try to assess their \textit{intensity} under the form of numerical scores. Some use the distance between the two concerned character occurrences, expressed as a number of some smaller narrative unit separating them (or, more exactly, some decreasing function of this quantity). For instance, Park \textit{et al}.~\cite{Park2013} use a narrative unit of ten sentences, and experiment with two distance functions: the numbers of words and sentences separating the two occurrences. Grayson \textit{et al}.~\cite{Grayson2016} count the number of tokens between occurrences. 

Others use the duration of the co-occurrence, e.g. the number of lines spoken in the scene~\cite{Suen2013}, or the number of seconds during which both characters jointly appear on-screen in a video~\cite{Tran2015, Tran2015a}. Rieck \& Leitte~\cite{Rieck2016} use the proportion of words spoken by the considered pair of characters, relative to the total number of words spoken during the scene. This amounts to giving more importance to the most talkative characters. Yeh \textit{et al}.~\cite{Yeh2012} take advantage of film-editing guidelines, especially the \textit{180 degrees} and \textit{shot alternation} rules (cf. Section~\ref{sec:IntroSpecifFiction}): their score is the number of consecutive shots showing alternatively both concerned characters.

\paragraph{Signed Scores}
In addition to the intensity of an interaction, a few authors try to detect its \textit{polarity}, i.e. whether or not it is friendly or hostile, by leveraging the context of the co-occurrence. This results in a \textit{signed} score (i.e. it can be negative). Some authors adopt an \textit{indirect top-down} approach consisting in assessing the general polarity of the concerned narrative unit, and extending it to all interactions occurring in it. For novels, Min \& Park apply an off-the-shelf tool to perform sentiment analysis and measure the polarity of each narrative unit constituting the plot~\cite{Min2016a}. For movies, Ding \& Yilmaz train a support vector regressor into estimating the polarity of a scene based on a set of visual and auditory features~\cite{Ding2010}. The same authors also experiment with affinity learning to take advantage of \textit{visual concepts}, a higher-level information describing the scene \textit{context} through the objects and environments it involves~\cite{Ding2011a}. 

Alternatively, the \textit{indirect bottom-up} approach consists in assessing the emotional state of the individual characters, instead of the whole narrative unit, and obtaining the polarity of an interaction by combining the states of the concerned characters at the time it occurs. Lee \textit{et al}. do so for movies~\cite{Lee2018}: they use off-the-shelf tools (completed by manual correction) to assess the emotional states of characters based on both conversational (movie script) and visual (facial expressions) cues. In comics, a similar approach could be applied by leveraging graphical elements such as effect lines, characters' facial expressions and poses, and conventional symbols used to reflect emotions. However, this is not possible yet, as only a handful of articles propose methods to detect such elements~\cite{Augereau2018}.

Finally, certain authors propose a \textit{direct} way to monitor the polarity of an interaction. For novels, Chaturvedi \textit{et al}.~\cite{Chaturvedi2016} use a Markov model to represent the chronology of the interactions between a given pair of characters, and detect friendly vs. hostile phases. For this purpose, they leverage a collection of features taking into account the content of the text associated to the co-occurrences (e.g. type of action, structure of the sentence, grammatical role of the characters, lexical aspects). This model is itself embedded in a semi-supervised framework allowing to train over a partially annotated corpus of novel summaries. They later propose an extended version of their tool in which the nature of the interactions is not necessarily binary, and is learned in an unsupervised way~\cite{Chaturvedi2016a, Chaturvedi2017}. In~\cite{Iyyer2016}, Iyyer \textit{et al}. use a relatively similar principle, except their model is a neural network, which is able to learn relationship descriptors in an unsupervised way. Instead of a single \textit{friendly}/\textit{hostile} score, each co-occurrence is therefore represented by a set of descriptor values, each one with its own semantics (e.g. \textit{love}, \textit{violence}, \textit{education}).

\subsection{Conversations}
\label{sec:InterDetConv}
A number of authors focus only on verbal interactions between two characters, i.e. when one explicitly \textit{talks} to another, in order to later extract so-called \textit{conversational networks}. Technically, the process of detecting them is generally harder than for co-occurrences, as one has to detect that a conversation is taking place, as well as to distinguish the involved speakers and addressees. A fundamental difference between co-occurrence and verbal interaction is that the latter is naturally \textit{unilateral}: one character talks, the other listens. This generally leads to the extraction of \textit{directed} networks~\cite{Park2011}.

As explained before, verbal interactions can be assumed to be subsumed by co-occurrences, as two characters need to co-occur in order to converse, but can co-occur without necessarily talking to each other~\cite{Bost2016}. One could therefore suppose that only focusing on conversations leads to some information loss. However, certain authors argue that this is not the case, as many aspects of interpersonal relationships~\cite{He2013a}, if not most~\cite{Waumans2015}, are conveyed through conversation. The validity of this argument actually depends on the type of the considered narrative, or even on the specific narrative itself. Several authors note that conversational networks are suitable only for narratives rich in verbal interactions~\cite{Lee2012f, Min2016a}. Lee \& Yeung~\cite{Lee2012f} give the counterexample of the \textit{Book of Genesis}, in which the tense relationship between Abraham's wife and servant is mentioned frequently in the text, while they never speak to each other. Ardanuy \& Sporleder~\cite{Ardanuy2015} identify plays as the most appropriate form of narrative for conversational networks, as they are essentially scripted dialogue, by comparison with novels, in which most of the action takes place off-dialogue (e.g. McCarthy's \textit{The Road}, Yourcenar's \textit{Mémoires d'Hadrien}). Visual media such as comics tend to rely only lightly on dialogues.

As detecting verbal interactions generally requires identifying \textit{who} talks to \textit{whom}, the appropriate method depends on the nature of the narrative. Like before, we distinguish textual (Sections~\ref{sec:InterDetConvText} and visual narratives (Section~\ref{sec:InterDetConvVideo}). We then turn to methods proposed to estimate the intensity and direction of verbal interactions (Section~\ref{sec:InterDetConvScoreDir}).

\subsubsection{Textual Narratives}
\label{sec:InterDetConvText}
In text, one can distinguish utterances (quotations of a character) from proper narration parts (description of the action occurring). Verbal interactions can take one of two forms: \textit{direct} speech, which consists in explicitly quoting the utterance, and \textit{indirect} speech, which consists in reporting what the speaker said (e.g. ``Arthur told his knights to go home''). Conversational networks are built upon the former, whereas the latter belongs to the class of direct interaction considered in Section~\ref{sec:InterDetDirAct}.

Identifying verbal interactions in fictional free text is a difficult task. First, the text generally contains much more proper narration parts than utterances, and both are tightly intertwined (e.g. speech verbs connecting utterances during a conversation). Second, the typographical conventions used to identify utterances in the text are not always respected (not all spoken text is quoted), and are sometimes ambiguous (e.g. sneer quotes). Third, the speakers are often not identified explicitly ~\cite{Glass2006}, and when they are, this information can be embedded in the narration. Therefore, automating the detection of verbal interactions requires the resolution of several NLP-related problems: detecting utterances, assigning a speaker to each of them, and possibly also identifying its listeners. Maybe for this reason, certain authors decide to proceed manually~\cite{Falk2016}.

By comparison, semi-structured text is easier to handle, as typography generally allows distinguishing utterances from the rest of the text, and speakers are explicitly indicated. Thus, the only remaining problem is to identify addressees. As for novels, this task is difficult to automate, which is why certain authors perform it manually~\cite{Moretti2011a, Rydberg2011, Muhuri2018}.

\paragraph{Quote Detection}
It is relatively straightforward to detect quotes when the text is clean, by simply relying on the presence of quotation marks~\cite{Elson2010, Celikyilmaz2010, Waumans2015} and/or using simple rules~\cite{Yeung2017}. It is worth noting that some complications can appear, depending on the file encoding: for example, Unicode includes tens of distinct glyphs likely to be used as quotation marks.

But quotes are not always used correctly, or the considered language can be without typographic quoting convention, e.g. old Chinese punctuation style~\cite{Lee2016d}. In this case, the authors rely on the presence of quotative verbs, and on the sentence structure. Zhang \textit{et al}.~\cite{Zhang2003a} train a decision tree based on morphological and typographical features, in order to recognize quoted speech and detect speaker changes between two consecutive utterances. Mamede \& Chaleira~\cite{Mamede2004} propose twelve heuristics to identify both direct and indirect utterances in Portuguese. These are based on typographical clues, verb tense, presence of certain pronouns, temporal adverbs, and interjections. 

\paragraph{Quote Attribution}
The literature contains a number of articles dealing with quoted speech attribution for general texts~\cite{Elson2012}, but we focus on the methods proposed specifically to handle fictions. For a given utterance, one can distinguish two cases: either the speakers are mentioned explicitly (be it by their name or some anaphora), or they are implicit and can be inferred from the context. 

The most widespread method to deal with \textit{explicit} speakers relies on detecting the speech verb associated to the concerned utterance, and considering its subject as the speaker. Certain authors propose to manually define~\cite{He2013a, Waumans2015, Glass2007} or automatically learn appropriate rules~\cite{Glass2006}, other look for patterns~\cite{Krug2015, Lee2016d}. In the absence of a speech verb, certain authors proceed similarly with nearby action verbs~\cite{Mamede2004}, whereas others look for the nearest character occurrence (excluding the utterance itself)~\cite{Zhang2003a}.

The first step in identifying \textit{implicit} speakers is generally to select a subset of candidates among all known characters. For this purpose, one first has to estimate the boundaries of the conversation: this can be done by leveraging long proper narration parts separating sequences of utterances. For this purpose, Waumans \textit{et al}.~\cite{Waumans2015} use empirically determined distance and length thresholds, and also include the narration located right before and after the conversation. The candidates correspond to the characters present in the obtained portion of text. 

A number of authors then deal with implicit speakers by assuming that the characters involved in the considered conversation respect certain rules relative to conversation turns (a.k.a. the \textit{conversation turn assumption}) such as: consecutive utterances are not spoken by the same character, or one speaker answers to the previous one. It is then possible to leverage the explicit speakers identified before to infer the missing ones. Certain authors identify speaker alternation patterns and use them to define rules~\cite{Waumans2015, Krug2015, Lee2016d}, or to learn them in a supervised way~\cite{Glass2006}. He \textit{et al}. translate these patterns into features describing the position and frequency of character occurrences relative to the considered utterance. These features (as well as Celikyilmaz \textit{et al}.'s ACTM, see below) are then used to train a classifier~\cite{He2013a}. Some authors additionally use gender compatibility to improve utterance attribution~\cite{He2013a}.

Elson \& McKeown train a classifier~\cite{Elson2010a, Elson2010} to deal with implicit and explicit speakers at once. They use standard low-level features as well as the syntactic category of the utterance. The latter is related to its structure and context in the paragraph containing it (e.g. being followed by a speech verb and a character name). They also rely on the \textit{conversation turn assumption}, as the classifier is aware of the last identified speaker. O'Keefe \textit{et al}.~\cite{OKeefe2012} note that this last point is not a realistic assumption, and consider instead the problem as a sequence labeling one. Using the same features as Elson \& McKeown, they experiment with various sequence decoding methods. 

Certain authors focus on the utterance content, in order to distinguish speakers based on the topic they talk about. Chaganty \& Muzny~\cite{Chaganty2015} propose a supervised method based on neural networks and use GloVe vectors to represent words~\cite{Pennington2014}, but they do not obtain conclusive results. Celikyilmaz \textit{et al}.~\cite{Celikyilmaz2010} build an Actor-Topic model (ACTM), a Bayesian model allowing to predict the probability that a given character spoke a given utterance, depending on the topics mentioned in this utterance and those mentioned before by this character (and the others). Each utterance is assigned its most probable character according to this model.

\paragraph{Addressee Identification}
The addressee is sometimes mentioned \textit{explicitly} in the utterance, when the speaker directly calls him. Certain authors define specific rules to take advantage of this situation~\cite{Mutton2004}. Krishnan \& Eisenstein~\cite{Krishnan2015}, in particular, focus on the level of formality of these verbal interactions. He \textit{et al}.~\cite{He2013a} train a classifier to detect these cases, based on features related to punctuation and typical forms of interjection. 

When the addressee is \textit{implicit}, one uses the \textit{conversation turn assumption} already leveraged for speaker identification. Certain authors define rules assuming that a speaker talks to the preceding speaker in the conversation~\cite{Nalisnick2013, Kwon2017}, or the few preceding speakers~\cite{Suen2013}, and/or to the following one~\cite{Mutton2004, Pope2016, Masias2017}. In the case of conversations involving more than two characters, certain authors assume that a speaker talks to everyone present~\cite{Mutton2004, Fischer2015a, Bost2016b}, which can also be considered as co-occurrence with an additional constraint (not just being present, but also speaking).

The two-stepped supervised method proposed by Yeung \& Lee~\cite{Yeung2017} applies a similar conversation turn-based principle to assign simultaneously both speakers and addressees to utterances. They first train a CRF-based classifier to identify explicit speakers and listeners, using morphological, grammatical and positional features. They then train another CRF classifier to identify the conversation boundaries, and use the \textit{conversation turn assumption} to fill the missing speakers and listeners from the first step.

\subsubsection{Visual Narratives}
\label{sec:InterDetConvVideo}
\paragraph{Comics} 
We could not find any article dealing with the extraction of conversational networks from comics, mainly because this requires solving a number of lower-level problems for which there are not much literature~\cite{Rigaud2015a} and no efficient solutions yet~\cite{Augereau2018}. The first ones are panel detection and ordering, already mentioned in Section~\ref{sec:InterDetCoocNarrUnit}. Then, it is necessary to detect speech bubbles and captions. This problem is difficult, because the shape and position of bubbles widely vary depending on the artist, culture, context, meaning, and a number of other factors. Existing approaches range from the exploitation of low-level color-based features (e.g. detecting white blobs) to adaptive outline detection methods~\cite{Rigaud2015a}.

\begin{figure}[htb!]
	\centering
	\begin{tabular}{c@{\hspace{1mm}} c@{\hspace{1mm}} c@{\hspace{1mm}} c}
    	\includegraphics[height=1.8cm]{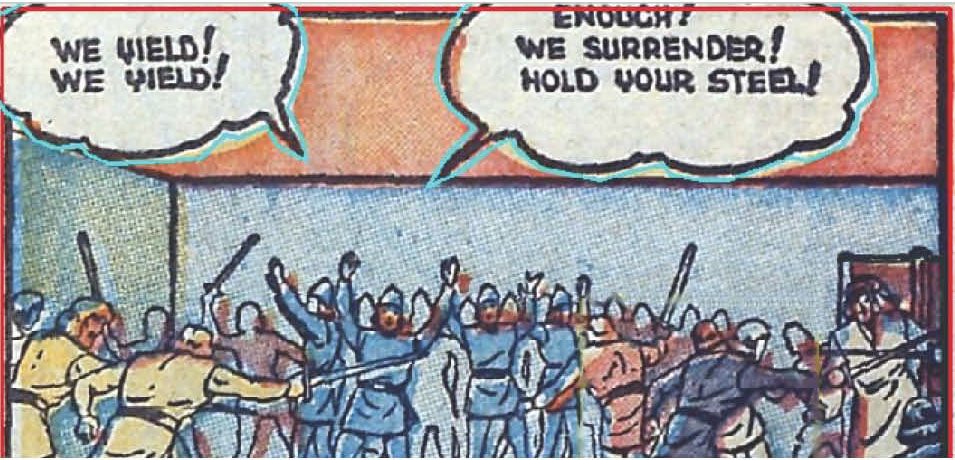} & 
    	\includegraphics[height=1.8cm]{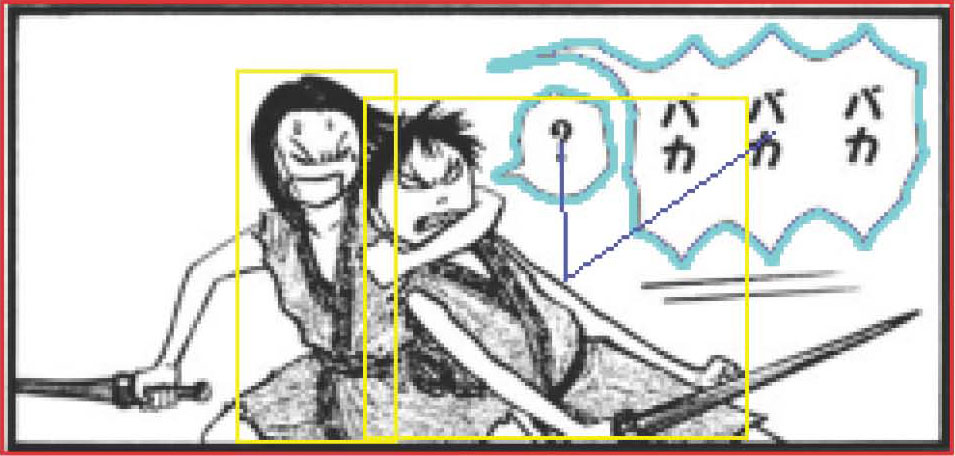} &
    	\includegraphics[height=1.8cm]{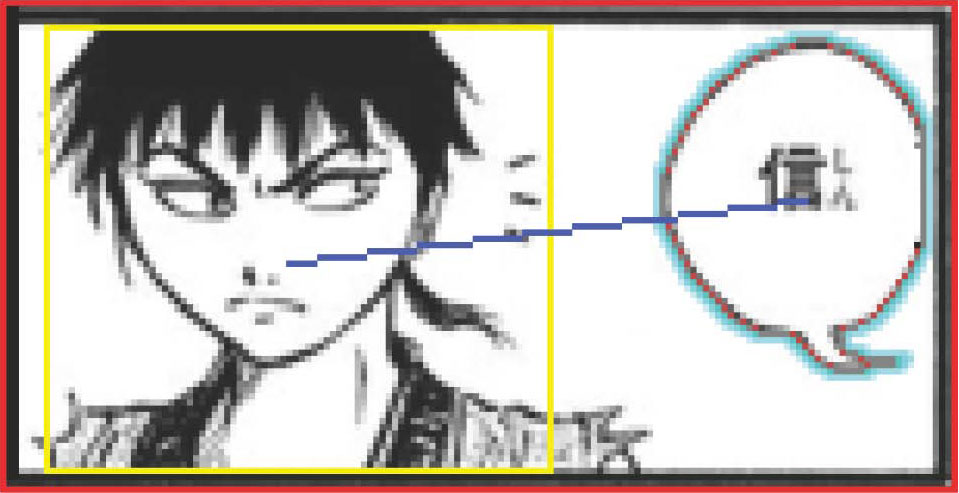} &
    	\includegraphics[height=1.8cm]{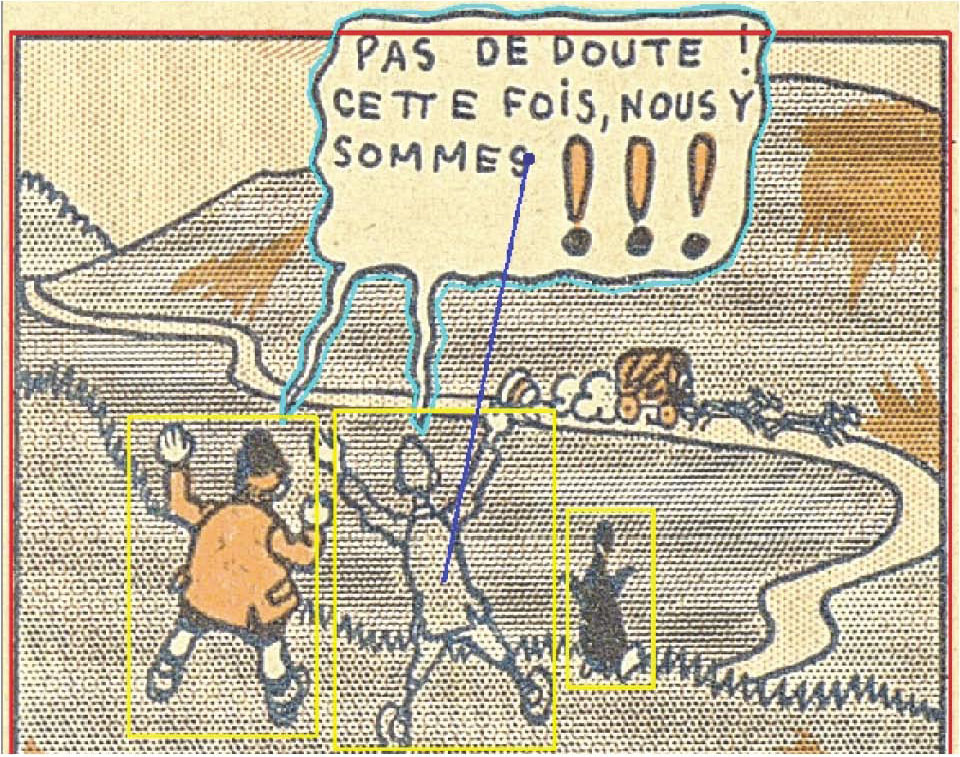} \\
    	a) & b) & c) & d) \\
	\end{tabular}
	\caption{Examples of distinct cases of speech bubble use in comics~\cite{Rigaud2015a}: a) unidentifiable speakers; b) in-between speakers; c) out-of-panel speaker; d) multiple speakers.}
	\label{fig:bubbles}
\end{figure}

One also needs to retrieve the shape of the bubbles, as it generally conveys a specific meaning: smooth for speech, cloudy for thoughts, spiky for screams. But there is no definitive convention, for instance certain artists use captions to show the characters' thoughts. The location of the bubble is also important. First, relative to the other bubbles, it directly affects the reading order, and is needed to extract properly a conversation. Second, relative to the characters, it indicates who speaks. Bubbles often possess a tail directed at their speaker, which helps solving the speaker attribution problem. However many different cases exist, see Figure~\ref{fig:bubbles} for instance. Rigaud \textit{et al}.~\cite{Rigaud2015a} adopt a graph-based approach leveraging the distance between bubbles and characters, and the angle of the tail. But again, using a tail is not a universal convention, which makes it harder to perform this task in general.

Finally, the last step is optical character recognition (OCR). Comics can be very challenging for standard OCR models\cite{Augereau2018}, for a number of reasons: 1) many authors write by hand; 2) use of a variety of fonts, sizes, and colors, sometimes in the same bubble, sometimes mixed with pictographs; 3) the environment is very noisy due to the drawings surrounding the text, which is sometimes integrated to the background (e.g. sound effects). In conclusion, there are a number of open low-level problems to solve before being able to extract conversational networks from comics.

\paragraph{Videos} 
There are not many articles describing the extraction of conversational networks based on audiovisual narratives either. The problems are roughly the same as for the other media: detecting utterances, identifying their speaker, and possibly their addressees.

When using the video stream, one has to detect which parts of a face track correspond to a speaking character. Certain authors use off-the-shelf tools to identify lip motion~\cite{Park2009, Zhang2009e}, which allows simultaneously detecting utterances and assigning them to speakers. When using the audio stream, the speaker diarization tools described in Section~\ref{sec:CharIdUnifVideo} also solve the two same problems simultaneously. In theory, it is also possible to apply Automatic Speech Recognition tools to the audio stream in order to obtain a transcription and get access to the utterance content, or to use directly the script if it is available, however in practice we did not find any use of these approaches in the fictional network extraction literature.

\begin{figure}[htb!]
	\centering
	\includegraphics[height=1.9cm]{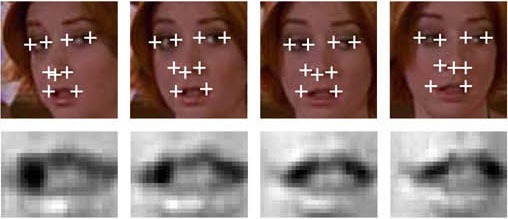}
	\hfill
	\includegraphics[height=1.9cm]{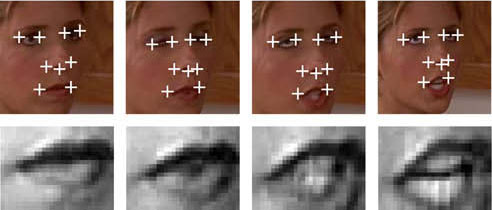}
	\hfill
	\includegraphics[height=1.9cm]{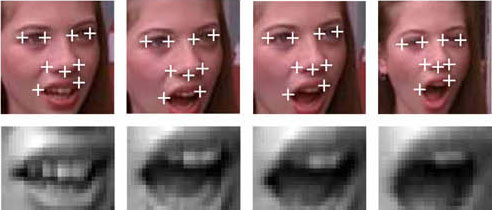}
	\caption{Examples of lip motion detection for three different characters of the TV series \textit{Buffy the Vampire Slayer}~\cite{Everingham2009}.}
	\label{fig:lips}
\end{figure}

Regarding the identification of addressee, one approach is to use the \textit{conversation turn assumption} that we already mentioned for text. For TV series, for instance, Bost \textit{et al}.~\cite{Bost2016, Bost2016b} consider that a speaker talks to another character if the latter speaks before and after the considered utterance. For an utterance starting or ending the conversation, they just consider the following or preceding speaker. When there are more than two speakers involved in the conversation, they use temporal proximity to determine the addressees. Alternatively, it is also possible to take advantage of lip motion detection: one can consider that characters appearing without moving their lips are addressees~\cite{Park2011}. However, this approach has the major drawback of missing addressees that are not shown on-screen.

\subsubsection{Score and Direction}
\label{sec:InterDetConvScoreDir}

\paragraph{Score}
The ways scores are handled for verbal interactions are quite similar to those already described for co-occurrences (Section~\ref{sec:InterDetCoocScore}). Some authors adopt a Boolean approach, only identifying whether a verbal interaction is occurring or not. This can be for the sake of simplicity~\cite{Mutton2004}, or by lack of technical resources ~\cite{Moretti2011a}. A number of authors use a numerical score to represent the intensity of each interaction. The simplest approach is to rely on the length of the interaction, be it expressed in words in the case of text~\cite{Suen2013, Jannidis2016}, or some temporal unit in the case of videos~\cite{Park2009, Bost2016}. Some authors prefer to compute the distance separating the utterances of the two concerned users, expressed in numbers of utterances, and apply a decreasing function, e.g. linear~\cite{Suen2013} or stepwise~\cite{Pope2016}.

Like for co-occurrences, certain authors take advantage of the content of the conversation to compute signed scores reflecting the polarity of the exchange. In plays, Nalisnick \& Baird~\cite{Nalisnick2013, Nalisnick2013a} use an existing lexicon designed for sentiment analysis, in order to associate a polarity (positive or negative) to certain words. The score of an utterance is the sum of these values over all of its constituting words.

\paragraph{Direction}
Certain authors consider verbal interactions as symmetric~\cite{Moretti2011a, Mutton2004, Elson2010, Waumans2015}, in the sense that both the speaker and the addressee are interchangeable. It is an important simplification though, as the information flows from the speaker towards the addressee, making the interaction asymmetric by nature: many authors prefer to consider that the speaker acts on the addressee~\cite{Park2009, Lee2016d, Kwon2017, Muhuri2018}. As we have seen before, the addressee is generally much more difficult to identify than the speaker, which could explain this choice. Certain authors go further in the simplification, and consider that all characters participating in a conversation speak to each other~\cite{Waumans2015}, which is conceptually very close to simply detecting co-occurrences.

Interestingly, Park \textit{et al}.~\cite{Park2009, Park2011} consider that a verbal interaction can involve a \textit{single} character as both the speaker and the addressee, in case of \textit{soliloquy} (i.e. when one talks to oneself). In terms of graphs, this results in loops, or self-edges (an edge connecting a vertex to itself). Note that such graphs are relatively unusual in the domain of complex network analysis, and most descriptive tools are not designed to take advantage of this information.

\subsection{Mentions}
\label{sec:InterDetMent}
Certain authors focus only on explicit mentions of characters during conversations. Compared to the conversational approach described in Section~\ref{sec:InterDetConv}, the difference is that they consider that there is an interaction not when one character speaks \textit{to} another, but when he speaks \textit{about} another. Thus, like for conversational interactions, this requires detecting utterances and their speakers, but not their addressees. Also like them, this approach is appropriate for speech-oriented media. Once the utterances are identified, one only needs to list which character occurs within them: this is trivial, since character identification is the object of the first step of the extraction process (Section~\ref{sec:CharacterIdentification}). This is the approach adopted to study novels, e.g.~\cite{Elson2010} (as a baseline, the main focus being conversational interactions), and movie~\cite{Gorinski2015} or TV~\cite{Deleris2018} scripts. The approach from~\cite{Choi2007} also uniquely relies on mentions, but these are non-verbal: this is a very specific case, as the studied work is a biographical dictionary. The authors consider that there is an interaction between the character described by a given dictionary entry and all the characters mentioned in the body of this entry.

As with other types of interactions, it is possible to compute a score to characterize the intensity of a mention-based interaction, although most authors do not in this case. Elson \textit{et al}.~\cite{Elson2010} count the number of times the character is mentioned in the concerned utterance, and divide by its length. Implicitly, they consider that the interaction is more intense if the utterance is short. Kwon \& Shim~\cite{Kwon2017} take advantage of the utterance content to compute signed scores in scripts. They look for utterances expressing the opinion of the speaker regarding the mentioned character. They use an off-the shelf sentiment analysis tool to detect the polarity of the utterance, and perform a manual pass to improve the result.

As a mention is by nature an asymmetric interaction, in the sense that some character performs an action involving another one~\cite{Choi2007, Gorinski2015, Deleris2018}, most authors consider them as such. However, a few authors choose to ignore this information~\cite{Mutton2004, Elson2010}, either because they do not consider it as useful for the task at hand, or to simplify the processing.

\subsection{Direct Actions}
\label{sec:InterDetDirAct}
Instead of focusing only on verbal interactions, certain authors take into account all forms of direct actions that one character can perform on another (e.g. thinking about someone), or that two characters can perform jointly (e.g. fighting). By comparison with conversational interactions, detecting \textit{general} actions allows handling narratives in which most of the interactions takes place off-dialogue (see Section~\ref{sec:InterDetConv}). However, the task is even more difficult, as one needs to identify the action taking place, as well as the characters performing and undergoing it. 

Some authors dealing with textual narrative proceed manually, possibly because they focus on certain classes of actions, semantically speaking: one character supporting another~\cite{Bossaert2013}, physical encounters~\cite{Holanda2017}, actions that influence the development of the story~\cite{Venturini2016}. In~\cite{Cipresso2016}, Cipresso \& Riva are interested in identifying the interactions between movie characters according to four basic emotions: anger, fear, sadness and joy. For this purpose, they rely on a survey conducted over a sample of eleven persons watching the movie and detecting these interactions manually. When dealing with the \textit{Game of Thrones} TV series, Stavanja \& Klemen~\cite{Stavanja2019} focus only on the action of murdering someone, in order to predict who the next kill will be.

\begin{figure}[htb!]
	\centering
	\begin{tabular}{c@{\hspace{6mm}} c@{\hspace{6mm}} c}
    	\includegraphics[height=3cm]{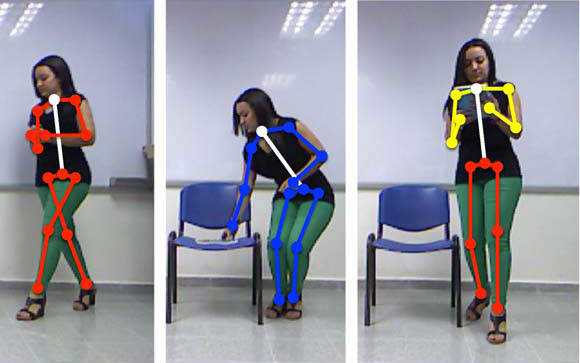} & 
    	\includegraphics[height=3cm]{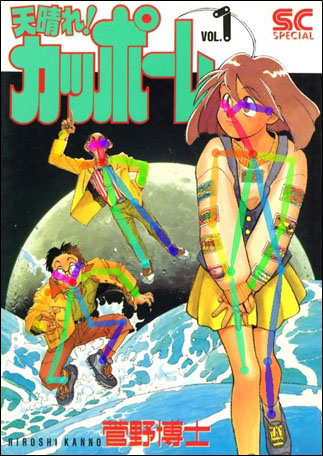} &
    	\includegraphics[height=3cm]{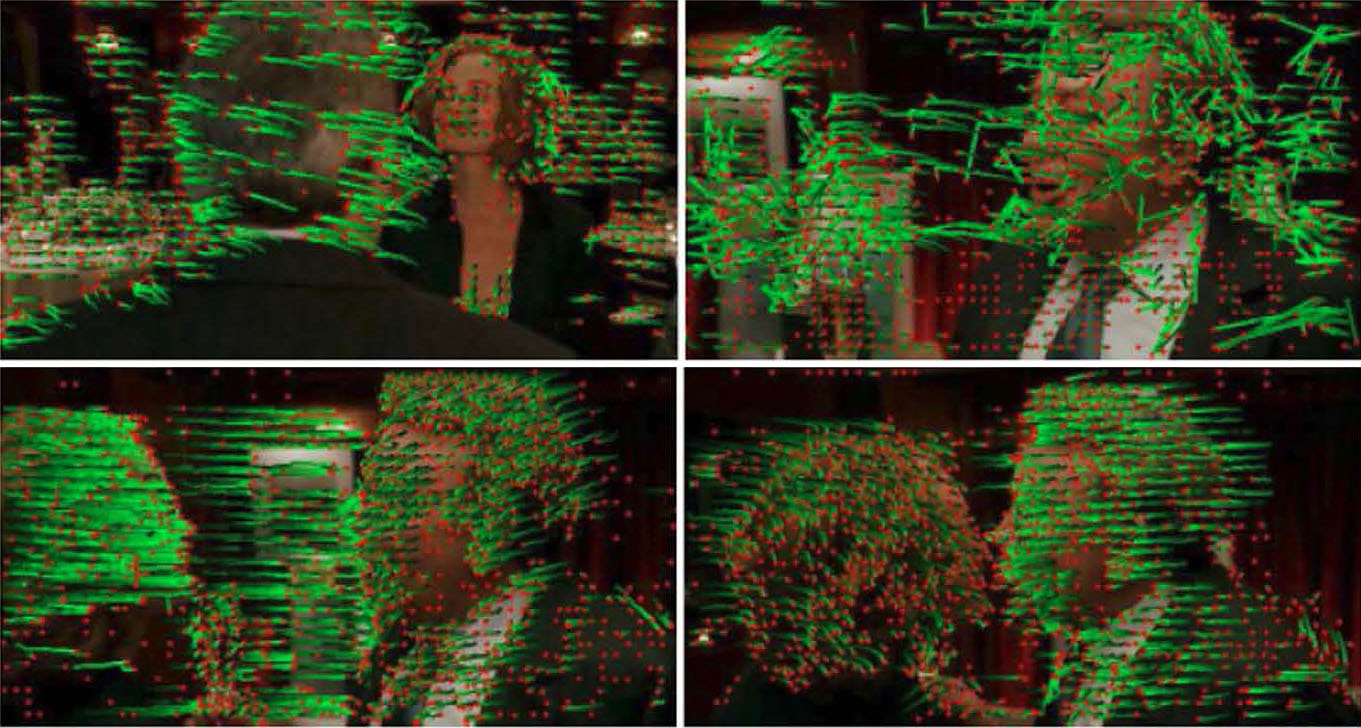} \\
    	a) & b) & c) \\
	\end{tabular}
	\caption{Examples of identification of: a) poses in videos~\cite{Lillo2014}; b) poses in comics~\cite{Augereau2018}; c) human motion in videos~\cite{Vrigkas2015}.}
	\label{fig:motion}
\end{figure}

The automatic approaches proposed in the literature are all designed for textual narratives. Indeed, processing visual narratives implies recognizing actions in videos or images, which is a difficult problem called or \textit{human-human interaction detection}~\cite{Vrigkas2015}. Generic methods exist for videos~\cite{Vrigkas2015, Weinland2011, Poppe2010}, some of which have been applied to works of fiction (mainly movies). However, the literature does not seem to contain any article leveraging them to extract character networks. For comics, the problem can be considered as even harder due to the static nature of the medium: state-of-the-art works only focus on the lower-level task of \textit{pose detection}~\cite{Augereau2018} (cf. Figure~\ref{fig:motion}).

Most automatic methods for text first employ parsing techniques to detect so-called SVO triplets (subject-verb-object), before filtering them in order to retain only those involving characters as subjects and objects~\cite{Trovati2014, Srivastava2016}. Certain authors take advantage of external resources such as \textit{FrameNet}~\cite{Baker1998} to additionally filter the remaining triplets by focusing on certain classes of verbs, semantically speaking. To summarize, FrameNet is a database containing \textit{frame semantics}, i.e. structured representations of situations involving various agents and objects, and semantic relationships between them (e.g. generalization) allowing to perform inference. Certain authors use predefined classes of interest, e.g. social interactions~\cite{Lee2012f}, while other estimate them \textit{ad hoc}, e.g. the fifty most frequent~\cite{Sudhahar2013}.

In~\cite{Agarwal2012, Agarwal2013a, Agarwal2014a}, Agarwal \textit{et al}. detect interactions using their own tool, described in~\cite{Agarwal2010}. It is based on a classifier fed with essentially grammatical features, which allow it to detect how verb-related connections exist between certain characters. This tool was originally designed and trained to handle newswire text, so Agarwal \textit{et al}. have to manually control its output when applied to novels. They treat a conversation as an interaction, so their interaction network subsumes the conversational network, as shown experimentally in~\cite{Jayannavar2015}.

Nijila \& Kala~\cite{Nijila2018} propose a supervised approach, not relying on SVO triplets but on simple co-occurrence of characters in the same sentence. They train a Convolutional Neural Network (CNN) to predict the class of action associated to each sentence involving at two characters. Of course, this method requires an annotated corpus, which they constitute manually.

Certain types of actions can be considered as naturally \textit{unilateral}, as one character acts on another one, whereas other are \textit{bilateral}, as several characters act together. In terms of graph extraction, this means extracting directed vs. undirected edges. Certain authors consider only unilateral~\cite{Sudhahar2013} or bilateral actions~\cite{Lee2012f}, and ignore the rest. Others use both, but consider all of them as bilateral, in order to avoid the identification of the subject and object characters, and therefore significantly ease the process .~\cite{Trovati2014}. Agarwal \textit{et al}.~\cite{Agarwal2012} distinguish between two types of actions: unilateral ones in which the object character is not aware of the action occurring (e.g. thinking about someone), which they call \textit{observations}, vs. all other actions. They handle them separately, in order to compare them relative to the task at hand.

A very few authors compute a score to represent the intensity of the interaction. Trovati \& Brady manually identify three categories of verbs~\cite{Trovati2014}: \textit{friendly}, \textit{hostile}, or \textit{unknown}. They ignore interactions described by verbs belonging to the latter, and assign a signed score to the others, depending on the category. Srivastava \textit{et al}.~\cite{Srivastava2016} use sentiment analysis to take advantage of the semantics of the textual context, and associate a polarity to the interaction (\textit{cooperative} vs. \textit{adversarial}).

\subsection{Affiliations}
\label{sec:InterDetAffil}
Unlike the other types of interactions listed in this document, affiliations do not correspond to actions but rather to states. Among others, affiliations include: being blood-related, being married, and belonging to the same social group. They are explicitly mentioned in the narrative, which means that they can appear either in the conversation or in the narration. It is clearly the less popular approach, as only a few authors leverage this type of interactions. Moreover, the methods proposed in the literature only concern textual narratives.

Srivastava \textit{et al}.~\cite{Srivastava2016} leverage an \textit{a priori} selection of keywords (e.g. father, wife) to identify family relationships explicitly mentioned in the text. Krug \textit{et al}.~\cite{Krug2017} propose a semi-supervised method to train a maximum entropy classifier in detecting family, romantic, professional and social relationships in literary texts. Starting from a set of annotated instances, it uses uncertainty-based active learning to select appropriate examples during training and ask the user to label them.

In~\cite{Lee2012f}, Lee \& Yeung detect affiliations using a heuristic approach, based on a dependency parsing and the linguistic resource \textit{FrameNet}~\cite{Baker1998}. They distinguish two cases: the relations stated \textit{directly} (``Sherlock is the brother of Mycroft''), and the ones mentioned \textit{in passing} (``Watson was referring to Sherlock's brother Mycroft''). The authors treat both of them by looking for predefined patterns, designed using a development dataset. More specifically, they look for sentences involving several characters and fitting one of the manually selected semantic frames. Kokkinakis \& Malm~\cite{Kokkinakis2011} apply a relatively similar template-based approach to Swedish novels.

Barbosa, Kondrak \textit{et al}.~\cite{He2013a, Makazhanov2014} adopt an utterance-based approach to extract family relations from novels. After having assigned each utterance to a speaker and a listener, they take advantage of family relationships explictly mentioned in the text (e.g. \textsc{Mother, did you know?}). They then infer additional relations based on manually defined rules. For instance, if two female characters have the same parents, the system can infer that they are sister.

\subsection{Hybrid Approaches}
\label{sec:InterDetMult}
The five different definitions of the notion of interaction presented in this section are not mutually exclusive, and can be combined in various ways. For instance, Kyrios \textit{et al}. consider all types of interactions except mentions and affiliations~\cite{Kydros2015a}; Mutton~\cite{Mutton2004} and Deleris \textit{et al}.~\cite{Deleris2018} combine conversations and mentions; Lee \& Yeung detect conversations, affiliations, and direct actions~\cite{Lee2012f}. Srivastava \textit{et al}. use co-occurrences, affiliations, and direct actions, and they also consider how similarly characters are described in the text~\cite{Srivastava2016}. When studying Edgeworth's \textit{The Absentee}, Falk \textit{et al}. detect not only oral conversations, but also \textit{written} ones~\cite{Falk2016}, which belong to the class of direct actions in our nomenclature. He \textit{et al}. use two distinct definitions, but in conjunction rather than disjunction: they use affiliations only if expressed in conversations~\cite{He2013a}.

Conducting manually the extraction process can be very costly, but this provides some advantages in return: some consider the obtained results as more reliable compared to automatic tools~\cite{MacCarron2012, Bazzan2018}, and this allows more flexibility when defining what an interaction is. Certain authors leverage this flexibility to simultaneously consider all forms of interactions~\cite{Miranda2013, Sparavigna2013, Kydros2015a} or distinguish between friendly and hostile interactions~\cite{MacCarron2012, MacCarron2013, MacCarron2014a}. Others define additional constraints, e.g. focusing only on face-to-face interactions~\cite{Prado2016}, or visual and physical contacts~\cite{Bazzan2018}.

Mixing several types of interactions raises the problem of combining their laterality, as some may be unilateral and others bilateral. Kydros \& Anastasiadis~\cite{Kydros2015a} proceed by replacing bilateral interactions such as co-occurrences by two distinct unilateral interactions involving the same characters, but possessing opposed directions.

The literature is lacking proper studies comparing the various interaction definitions, through their effect on the extracted network and possibly the resolution of the problem at hand. Jayannavar \textit{et al}.~\cite{Jayannavar2015} study various combinations: Agarwal \textit{et al}.'s observations (cf. Section~\ref{sec:InterDetDirAct}) only, other direct actions only, both forms of direct actions, and finally conversations as in~\cite{Elson2010}. They conclude the conversational interactions are indeed a subset of the more general ones. In~\cite{Edwards2018}, Edwards \textit{et al}. compare networks based on manually identified direct actions with networks based on automatically detected co-occurrences, as well as conversations and mentions. They conclude that networks obtained automatically are good approximations with regards to density, centrality, and weight distribution, but not transitivity (see Section~\ref{sec:DescriptiveTools} for a description of these topological measures in the context of character networks).

\section{Graph Extraction}
\label{sec:GraphExtraction}
At this stage of the process, the characters and their interactions have been identified. The last step consists in building the character network based on this information. This requires making two methodological choices: how to define vertices and edges.

\paragraph{Vertices}
In almost all the approaches presented in the literature, characters are represented as individual vertices. However, a few authors alternatively consider certain characters \textit{collectively}, and represent them as a single vertex. This can be due to several reasons. The first is that some groups of people are mentioned in an indistinguishable way in the raw material, for instance: townfolks in Twain's \textit{Huckleberry Finn}~\cite{Knuth1993}, Olympian gods and Greek soldiers in Homer's \textit{Iliad}~\cite{Venturini2016}, by-passers in certain plays~\cite{Kydros2015, Muhuri2018} and novels~\cite{Falk2016}. Second, it can also be that some characters always appear simultaneously in the plot: considering them collectively can be viewed as some form of simplification. However, this can have an effect on many of the topological measures computed to describe the network, unless the multiplicity of the vertex is encoded in some way (e.g. vertex weight) and used when processing said measures. Third, it is possible that one wants to model the narrative at the level of groups rather than individuals, as Liu \& Albergante~\cite{Liu2017d} do for the \textit{Game of Thrones} TV series, when they consider what they call \textit{meta-entities} (houses, armies, religious groups) instead of characters.

Handling such collective vertices is difficult when their composition evolves over time. Moreover, it is possible for the same character to appear in the work of fiction both as an individual and as a part of a group. For these reasons, most authors choose to simply ignore such groups altogether, as they are generally secondary~\cite{Agarwal2012, Kydros2015}.

\begin{figure}[htb!]
	\centering
	\begin{tabular}{c@{} c@{} c@{} c}
    	\includegraphics[width=0.23\textwidth]{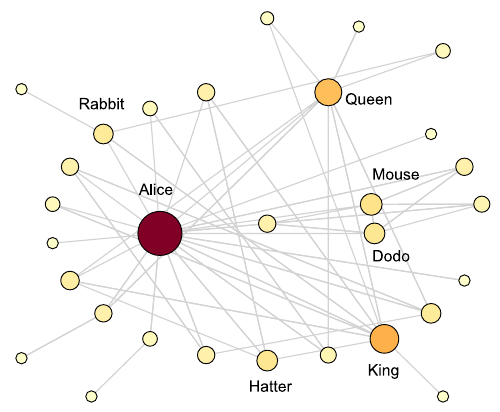} & 
    	\includegraphics[width=0.23\textwidth]{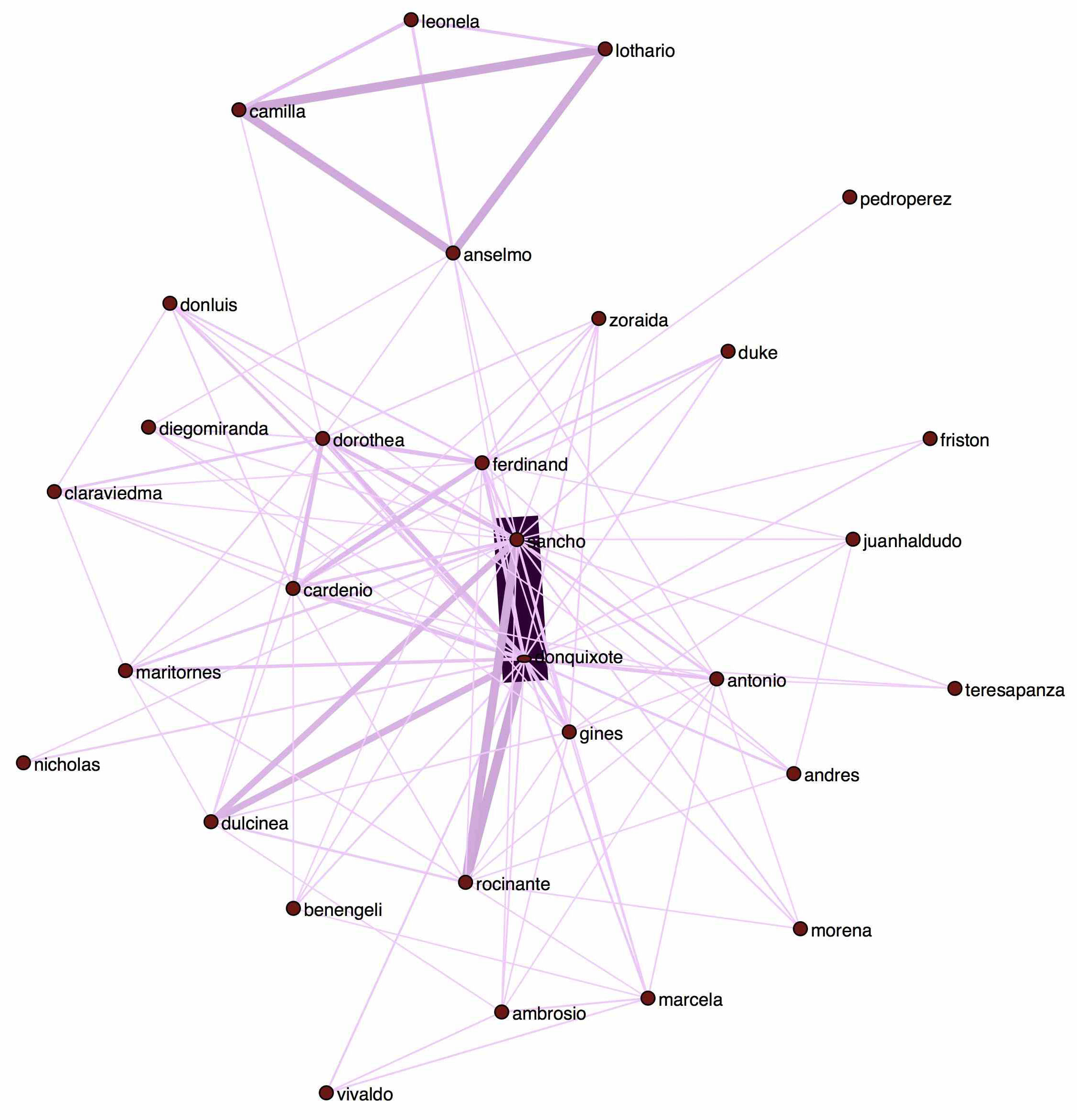} & 
    	\includegraphics[width=0.23\textwidth]{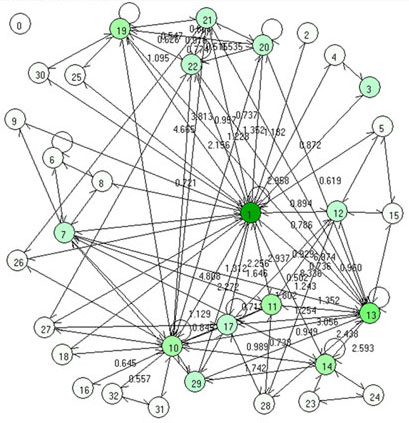} & 
    	\includegraphics[width=0.23\textwidth]{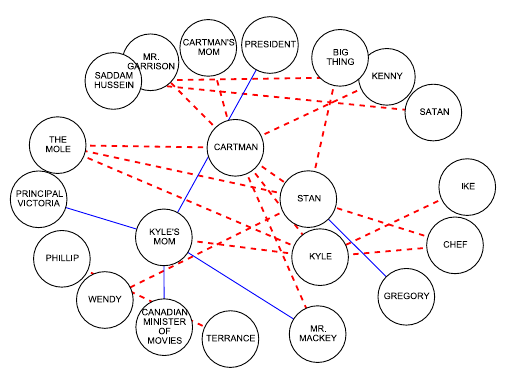}\\
    	\small a) & b) & c) & d) \\
	\end{tabular}
	\caption{Examples of the different types of edges found in the literature: a) \textit{simple} character network of \textit{Alice in Wonderland}~\cite{Prado2016}; b) \textit{weighted} character network of \textit{Don Quixote}~\cite{Sack2013}; c) \textit{directed} character network of the movie \textit{Avatar}~\cite{Jung2013a}; d) \textit{signed} character network of the movie \textit{South Park: Bigger, Longer \& Uncut}~\cite{Krishnan2015}.}
	\label{fig:edges}
\end{figure}

\paragraph{Edges}
As illustrated by Figure~\ref{fig:edges}, the methods proposed to define edges differ much more from one author to the other, which is why we focus on them in this section. One has to consider several aspects of the interactions, which must be translated in graph-related concepts: their laterality, score, polarity, and temporality. The general approach is to represent unilateral interactions by directed edges, and bilateral ones either by undirected edges or by pairs of reciprocal directed edges. The scores computed by certain authors to measure interaction intensity are modeled by edge weights, which can be signed to represent the polarity of the interaction (friendly vs. hostile). This results in graphs which can be (un)directed, (un)weighted and (un)signed. A signed network has both \textit{positive} and \textit{negative} edges, and is used to model antagonistic relationships~\cite{Heider1946}.

Interestingly, certain authors decide to ignore information that could be modeled as edge directions or weights, even if they have it at their disposal. This can be due to some methodological priorities, for instance, Yose \textit{et al}. state that using directed edges to model unilateral interactions would result in a loss of statistical power when testing certain assumptions on the resulting network. The choice can also be caused by the will to adopt a simple approach. For example, Yose \textit{et al}. justify their use of unweighted networks by the fact they are interested only in the presence of a relation between two characters, and not by its intensity~\cite{Yose2017}.

The most important methodological issue one has to solve when extracting edges is to decide how to handle time. It is possible to consider the complete set of interactions occurring between two characters of interest over the narrative, and create a single edge summarizing all of them: this results in \textit{static} networks (Section~\ref{sec:GraphExtrStatic}). However, this complete temporal integration leads to some information loss. Certain authors propose methods aiming at solving this issue, and produce \textit{dynamic} networks (Section~\ref{sec:GraphExtrDynamic}) instead.

\subsection{Static Networks}
\label{sec:GraphExtrStatic}
In most of the approaches presented in the literature, the extracted character network is \textit{static}, i.e. it represents the interactions between the characters for the period of time corresponding to the whole narrative. As interactions can occur anywhere in the storyline, obtaining such a network requires some form of temporal integration. On the one hand, the most widespread approach is to consider the set of successive interactions between two characters and derive an edge by applying a simple mathematical function, e.g. by counting them (Section~\ref{sec:GraphExtrStatCount}). On the other hand, certain authors propose more advanced methods based on the explicit modeling of inter-character relationships (Section~\ref{sec:GraphExtrStatModel}).

\subsubsection{Count-Based Approaches}
\label{sec:GraphExtrStatCount} 
The most widespread methods to derive an edge from a series of interactions are relatively simple, relying on the existence or number of such interactions. Certain authors then discard some of the resulting edges, which they deem unreliable. Finally, a different extraction approach consists in extracting networks whose vertices represent groups of characters, and edges correspond to various forms of overlap between them.

\begin{figure}[htb!]
	\centering
	\begin{tabular}{c c}
    	\includegraphics[height=5cm]{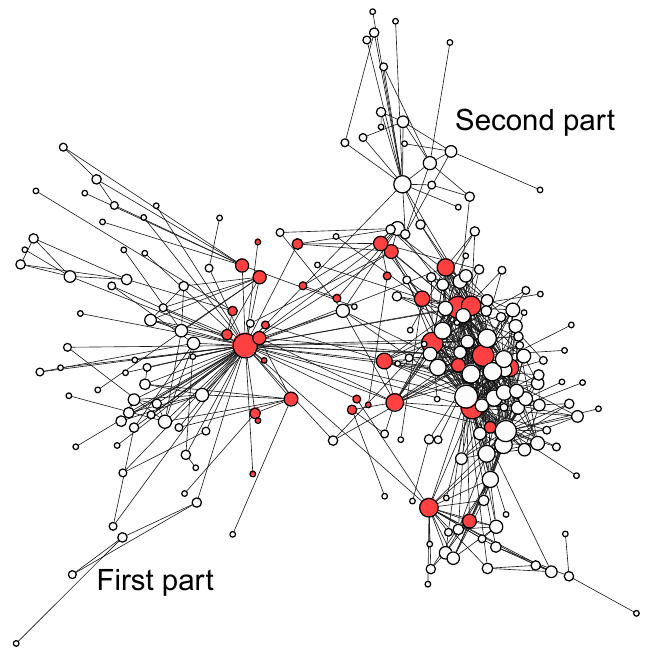} & 
    	\includegraphics[height=5cm]{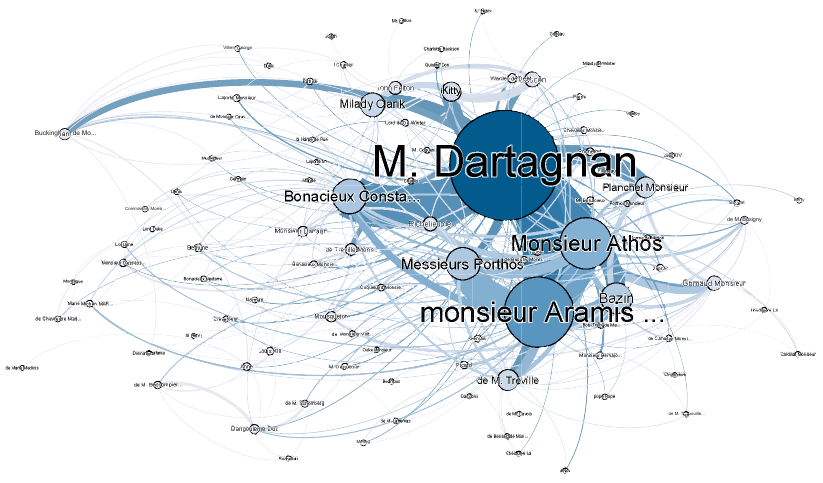}\\
    	\small a) & b) \\
	\end{tabular}
	\caption{a) \textit{unweighted} character network of Rousseau's \textit{Les confessions}~\cite{Rochat2014a}, integrated over the two parts of the autobiographic novel (red nodes denotes characters appearing in both parts). b) \textit{weighted} character network of \textit{Les trois mousquetaires}~\cite{Dekker2019}, integrated over the whole novel.}
	\label{fig:integration}
\end{figure}

\paragraph{Interaction aggregation.}
The simplest form of temporal integration, and also the most widespread, results in an \textit{unweighted} network (cf. Figure~\ref{fig:integration}a). It consists in creating an edge between two vertices if at least one interaction is detected between the corresponding characters over the whole narrative, e.g.~\cite{Moretti2011a, Stiller2003, Min2016, Ribeiro2016, Holanda2017}. When interactions are described by their polarity, it is possible to produce a \textit{signed} network: the sign of an edge is obtained by considering the majority sign among the set of interactions between the two concerned characters~\cite{Trovati2014}.

It is straightforward to generalize this approach to produce a \textit{weighed} network, in which the weight of an edge reflects the overall intensity of all the interactions occurring between the two considered characters (cf. Figure~\ref{fig:integration}b). Certain authors use a frequency-based weight, the most straightforward being the number of times they interacted~\cite{Rochat2014, Mutton2004, Elson2010, Lee2016d}, or the proportion of interactions~\cite{Park2009, Sudhahar2013}. Amancio~\cite{Amancio2015b} applies a probabilistic normalization to the number of interactions, which favors the stronger relationships. Sang \textit{et al}.~\cite{Sang2011, Sang2012} consider that their weights are noisy, and prefer to turn them into ordinal values by replacing them by their respective ranks.

If the interactions are already associated to some numerical score describing their intensity, it is possible to combine them (instead of just counting the interactions). Certain authors sum the scores of all the interactions detected for the considered pair of characters~\cite{Jung2013a, Grayson2016, Pope2016}, or take their average value~\cite{Park2013, Seo2014, Muhuri2018}. Alternatively, others normalize the sum, for instance by dividing by the total network weight~\cite{Park2011}, or through some non-linear function~\cite{Seo2013}.

A very few authors extract co-occurrence networks that do not take the form of \textit{unipartite} graphs as above, but rather of \textit{bipartite} ones~\cite{Alberich2002, Weng2007, Grandjean2015}, which contains two distinct types of vertices: characters vs. narrative units. Each edge connects a character to a narrative unit, and represents the occurrence of this character during this unit. However, maybe because there are many more tools designed to study unipartite graphs, they eventually perform a projection over the character dimension to make their graphs unipartite. In the end, this is equivalent to directly aggregating the interactions as previously described.

\paragraph{Edge Filtering.}
Certain authors consider that the weakest relationships amount to noise, and choose to remove edges with low weights. This can be done through a fixed threshold, e.g. at least three~\cite{Rochat2014, Sack2012} or five~\cite{Iyyer2016} co-occurrences to keep an edge, whereas Stiller \& Hudson just remove isolates~\cite{Stiller2005}. Park \textit{et al}. use a normalized weight, for which they empirically define a minimal threshold~\cite{Park2013}. Sudhahar \& Cristianini~\cite{Sudhahar2013} do not deal with a single narrative, but with a \textit{collection} of related stories, at once. They filter relationships depending on the number of documents in which they appear. The same method could be applied to subdivisions of a single fiction work. 

Amancio~\cite{Amancio2015b} assesses the statistical relevance of a relationship through a null model. He works on co-occurrences, so his approach consists in comparing the edge weight (number of co-occurrences between the concerned characters) with the same weight computed over a randomly generated text. In~\cite{Seo2013}, the authors transform their weights so that they represent distance in place of intensity. They then extract the minimum weighted spanning tree of their co-occurrence network, and use it in place of the original network, therefore discarding the rest of the edges.

\paragraph{Networks of character groups.}
Certain authors aggregate interactions, and more precisely co-occurrences, not only temporally, but also over \textit{groups of characters}. Tsai \textit{et al}.~\cite{Tsai2013} extract a direct network in which a vertex corresponds to a group of characters co-occurring during the same scene, and possibly several ones. An edge going from one vertex to the other represents an inclusion between the character sets they represent. As described in Section~\ref{sec:AppSummary}, this type of network is designed specifically to summarize the plot. Grener \textit{et al}.~\cite{Grener2017} proceed similarly, but an edge represents any non-empty intersection between two groups (instead of strict inclusion). Hettinger \textit{et al}.~\cite{Hettinger2015} model temporal proximity rather than group similarity, by connecting two groups appearing successively in the plot.

\subsubsection{Model-Based Approaches}
\label{sec:GraphExtrStatModel} 
Certain authors propose specific models to represent the relationships between two characters, taking advantage of the sequence of interactions between them, but also possibly of additional information such as the context of these interactions in the narrative. All these models rely either on co-occurrence- or conversation-related interactions.

\paragraph{Co-occurrences.}
Elson \textit{et al}.~\cite{Elson2010} describe characters by vectors constituted of their numbers of occurrences during each unit of the considered narrative. They compute the correlation matrix over this representation of the character set, and use it as a weighted adjacency matrix to obtain what could be called an occurrence similarity network.
Several authors propose to compare the context in which the characters occur in terms of topics. Kokkinakis \& Malm~\cite{Kokkinakis2011} do so in Swedish novels: they characterize contexts through bags-of-words, and detect similar contexts using cluster analysis. They connect characters occurring in similar context, but also combine this information with other features, such as affiliations (see Section~\ref{sec:InterDetAffil}). Hutchinson \textit{et al}.~\cite{Hutchinson2012} first perform a Latent Semantic Analysis (LSA) to identify the lexical context of character occurrences in novels, followed by a Singular Value Decomposition (SVD) for dimension reduction. In the resulting matrix, each character is represented by a real vector. They then use the cosine similarity to compute the weight associated to each pair of characters. The result is a topic similarity network. Nan \textit{et al}.~\cite{Nan2015} proceed similarly with TV series, by combining visual (objects surrounding the characters at the time of the occurrences) and textual information (subtitles).

Ding \& Yilmaz~\cite{Ding2010} assume the existence of two conflicting groups of characters in the plot. As explained in Section~\ref{sec:InterDetCoocScore}, at the interaction detection step they leverage the narrative to compute a score representing the polarity of each scene. They then propose a Bayesian model taking into account the sequence of scene polarity scores and each character's group membership. It allows estimating the overall signed weight representing the polarity of the relationship between any two characters for the whole narrative, resulting in a signed weighted network.
In~\cite{Srivastava2016}, Srivastava \textit{et al}. define a supervised approach to estimate the polarity of the relationships. They first extract an unsigned graph based on co-occurrences, affiliations, and other interactions. They then train a classifier to predict the polarity of the relations, based on a number of textual features describing the context of the interactions, as well as structural features related to the notion of \textit{structural balance} (categories of signed triads, cf. Section~\ref{sec:ToolsTriangles}).

\paragraph{Conversations.}
Celikyilmaz \textit{et al}.~\cite{Celikyilmaz2010} focus on the content of the utterances. They propose a Bayesian model describing each character in terms of the topic about which he talks. From this, they extract build a topical similarity network: a fully connected graph in which the weight of the edge connecting two characters reflects how much they talk about the same things. According to the authors, this allows identifying hidden relationships between characters.

In~\cite{Krishnan2015}, Krishnan \& Eisenstein, take advantage of \textit{how} two characters address each other over all their conversations, to estimate the nature of their relationship: \textit{informal} vs. \textit{formal}. The resulting network is therefore signed, positive and negative edges representing informal and formal relationships, respectively. In addition to linguistic content, their probabilistic model is able to take into account the structure of the graph, for example by assuming it is consistent with structural balance.

\subsection{Dynamic Networks}
\label{sec:GraphExtrDynamic}
Static networks present an obvious limitation, as identified by numerous authors in the context of character networks~\cite{Rochat2014, Oelke2013, Ardanuy2014, Bost2016, Tran2017b}: they result in a significant information loss, as they completely hide the chronology of the interactions. Yet, the order in which events occur is crucial to characterize a story, and for the writer it is generally at the core of the writing process. Moreover, the relationship between two characters is likely to evolve with the plot. This can lead to poor performance when solving certain problems based on a static network, or even completely prevent any resolution. Even from the descriptive point of view, Prado \textit{et al}.~\cite{Prado2016} show empirically that the most central characters detected in static vs. dynamic networks can differ dramatically.

Certain authors try to cope for this by allowing multiple edges between vertices. For instance, in their signed network, Yose \textit{et al}.~\cite{Yose2017} can have both a positive edge and a negative one connecting the same pair of characters, in order to model a relation evolving over time. Although simple, this is a very imperfect and incomplete solution. Another possibility is to extract a dynamic network. It takes the form of a sequence of character graphs called \textit{times slices}, each one corresponding to a certain sub-period of the story. Like a static graph, a time slice is obtained through temporal integration, as it represents all the interactions occurring over a period of time. But this integration is performed at a much smaller time scale, which allows limiting the information loss. We distinguish two types of approaches: those using a fixed-size window to obtain these time slices (Section~\ref{sec:GraphExtrDynWindow}) vs. the others (Section~\ref{sec:GraphExtrDynOthers}).

\subsubsection{Temporal Window}
\label{sec:GraphExtrDynWindow}
\begin{figure}[htb!]
	\centering
	\includegraphics[width=\textwidth]{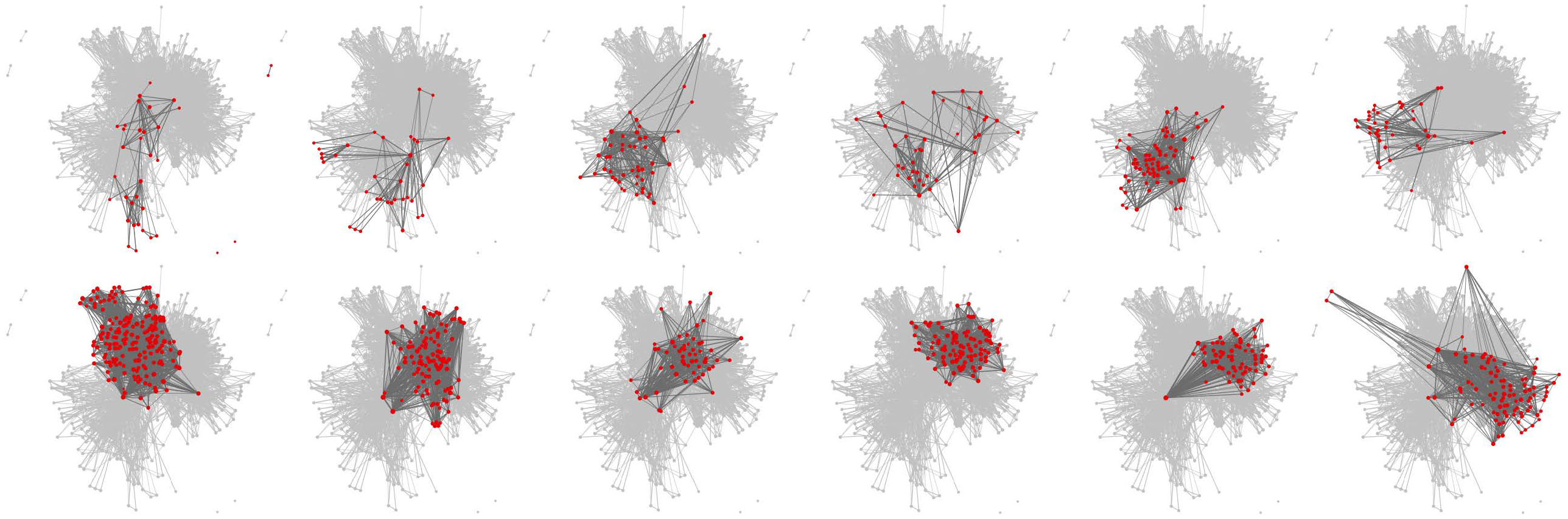}
	\caption{\textit{Dynamic} version of the character network of Rousseau's \textit{Les confessions}~\cite{Rochat2014a}, whose \textit{static} version is presented in Figure~\ref{fig:integration}a. Each time slice corresponds to a specific chapter.}
	\label{fig:window}
\end{figure}

The fixed-size temporal window approach is clearly the most widespread. The notion of window depends on both the nature of the considered narrative and the utilization of the extracted network. The literature exhibits a variety of window sizes. For novels, most authors use one chapter~\cite{Agarwal2012, Ardanuy2014, Grayson2016, Prado2016}, as in Figure~\ref{fig:window}. Grener \textit{et al}.~\cite{Grener2017} use a $1,000$-word window, whereas Seo \textit{et al}.~\cite{Seo2014} arbitrarily split the novel in ten equal-sized pieces. In theater scripts, Kwon \& Shim~\cite{Kwon2017} use a one-act window. For movies, authors use one scene~\cite{Ding2011a}, or automatically detected segments roughly corresponding to scenes~\cite{Lee2018}. For TV series, authors use one scene~\cite{Bost2016}, one episode~\cite{Liu2017d, Bazzan2018} or one season~\cite{Bazzan2018}. Furthermore, certain authors use an overlap between consecutive windows, in order to preserve continuity as much as possible, e.g. for novels: fifteen pages with a fourteen-page overlap~\cite{Rochat2014}, $100$ sentences with a ten-sentence overlap~\cite{Oelke2013}. Weights can be processed for each slice, similarly to what is done for static networks (cf. Section~\ref{sec:GraphExtrStatCount}).

\begin{figure}[htb!]
	\centering
	\includegraphics[width=\textwidth]{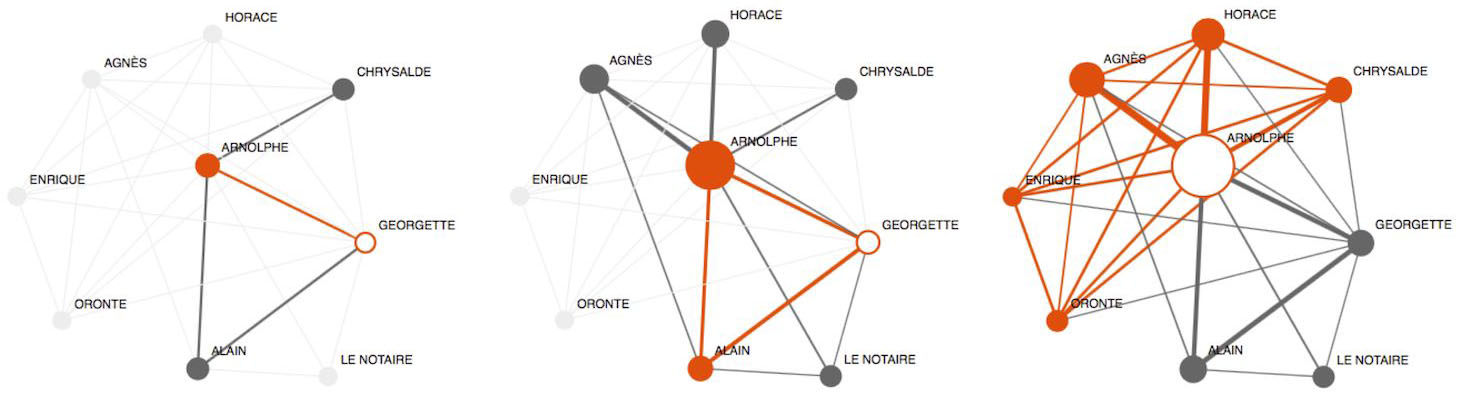}
	\caption{Dynamic \textit{cumulative} character network of Molière's \textit{L'école des femmes}~\cite{Xanthos2016}. By construction, the size of the network increases over time.}
	\label{fig:cumulative}
\end{figure}

Certain authors extract a dynamic but \textit{cumulative} network~\cite{Nalisnick2013, Waumans2015, Min2016, Tran2017b}. This means that, to obtain the graph representing a given moment in the narrative, they do not use only the corresponding time slice, but rather the portion of the narrative going from the very start to the considered moment, as illustrated by Figure~\ref{fig:cumulative}. Other than that, the principle is the same as before: they perform a temporal integration over this period, and do so for every moment in the narrative to get a series of graphs constituting their dynamic network. As before, some authors consider unweighted graphs~\cite{Waumans2015}, others use the numbers of interactions~\cite{Min2016, Xanthos2016, Tran2017b} or the total interaction scores~\cite{Tran2017c} as edge weights. Nalisnick \& Braid~\cite{Nalisnick2013} proceed similarly, but as their interaction scores are signed, the weights of the edges do not necessarily increase with time. On the contrary, one of the benefits of their approach is that a moment in the plot corresponding to a sign inversion for a given relationship is supposed to be narratively important for the concerned characters.

Generally speaking, using a fixed-size window to discretize a time series requires knowing which window size to use. This is no trivial task, and it was shown that this parameter can have a very strong effect on the extracted network, and consequently on the process conducted on this network~\cite{Clauset2007a}. With character graphs in particular, on the one hand too large windows will miss many details, as relationships in works of fiction can evolve quickly, and on the other hand too small windows will result in very unstable networks, which may hide relevant information~\cite{Oelke2013}. Bost \textit{et al}.~\cite{Bost2016} have shown this experimentally, and also that a window too small can lead to mistaking irrelevant events as important ones. Finally, a window too large may also cause problems when later studying the network: when the time slices differ much in terms of involved characters, it makes little sense to compare them through topological measures, as these are generally defined in a relative way~\cite{Prado2016}. Grayson \textit{et al}.~\cite{Grayson2016} propose to increase the window size until the graph density reaches a plateau, but this can produce very dense networks, likely to be uninformative.

\subsubsection{Other Methods}
\label{sec:GraphExtrDynOthers}
In addition to the difficulties of estimating the best window size, there is no reason to suppose that this size should even be fixed: the tempo of the narrative is likely to change in any way, e.g. accelerating during action scenes and slowing down during emotional ones. A very few authors try to take this point into account.

Instead of considering fixed-size time slices, Mutton~\cite{Mutton2004} uses an event-based approach to extract the sequence of graphs constituting his dynamic network. Each modification (creation or deletion of an edge, revision of a weight) results in the addition of a new graph in the sequence, representing the state of the character relationships after the modification. In some sense, this is the smaller possible time slice. Like some methods from Section~\ref{sec:GraphExtrDynWindow}, he adopts a cumulative approach. However, he also includes a temporal decay mechanism, which decreases the weights of an edge depending on the last time it was updated.

Another important point is that the narrative is linear, but not necessarily the story, or even the plot\footnote{The notions of plot (\textit{what} is told), narrative (\textit{how} it is told), and story (\textit{perceived}) are defined in the introduction.}. The narrative does not reflect the full state of all inter-character interactions at all times: it focuses on certain characters, presents under a sequential form some events actually occurring simultaneously, shows past events as flashbacks. A character can be absent from the narrative for a while, but still be active under the scene, performing actions that will be revealed later to the audience. Some authors consider it is necessary to model this hidden evolution to represent fully the character graph. Elsner~\cite{Elsner2012} do so by applying a linear interpolation to determine the weight of a relation at a given point in the narrative for which this relation does not appear explicitly.

In~\cite{Bost2016, Bost2016b}, Bost \textit{et al}. propose \textit{narrative smoothing}, an interpolation method allowing both to solve this problem and to mitigate the fixed-size window issue. Starting from the list of verbal interactions occurring at each scene, and weighted depending on their duration, they apply the following transformation to get their smoothed dynamic network. For a given pair of characters, if a verbal interaction occurs at the considered scene, the smoothed weight is the raw weight (duration). Otherwise, they take into account the last and next interactions of the considered characters: the raw weight is decreased depending on how much the characters interact with \textit{others} in the meantime. Note that this decrease does not depend on the amount of time between two interactions of the considered characters, in order to avoid the fixed-size problem. This method allows estimating the weight of the relationship between any two characters at any moment of the narrative, even if there are not interacting at this time.

\section{Analysis and Applications}
\label{sec:AnalApps}
In this section, we describe how character networks can be used. We first review standard tools widely used to describe general networks, and discuss their interpretation in the specific case of character networks (Section~\ref{sec:DescriptiveTools}). Then, we turn to real-world applications, and review the high-level problems solved thanks to character networks in the literature (Section~\ref{sec:Applications}).

\subsection{Descriptive Tools}
\label{sec:DescriptiveTools}
The literature contains a number of tools designed to characterize and study complex networks in general. Technically, most of them can be applied to character networks. However, their output is not necessarily relevant to this specific type of network, and when it is, some of them are used to objectively describe these networks without providing any specific interpretation (e.g. centrality measures for automatic classification~\cite{Waumans2015} or motifs for model fitting~\cite{Bonato2016}). In this section, we review the topological measures used in the literature to describe character networks, focusing only on those for which authors propose a character network-related interpretation. We group them depending on the general approach they rely upon to describe the network.

\subsubsection{Graph Density and Edge Weight}
\label{sec:ToolsDensityWeight}
The most basic tools rely on counting the edges constituting the graph, or studying their weights (in the case of weighted graphs).

\paragraph{Graph Density.}
The density is the proportion of existing edges in the network, i.e. the ratio of its observed to possible edges. It is frequently used in the literature (together with other measures such as the number of vertices or edges) to compare networks extracted from several works, e.g.~\cite{Elson2010, Rochat2015, Jayannavar2015, Holanda2017}. A generalization of the density exists for weighted networks~\cite{Grayson2016}.

Certain authors directly and explicitly tie the density to some aspects of the narrative. Rochat \& Triclot~\cite{Rochat2017} use it to identify \textit{behind closed doors} stories, which they associate to a high density (in addition to a structure devoid of any real center). Stiller \textit{et al}.~\cite{Stiller2003, Stiller2005} study the evolution of the density as a function of the number of characters in Shakespeare's plays. They show that it decreases, which they tie to the will of the writer not to involve too many characters in his subplots, in order to keep the story understandable. Together with a moderate transitivity, Sack~\cite{Sack2012} interprets low density as a marker of social dissociation, i.e. presence of largely disconnected social groups. 

It is worth noting that, despite the apparently plot-related interpretation of density, this measure is also used to characterize stylistic aspects of writing. For instance, Voloshinov \& Gozhanskaya~\cite{Voloshinov2008} use the density to discriminate plays by different playwrights.

\paragraph{Edge Weight.}
By construction, a weight (the numerical value associated to an edge in weighted networks) represents the intensity of the relation modeled by the concerned edge. In the literature, it is considered from a comparative perspective, according to two modes: between characters and over time.

\begin{figure}[htb!]
	\centering
	\begin{tabular}{c c}
	    \includegraphics[height=4.75cm]{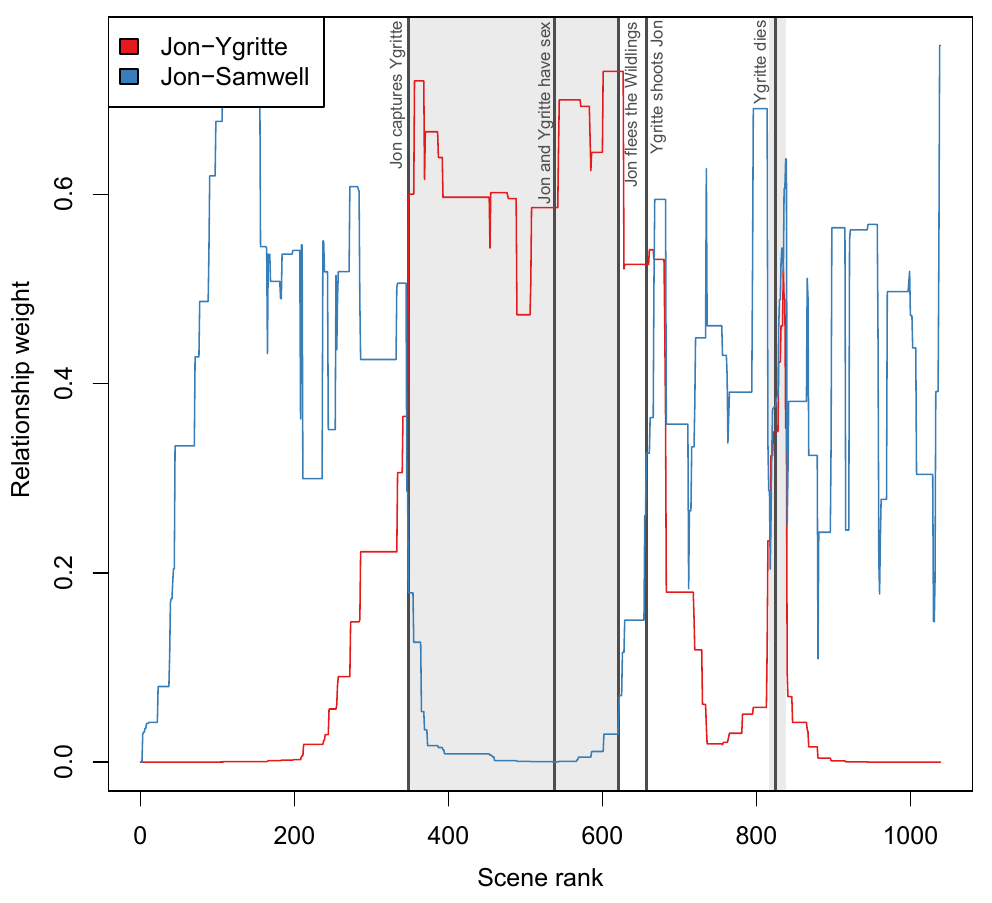} &
	    \includegraphics[height=4.75cm]{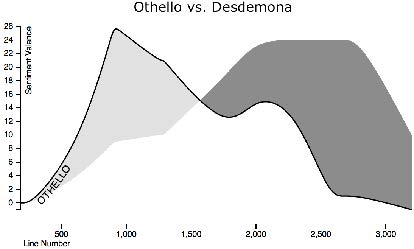} \\
	    a) & b) \\
    \end{tabular}
	\caption{a) Evolution of the link weights for three important characters in the \textit{Game of Thrones} TV series~\cite{Bost2016b}; b) Evolution of the signed weight of a relationship of interest in \textit{Othello}~\cite{Nalisnick2013}.}
	\label{fig:weights}
\end{figure}

Certain authors distinguish between weak and strong edges, depending on their values. Sack~\cite{Sack2012} observes that a low proportion of the latter indicates infrequent and superficial interactions (as their weights correspond to interaction counts). These are often taking place between the protagonist and punctual minor characters, in order to provide variety, color, and some background to the story~\cite{Sack2014}. On the contrary, a large proportion of strong edges is indicative of complex and durable relationships, developed over a long period of time~\cite{Sack2014}. Suen \textit{et al}.~\cite{Suen2013} focus only on the strongest edges: they propose to compare the top two weights of the network, in order to assess how much the plot focuses on the most important relationships relative to the others. It is also possible to consider the whole distribution, as Gleiser does for comics~\cite{Gleiser2007}: he identifies a power law, and interprets it as revealing the presence of a few relationships largely dominating the narrative space (e.g. \textsc{Spider Man} and his wife).

Instead of comparing the weights of the edges connecting different characters, some authors focus on a relationship and study how its weight evolves through time (cf. Figure~\ref{fig:weights}). This is what Bost \textit{et al}.~\cite{Bost2016, Bost2016b} do for TV series, interpreting significant changes as important event occurring in the narrative. Nalisnick \& Braid~\cite{Nalisnick2013} adopt the same principle, with the difference that their weights are signed. They interpret a sign change as an event of importance. For instance, in Shakespeare's \textit{Hamlet}, the moment where the sign of the relationship between the eponymous character and his mother switches from negative to positive corresponds to a crucial revelation regarding the death of their father/husband.

\subsubsection{Degree and Strength}
\label{sec:ToolsDegree}
The \textit{degree} of a vertex is the number of edges attached to it. It is sometimes normalized to get a value in $[0;1]$, as in~\cite{Stiller2005}. If the considered network is directed, one can distinguish the \textit{incoming} from the \textit{outgoing} degrees, which correspond to the number of incoming and outgoing edges attached to the considered vertex, respectively. The \textit{strength} is the weighted generalization of the degree: the sum of weights over the edges attached to the vertex. A high strength can be due to few edges with high weight, but also to many ones with small weight, so both measures are not necessarily correlated.

\paragraph{Hubs.}
A \textit{hub} is a vertex with a high degree. In a character network, it is interpreted as a pivotal character, one that interacts with many others. Depending on how the network is extracted, this large number of interactions can be spread over time, or concentrated in a few events involving many characters. When considering strength, the idea is the same as for the degree, but considering the intensity (or whatever property conveyed by the weights) of the exchanges. Many authors use these measures to identify primary characters in the narrative. For instance, Hamlet has the highest degree in Shakespeare's \textit{Hamlet}~\cite{Moretti2011a}, and the hero always has the highest degree in the collection of stories and tales analyzed in~\cite{Sudhahar2013}. When studying Bengali Literature, Muhuri \textit{et al}.~\cite{Muhuri2018} observe that both protagonists and antagonists have a significantly higher strength than minor characters. If the narrator is a character, she is probably a hub, as she will tell the story from her perspective, and is therefore likely to intervene more~\cite{Rochat2014a}. Of course, it is possible for a character network to contain no hub at all, e.g. in the case of a complete network (all characters are interconnected), or to contain several hubs. Agarwal \textit{et al}.~\cite{Agarwal2012} propose a method to identify the narrator (see Section~\ref{sec:AppRoleDetection}).

In directed networks, the incoming and outgoing degrees or strengths are often seen as popularity and sociability markers~\cite{Muhuri2018}, respectively. It is also possible to interpret them in terms of dependency and force for change, as an incoming hub is someone towards which other characters turn to in case of need, and an outgoing one denotes a will to act on the environment. For instance, in their directed mythological network, Choi \textit{et al}.~~\cite{Choi2007} identify Zeus as an incoming hub (\textit{in charge} for the whole world) and Heracles as an outgoing one (the \textit{cause} of many great events). Sudhahar \& Cristianini~\cite{Sudhahar2013} make a similar observation for stories and tales, as heroes generally have a higher outgoing strength (they tend to be the agent of many actions). However, they also notice that strength (or degree) alone may not be enough to identify primary characters, since other authoritative characters (e.g. God, the King) can be as much (or even more) of a hub.

When a dynamic network is extracted, certain authors study the evolution of the degree to study the importance of the characters. For example, Rochat~\cite{Rochat2014a} shows that in Rousseau's \textit{Les confessions}, the transition from the first part of the novel to its second part corresponds to some sort of passing of power between two important characters. Agarwal \textit{et al}. observe that \textsc{Alice} has not always the highest degree in \textit{Alice in Wonderland}~\cite{Agarwal2012}. Min \& Park show that in \textit{Les miserables}, the focus alternates between \textsc{Valjean} and \textsc{Marius}. Bost \textit{et al}.~\cite{Bost2016, Bost2016b} leverage degree evolution to detect important events, such as the death of a character.

\begin{figure}[htb!]
	\centering
	\begin{tabular}{c@{} c@{} c}
	    \includegraphics[height=4.8cm]{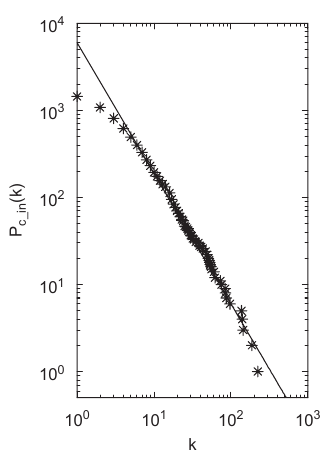} &
	    \includegraphics[height=4.8cm]{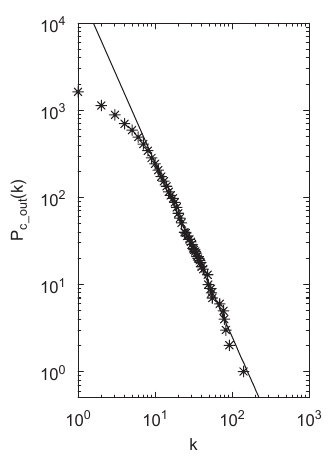} &
	    \includegraphics[height=4.8cm]{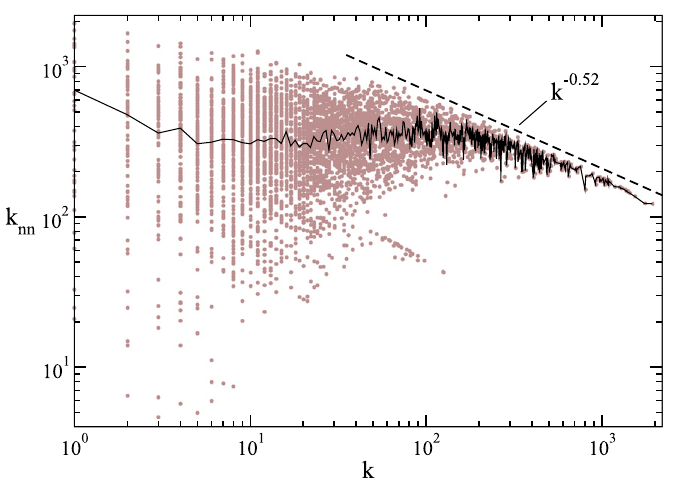} \\
	    a) & b) & c) \\
    \end{tabular}
	\caption{a) Complementary cumulative in-degree distribution for the \textit{Greek and Roman mythology} network~\cite{Choi2007}; b) Complementary cumulative out-degree distribution for the same network~\cite{Choi2007}; c) Mean degree of the neighbors of a vertex as a function of its degree, for the \textit{Marvel} network~\cite{Gleiser2007}.}
	\label{fig:degreedist}
\end{figure}

\paragraph{Distribution.}
Alternatively, instead of focusing on individual vertices, it is standard in the domain of complex network analysis to study the \textit{distribution} of the degree and/or strength. Authors generally observe either an exponential~\cite{MacCarron2012} or a power law distributed degree, for undirected~\cite{Alberich2002, Park2013, Sparavigna2013, Kwon2017} as well as directed~\cite{Choi2007, Kydros2015a} networks (cf. Figure~\ref{fig:degreedist}a and \ref{fig:degreedist}b). The power law is frequently found in other (non-fictional) social networks. In character networks, it tends to appear when the narrative is dominated by one or a few characters. A network displaying such a degree distribution is called \textit{scale-free}~\cite{Barabasi1999}, and this property has been quite widely observed in many types of real-world networks. This supposed ubiquity has been criticized lately though~\cite{Labatut2014b, Broido2018}, as properly determining if a sample follows a power law is difficult in practice.

With a power law distribution, the mean is not a very informative statistic to describe the degree, as it is not a characteristic value. However, authors traditionally like to indicate it~\cite{MacCarron2012, Kydros2015, Holanda2017}, sometimes with the minimal and maximal degree values~\cite{Alberich2002, Yose2017}, standard deviation~\cite{Sack2012}, variance~\cite{Suen2013}, or median~\cite{Tan2014a}. In~\cite{Elson2010, Jayannavar2015}, the average degree and strength are two among several features used to compare networks extracted from distinct works.

Other statistics are designed to summarize the degree distribution, especially its heterogeneity. Suen \textit{et al}.~\cite{Suen2013} propose the \textit{Single character centrality}, which is not unlike Freeman's \textit{Centralization}~\cite{Freeman1978}. It is based on the comparison of the network highest and second-highest degree/strength, normalized by its total degree/strength. By comparison, the centralization is defined generically, for any vertex centrality measure, and compares how central the most central vertex of the network is, relatively to the rest of the vertices. The Single character centrality allows quantifying how much the top character dominates the others. In order to assess the spread of the distribution, they also propose to use the entropy, and the degree/strength mean and variance for the network restricted to its top $10$ characters. The same statistics could, of course, be applied to any other centrality measure. Suen \textit{et al}. use all of these as features to train a classifier in distinguishing predefined classes of plots (cf. Section~\ref{sec:AppClassification}).

\paragraph{Assortativity.}
The \textit{degree assortativity}~\cite{Newman2002} measures how correlated the degrees of connected vertices are. Real-world social networks are usually assortative, i.e. their vertices tend to be connected to other vertices of similar degree~\cite{Newman2003b}, but this is not the case for many other types of real-world networks (e.g. technological or biological networks). Rochat uses assortativity~\cite{Rochat2014a} to characterize the overall relationship between primary and minor characters. In the case of Rousseau's \textit{Les confessions}, he finds a slightly disassortative network, indicating that the edges tend to connect a small set of primary characters present in most of the narrative, to a large set of minor characters merely passing by. Moreover, it also shows that major (resp. minor) characters do not tend to connect to each other. 

Similarly, Mac Carron \& Kenna~\cite{MacCarron2012} observe that certain fictional character networks tend to be disassortative. They make the same observation for epic narratives, yet these are not necessarily (completely) fictional. They assume this is due to the conflictual nature of their plots: a lot of characters are introduced just to be killed by heroes, resulting in many small degree vertices connected to large degree ones. Indeed, when ignoring this type of interactions, the networks are assortative. They use this observation to support the assumption that some aspects of these tales are real, whereas others are fictional.

\subsubsection{Shortest Path and Distance}
\label{sec:ToolsDistance}
The vertex-to-vertex distance, a.k.a. the geodesic distance, or \textit{distance} for short, is very popular to describe networks in general. It corresponds to the length of the shortest path between two vertices. Several popular topological measures are based on the notions of shortest path or distance.

\paragraph{Distance-Based Statistics.}
As a part of the \textit{small world} property~\cite{Watts1998}, it is often observed that the distance averaged over all pairs of vertices (or \textit{average distance}) is relatively small in real-world networks, and increases only slowly (logarithmically, to be precise) with the number of vertices, in growing networks. In such small world networks, the \textit{diameter}, which corresponds to the maximal distance in the network, has roughly the same behavior and is therefore alternatively studied. Most authors examining the average distance or diameter in character networks find small values, e.g.~\cite{Alberich2002, Choi2007, MacCarron2012, Bonato2016}.

\begin{figure}[htb!]
	\centering
	\begin{tabular}{c c}
	    \includegraphics[height=4.5cm]{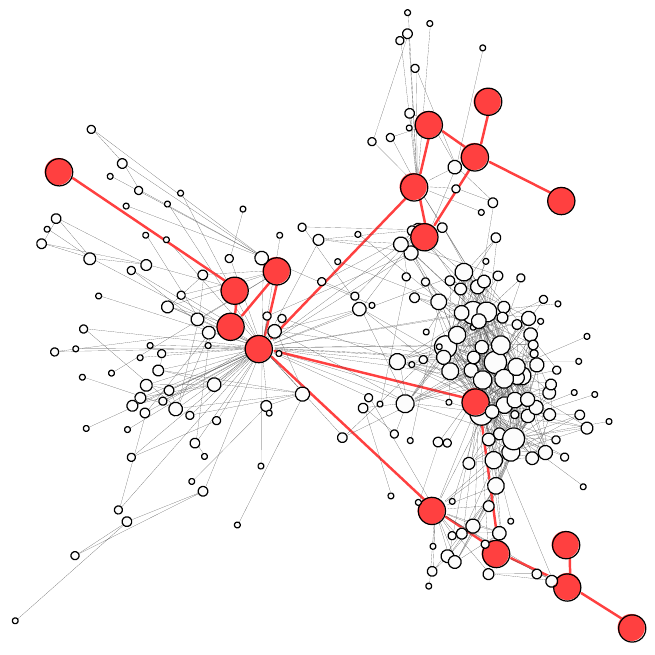} &
	    \includegraphics[height=4.5cm]{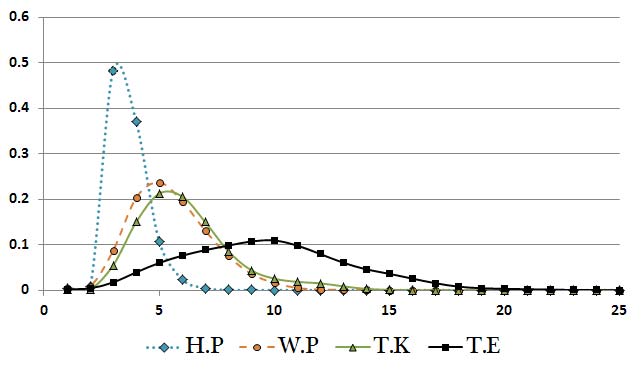} \\
	    a) & b) \\
    \end{tabular}
	\caption{a) All diameter paths (in red) of Rousseau's \textit{Les confessions} (cf. Figure~\ref{fig:integration}a)~\cite{Rochat2014a}; b) Distribution of character-to-character distance in 4 novels (\textit{War and Peace}, \textit{The Three Kingdoms}, \textit{Harry Potter and the Philosopher's Stone}, \textit{The Earth})~\cite{Seo2013}.}
	\label{fig:distance}
\end{figure}

In the context of character networks, both the \textit{average distance} and \textit{diameter} (maximal distance) are used to determine the compactness of the story (cf. Figure~\ref{fig:distance}). A large diameter, with respect to the network size, means that at least certain characters are narratively well separated, whereas a large average distance denotes a more general trend. Rochat~\cite{Rochat2014a} interprets a small diameter as a strong overlap between the characters, resulting in a narrative proximity between characters, even those with different profiles, roles, or appearing at different moments. He also studies the nature of the different shortest paths whose length corresponds to the diameter, and observes that in this case, they connect minor characters by passing through primary ones. Using Cervantes' \textit{Don Quixote} as an example, Sack~\cite{Sack2014} observes that a large diameter is indicative of an episodic narrative, i.e. one where the protagonist successively meets independent groups of characters. This large diameter then goes with a high transitivity (Section~\ref{sec:ToolsTriangles}), reflecting the locally cliquish structure of the network. Incidentally, such a high transitivity is also a prerequisite to Watt \& Strogatz's \textit{small world} property~\cite{Watts1998}. 

\paragraph{Betweenness Centrality.}
The \textit{betweenness} of a vertex~\cite{Freeman1977} is related to the number of shortest paths going through this vertex. In a character network (cf. Table~\ref{tab:centralities}), it allows measuring how much a character acts as a broker or narrative bridge, i.e. connects separate parts of the plot~\cite{Rochat2014a, Bolioli2013}. Characters with low betweenness are considered as narratively dependent on more central ones~\cite{Rochat2014a}. However, Bolioli \textit{et al}. remark that the most central characters in terms of betweenness are not necessarily those appearing as the most important to the reader, as this centrality highlights their bridging role~\cite{Bolioli2013}. Muhuri \textit{et al}. also remark that a high betweenness is associated to a role of messenger, generally towards supporting characters~\cite{Muhuri2018}. In the collections of stories and tales studied by Sudhahar \& Cristianini, heroes have the highest betweenness centrality~\cite{Sudhahar2013}. The edge-betweenness measure is an edge-based variant of the measure\footnote{Edge-betweenness: number of shortest paths going through some edge of interest.}, and allows highlighting important narrative transitions~\cite{Rochat2014a}.

The whole network can be described through betweenness-based statistics, for instance by averaging it over all vertices, or by computing the betweenness-based centralization~\cite{Freeman1978}. In~\cite{Rochat2015}, Rochat computes this centralization for all $20$ novels of Zola's \textit{Les Rougon-Macquart} series of novels, in order to determine which ones are centered on a single protagonist, or revolve around several ones (a type of plot called \textit{polyfocalisation}). Suen \textit{et al}.~\cite{Suen2013} proceed like they do for the degree, i.e. using the maximum value, the difference between the $2$ top vertices, and the entropy to summarize the betweenness distribution.

\begin{table}[htb!]
	\centering
	\begin{tabular}{c c}
	    \includegraphics[height=4cm]{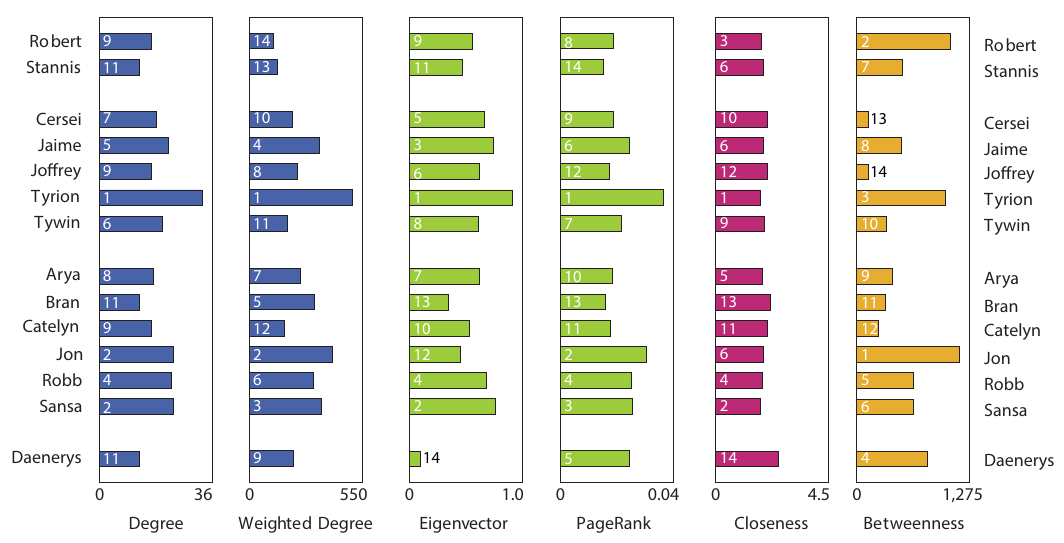} &
	    \includegraphics[height=4cm]{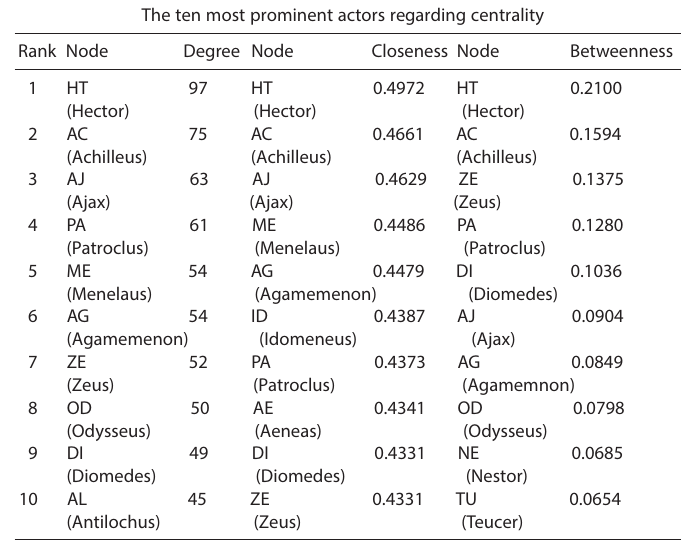} \\
	    a) & b) \\
    \end{tabular}
	\caption{a) Various centrality measures computed for the network of \textit{Game of Thrones}~\cite{Beveridge2016}; b) and for the network of \textit{The Iliad}~\cite{Kydros2015}.}
	\label{tab:centralities}
\end{table}

\paragraph{Closeness Centrality.}
The \textit{closeness} of a vertex~\cite{Bavelas1950} is the reciprocal of the total (or sometimes average) distance between this vertex and the rest of the network. In a character network (cf. Table~\ref{tab:centralities}), it can be considered as measuring how close a character is from all the others, narratively speaking~\cite{Rochat2014a}, i.e. in terms of their respective character-spaces. 

Alberich \textit{et al}.~\cite{Alberich2002} use the closeness to identify the center of their \textit{Marvel universe} network (which turns out to be Captain America). Rochat notes that the variance of this measure is usually low, which makes it harder to discriminate vertices. He prefers to use the \textit{harmonic} closeness (based on the harmonic mean)~\cite{Boldi2014} instead, which additionally allows handling disconnected networks. Rochat uses it to identify important minor characters that, although not proper protagonists, are still involved in non-anecdotic branches of the story.

Some authors alternatively use a related measure, which has a similar interpretation. The \textit{eccentricity} of a vertex is the distance to its farther vertex~\cite{Harary1969}. It can be used to determine the center of the network (vertex of lowest eccentricity), or determine how many centers it contains. For instance, in~\cite{Rochat2014a} Rochat identifies \textsc{Mme de Warens} as the only center in Rousseau's autobiographic novel \textit{Les Confessions}, as she is, by far, the least eccentric vertex (in the story, she is his mentor). In~\cite{Choi2007}, Choi \textit{et al}. consider both the maximal (eccentricity) and minimal distance to a few vertices of interest, to further describe their position.

\subsubsection{Connectivity}
\label{sec:ToolsConnectivity}
It is possible to characterize a network by studying its \textit{connectivity}, either directly or through derived measures. A graph is said to be \textit{connected} if there is a path connecting each pair of its vertices, and disconnected otherwise. A disconnected graph is consequently made of several separated \textit{components} (which are themselves maximally connected subgraphs). Real-world networks generally possess a so-called \textit{giant component}, which contains almost all the vertices.

\paragraph{Components.}
In the context of character networks, the meaning of the connectivity depends on several aspects of the extraction process: raw data, type of interactions modeled by the edges, and whether the network is static or dynamic. Suppose the network represents mentions or affiliations: it is very possible to have separated components, for example because the plot contains two opposed families. But if the edges model conversation or other interactions, authors generally observe a giant component~\cite{MacCarron2013}: indeed, for such edges, a disconnected network would mean there are completely separated subplots. 

This is unlikely, but possible though: for instance, Mac Carron \& Kenna~\cite{MacCarron2012} observe that the epic narrative \textit{Beowulf} contains two subplots taking place in the past, and completely disconnected from the main plot. They appear as two separate components, whereas the main plot is a giant component. It is also possible that the raw material is not a single work, but rather a collection of works. It is the case, for instance, with the comic books composing the Marvel universe, which leads to a disconnected network dominated by a giant component~\cite{Alberich2002}. It is also likely to see such disconnections if the network is dynamic: one can observe several separated components at some point, representing several subplots likely to merge later in the story. In the presence of a giant component, some authors choose to ignore isolates and small components~\cite{Kydros2015, Kydros2015a}, which are considered as negligible, as is generally done more generally in the literature when studying other types of complex networks.

\paragraph{Articulation Points.}
The importance of a vertex (or a group of vertices) in a graph can be studied by assessing how its removal affects the graph connectivity. This is related to the notion of \textit{articulation point}: a vertex whose removal makes the graph disconnected. Several authors adopt this approach: Moretti~\cite{Moretti2011a} to study Shakespeare's \textit{Hamlet}, Falk~\cite{Falk2016} for Edgeworth's \textit{The absentee}, and Mac Carron \& Kenna~\cite{MacCarron2012, MacCarron2013} for mythological works. Moretti removes various combinations of the most central vertices in terms of degree or closeness, in order to study which ones split the network. He notes that a protagonist is significant mainly because of the stability he brings to the network. Moreover, he shows through a series of examples that this quality is not completely reflected by these centrality measures, as the removal of similarly central characters differently affects the network. He favors transitivity to explain how the removal of these characters leads to different results, as some of them are part of a dense, highly transitive part of the network, whereas others are not. Mac Carron \& Kenna observe that removing the few top central vertices in terms of betweenness~\cite{MacCarron2012} or degree~\cite{MacCarron2013} is enough to disconnect their mythological networks, whereas they are robust to random deletion. The authors use this to highlight the importance of a few main characters: the story just falls apart without them.

In~\cite{Koschutzki2005}, Koschützki \textit{et al}. propose to measure the importance of a vertex based on how its removal affects a graph statistic of interest. For this purpose, one computes the statistic for the whole network, then for the same network without the vertex of interest. The difference between the two resulting values is called \textit{Vitality}. Rochat adopts this approach~\cite{Rochat2014a} when studying Rousseau's \textit{Les confessions}, using centralization as the graph statistic. Rochat applies it to the degree, closeness, betweenness and Eigenvector centralities, in order to study the vitality of the vertices. He also computes the vitality using the number of components as the graph statistic, which allows him to detect articulation points.

\subsubsection{Triads}
\label{sec:ToolsTriangles}
Several standard topological measures are based on the notion of triads, i.e. subgraphs constituted of three vertices. We focus here on \textit{transitivity} and \textit{structural balance}.

\begin{figure}[htb!]
	\centering
	\begin{tabular}{c c}
	    \includegraphics[height=4.3cm]{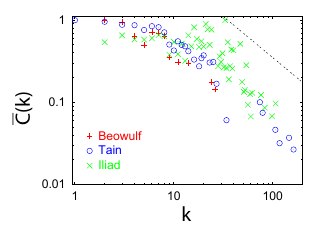} &
	    \includegraphics[height=4.3cm]{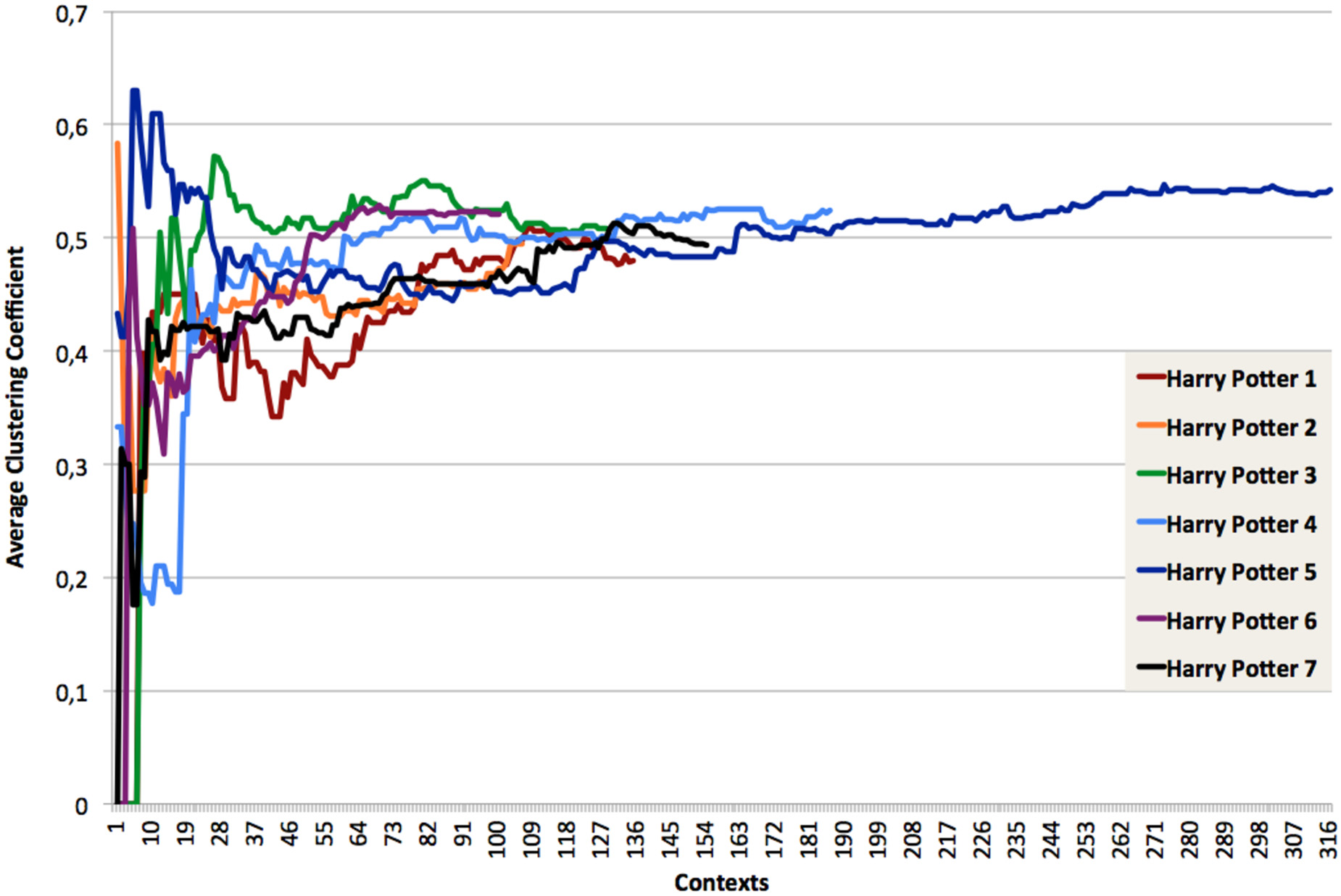} \\
	    a) & b) \\
    \end{tabular}
	\caption{a) Average local transitivity as a function of the degree, in 3 mythological narratives (\textit{Beowulf}, \textit{The Iliad}, the \textit{Táin Bó Cúailnge})~\cite{MacCarron2012}; b) Average local transitivity as a function of time, in the cumulative networks of each \textit{Harry Potter} novel considered separately~\cite{Waumans2015}.}
	\label{fig:transitivity}
\end{figure}

\paragraph{Transitivity.}
The \textit{transitivity}, a.k.a. \textit{clustering coefficient} is a measure defined at two different levels. At the \textit{global} one~\cite{Luce1949}, it is the proportion of existing triangles (i.e. closed triads) relatively to the possible ones (open triads). More formally, it is the probability, when randomly selecting one vertex and two of its neighbors, that these neighbors are themselves connected. Some authors adopt a relatively close approach by counting the number of triangles in the graph~\cite{Elson2010}, as illustrated by Figure~\ref{fig:transitivity}. At the \textit{local} level~\cite{Watts1998}, the transitivity is defined relatively to a vertex of interest: it is the proportion of its neighbors forming connected pairs. Put differently, it is the probability, when randomly picking two neighbors, that there is an edge between them. The \textit{small world} property mentioned before (Section~\ref{sec:ToolsDistance}) requires that, in addition to a small average distance, the considered network exhibits a high transitivity when compared to an equivalent random (Erdös-Rényi -- ER) network~\cite{Watts1998}. A number of authors observe such a high transitivity in character networks~\cite{Alberich2002, Stiller2003, MacCarron2012, Ardanuy2014, Bonato2016}. A weighted generalization of the transitivity exists, which is occasionally used, e.g.~\cite{Suen2013}.

When studying Shakespeare's plays, Stiller \textit{et al}.~\cite{Stiller2003} associate a high \textit{global} transitivity to the presence of separated social groups always appearing on stage together, thereby forming cliques. In a subsequent work~\cite{Stiller2005}, Stiller \& Hudson show that certain characters have a low \textit{local} transitivity though, because they appear in several such subgroups (so, many of their neighbors are not connected). These characters, which they call \textit{keystones}, are very important in the plot structure, as they provide some continuity by connecting subplots. Other authors observe a strong inverse correlation between degree and transitivity in various fiction works~\cite{Gleiser2007, Choi2007, MacCarron2012}: hubs tend to be keystones, whereas minor characters are well embedded. In~\cite{Muhuri2018}, Muhuri \textit{et al}. observe that the supporting characters of antagonists are more tightly-knit than those of protagonists, leading to higher transitivity for the former.

Certain authors use the local transitivity to determine the hierarchical nature of their network: \textit{Marvel universe}~\cite{Gleiser2007}, mythological works~\cite{Choi2007, MacCarron2012, MacCarron2013, Miranda2013}. They observe that the transitivity considered as a function of the degree takes the form of a power law. This means that vertices with a small degree have a much higher transitivity, and that hubs have a very small one. Narratively speaking, minor characters tend to form small dense clusters, resulting in high transitivity but low degree, whereas major characters act as bridges and connect these clusters, which results in high degree (many clusters to connect) but low transitivity (this clusters are mutually disconnected). In the case of Choi \textit{et al}.'s Greek/Roman mythological network~\cite{Choi2007}, the authors assume that this is mainly due to two reasons. The first is the largely genealogical nature of the relationships between characters, resulting in a tree-like structure. The second is the presence of classes of characters (gods, titans, monsters, etc.) shaping the narrative of the myths.

\paragraph{Structural Balance}
Like the transitivity, the notion of \textit{structural balance} is related to triangles (closed triads), but in \textit{signed} networks this time. There are several variants of the structural balance concept, the simplest being: a triangle is structurally balanced if it contains exactly zero or two negative edges~\cite{Cartwright1956}. Consider for instance a social network, in which positive and negative represent friendly and hostile relationships, respectively. In a balanced triangle, all three persons can be friends, or two can be friends while being both hostile towards the third. The case where all three person are mutually hostile to each other is considered as imbalanced, as two of them are expected to eventually unite against the third, resulting in the previous situation. Similarly, when one person is friend with two mutually hostile persons, he might start a fight with one of them, or both of them might become friends, also resulting in a balanced situation. Real-world social networks were initially expected to contain mainly balanced triangles, although this assumption has been put in question lately~\cite{Doreian2017}.

In the case of character networks, edge signs represent antagonistic relationships, generally friendly vs. hostile~\cite{Ding2010, MacCarron2014a, Nalisnick2013, Liu2017d}. Mac Carron \textit{et al}.~\cite{MacCarron2012, MacCarron2013, MacCarron2014a, Yose2017} study the number of imbalanced triangles in their mythological networks, and the low proportion they find is comparable with what is generally observed in real-world social networks. Sack~\cite{Sack2012, Sack2013, Sack2014} studies the evolution of the network \textit{balance}, a quantity related to the number of balanced triangles it contains, to identify several phases in the plot (progress towards stability, reversal, factions formation, resolution...). Their approach relies on the Aristotelian structure, and they assume that the character network is initially imbalanced, and that the plot drives it towards perfect balance (the story resolution). 

However, Liu \& Albergante~\cite{Liu2017d} make an opposite observation when studying the \textit{Game of Thrones} TV series, as they observe around 30\% of imbalanced triangles. They assume that the writers voluntarily keep the plot in this imbalanced state in order to make it more interesting for the viewer. When comparing episodes, they also remark that increases in the number of edge changes during an episode (be it the inversion of a sign or the addition/removal of an edge) are related to higher viewer figures. They assume that such changes reflect how much the plot progresses, and that an important evolution is likely to strongly engage the audience.

\subsubsection{Modular Structure}
\label{sec:ToolsModularStruct}
A character network can generally be broken down into a collection of disconnected or loosely connected subnetworks, corresponding to its modular structure. 

\paragraph{Community Structure.}
A network is said to possess a \textit{community structure} when it can be partitioned into subgraphs called communities, such that many edges fall inside these groups and few between them~\cite{Girvan2002a}. Community detection is an important problem in the domain of complex network analysis, and a number of methods have been proposed for this purpose (cf.~\cite{Fortunato2010} for a detailed review).

\begin{figure}[htb!]
	\centering
	\begin{tabular}{c c}
	    \includegraphics[height=5.5cm]{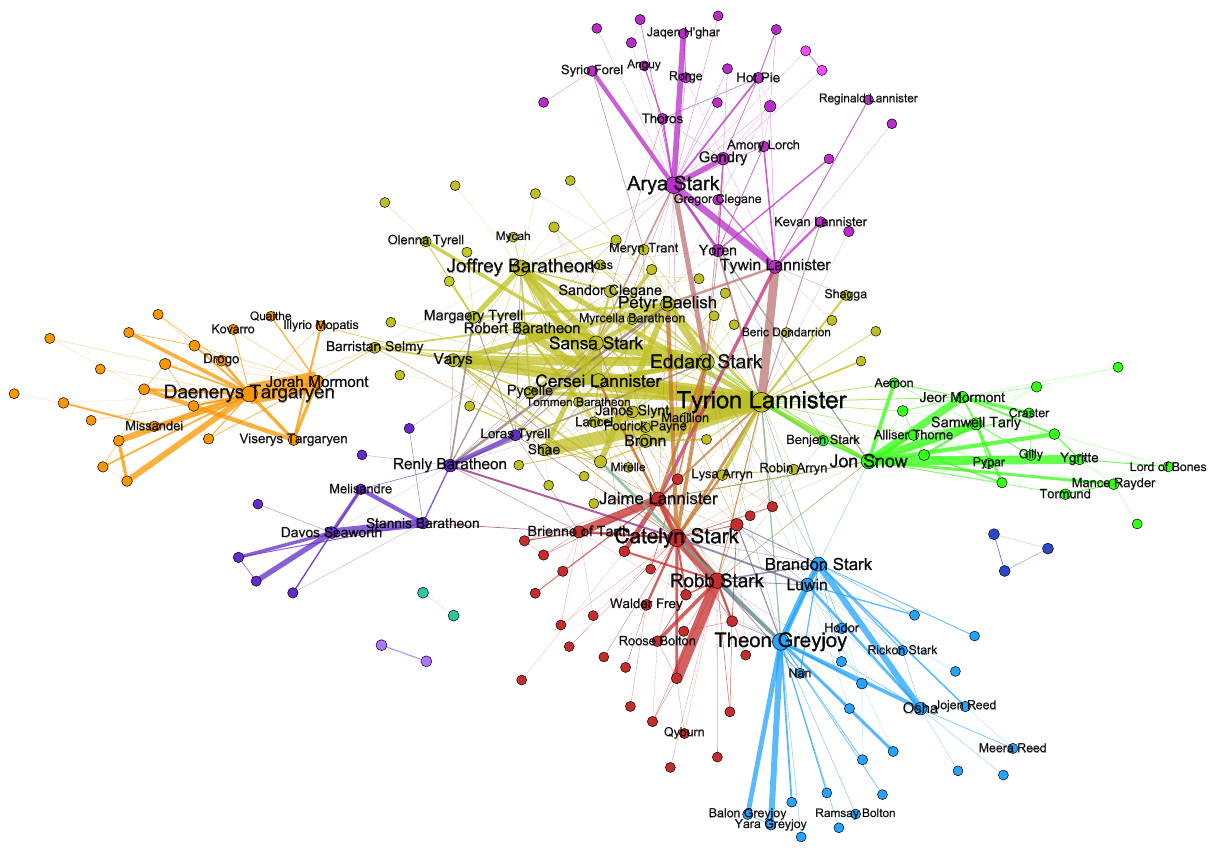} &
	    \includegraphics[height=5.5cm]{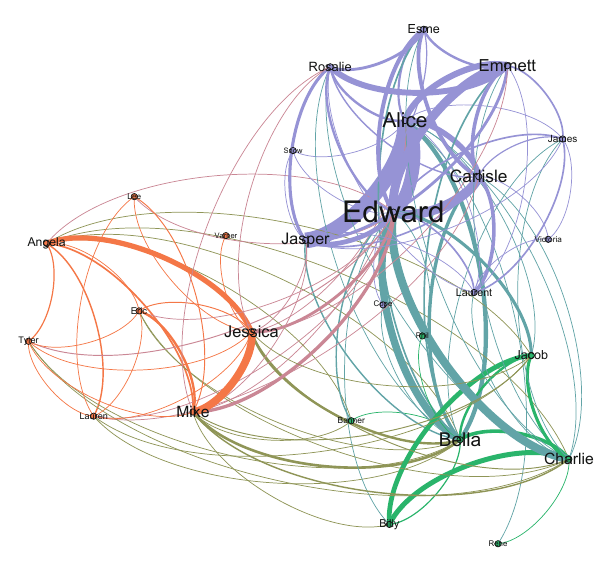} \\
	    a) & b) \\
    \end{tabular}
	\caption{a) Communities in the TV series \textit{Game of Thrones} (seasons 1 \& 2)~\cite{Bost2016}; b) Communities in the three novels of the series \textit{Twilight}~\cite{Bonato2016}.}
	\label{fig:communities}
\end{figure}

In character networks, communities are related to parts of the narrative, as illustrated by Figure~\ref{fig:communities}. They can be used to identify either individual subplots, or groups of subplots involving roughly the same characters~\cite{Newman2004e, Park2011, MacCarron2013, Bolioli2013, Kydros2015a}. For instance, in the epic narrative \textit{Beowulf}, Mac Carron \& Kenna~\cite{MacCarron2012} identify two subplots taking place in the past and appearing as two components, completely disconnected from the main plot. It is also likely to see such disconnections if the network is dynamic: one can observe several separated components at some point, representing several subplots likely to merge later in the story. But the different subplots are generally at least weakly connected, resulting in a community structure~\cite{Fortunato2010} instead of several distinct components. 

Several authors observe that such communities gather characters based on spatial and/or temporal criteria. This is for instance the case in Rousseau's \textit{Les confessions}~\cite{Rochat2014a}, Martin's \textit{Game of Thrones}~\cite{Beveridge2016}, and certain Icelandic sagas~\cite{MacCarron2013}. Jayannavar \textit{et al}.~\cite{Jayannavar2015} observe that, in networks extracted from 19th century British novels, the number of communities increases while their average size decreases when the network gets larger in terms of characters. This would be related to writing constraints, as it is not easy to handle large groups of interacting characters without confusing the reader.

For Lee \textit{et al}.~\cite{Lee2018}, in the case of movies, each community is centered on a main character, accompanied by sidekicks and other types of minor characters. However, this is not the consensus, as several authors note that primary characters are likely to belong to several communities, because they appear in a number of subplots~\cite{Park2011, Weng2007, Muhuri2018}. In particular, Muhuri \textit{et al}.~\cite{Muhuri2018} propose a custom measure called \textit{Diversity}, in order to quantify how well-connected a character is to the communities of the network. It is similar enough in principle to the \textit{Participation Coefficient}~\cite{Guimera2005a} and \textit{Diversity Score}~\cite{Dugue2015a} proposed to study other types of complex networks. They observe that main characters tend to be connected to multiple communities, unlike minor ones. Implicitly, all these authors identify the need to apply a community detection method able to identify \textit{overlapping} communities, however none of them apply such a tool.

\paragraph{Correlation Clustering.}
When the network is signed, its modular structure is detected by solving the \textit{Correlation clustering} problem~\cite{Bansal2002}, which consists in finding groups such that most positive and negative edges are located inside and between the groups, respectively. The result is interpreted in terms of antagonistic groups rather than subplots. Ding \& Yilmaz~\cite{Ding2010} adopt this approach to identify opposed factions in so-called \textit{adversarial} movies.

\subsubsection{Miscellaneous Approaches}
\label{sec:ToolsMisc}
More minor methods are used in the literature to described character networks, which we summarize in this last section.

\begin{figure}[htb!]
	\centering
	\begin{tabular}{c c}
	    \includegraphics[height=5cm]{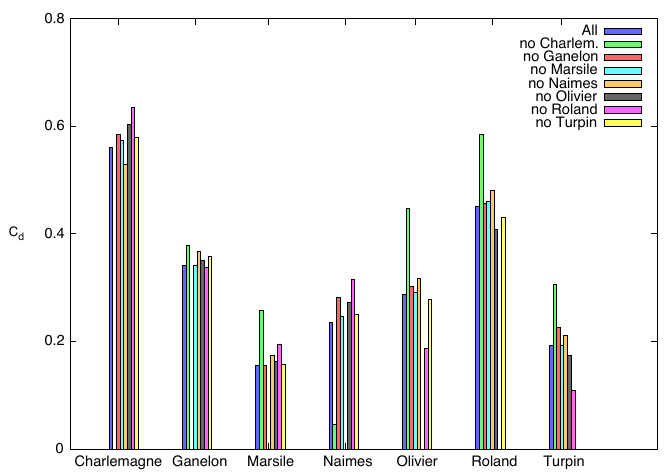} &
	    \includegraphics[height=5cm]{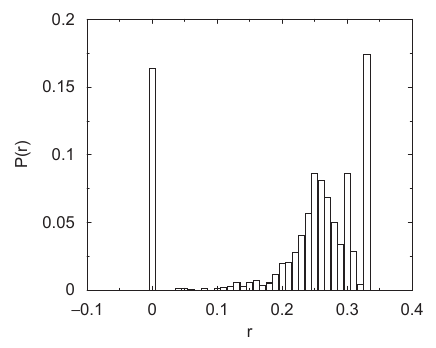} \\
	    a) & b) \\
    \end{tabular}
	\caption{a) Eigenvector centrality for the main characters of the song of heroic deeds \textit{La Chanson de Roland}~\cite{Prado2016}; b) Distribution of local cyclic coefficient for the \textit{Greek and Roman mythology} network~\cite{Choi2007}.}
	\label{fig:eigenvector}
\end{figure}

\paragraph{Spectral Measures.}
A family of vertex centrality measures is based on the spectrum of the graph adjacency matrix, or one of its related matrices (e.g. graph Laplacian), cf. Table~\ref{tab:centralities}a and Figure~\ref{fig:eigenvector}a. The \textit{Eigenvector centrality}~\cite{Bonacich1972} can be considered as a generalization of the degree centrality, in which instead of counting the neighbors of a vertex, one also considers their respective centrality values. Rochat~\cite{Rochat2014a} notes that this measure tends to be higher in the center of dense communities, such as those corresponding to subplots revolving around multiple characters. By opposition, if the subplot is highly centralized around a single character (such as Mme de Warens in Rousseau's \textit{Les confessions}), the Eigenvector centrality is lower. 

Algee-Hewitt~\cite{Algee-Hewitt2017} also notes that this measure is lower for characters who tend to be largely connected to peripheral characters, than for those who are connected to a few core characters. In addition, he exemplifies the interest of comparing several distinct centrality measures when studying the network. In plays, the existence of a group of characters with high Eigenvector centrality but low betweenness centrality is often related to the presence of a conspiracy in the plot: the group corresponds to the conspirators, and is mostly disconnected from the rest of the network.

In~\cite{Ding2010}, Ding \& Yilmaz study what they call \textit{adversarial} plots, which oppose two groups of characters. After having first bisected their network to estimate these two groups (cf. Section~\ref{sec:ToolsModularStruct}), they use the Eigenvector centrality to identify the leaders of these communities. These are then assumed to be the hero and the villain of the story. In~\cite{Prado2016}, Prado \textit{et al}. compute the \textit{Centralization}~\cite{Freeman1978} based on the Eigenvector centrality. They consider it allows measuring how close a graph is from a star structure. They associate this star structure to single character-centered plots, such as biographies. They also use a dynamic generalization of the Eigenvector centrality to describe the trajectory of individual characters.

\textit{HITS} is a generalization of the Eigenvector centrality to directed networks~\cite{Kleinberg1999}. It includes an \textit{Authority score} and a \textit{Hub score}, depending on whether one considers incoming or outgoing edges. Certain authors use them to identify the main characters in novels~\cite{Agarwal2012, Sudhahar2013}.

\paragraph{$k$-Cores.}
Cliques and various relaxations of this concept can be used to describe a graph. The notion of $k$-core refers to a maximal induced subgraph whose all vertices have a degree of at least $k$~\cite{Seidman1983}. It is obtained by iteratively removing isolates and leaves. Park \textit{et al}.~\cite{Park2013, Park2013a} use it to study the structure of novels, and distinguish what they call the \textit{kernel} from the more peripheral characters. They proceed similarly in a subsequent article~\cite{Seo2013}, but using a filtered version of their co-occurrence network, taking the form of a minimum weighted spanning tree.

The \textit{$k$-core number} (or \textit{coreness}, or \textit{degeneracy}) of a graph is the largest $k$ for which this graph contains a $k$-core. Rochat~\cite{Rochat2015} proposes to normalize it relatively to the size of the network, in order to measure the compactness of the group of main characters.

\paragraph{Cyclic Coefficient.}
The \textit{cyclic coefficient} was defined by Kim \& Kim to measure what they call the degree of circulation of a network~\cite{Kim2005}. It quantifies how cyclic vs. treelike a structure is. It is defined relatively to a given vertex. One first identifies the set of shortest cycles going through the vertex and each pair of its neighbors, then averages the reciprocals of their lengths. The resulting value ranges from zero (the graph is locally a tree) to $1/3$ (a clique). The authors propose to average this measure over $V$ in order to describe the whole graph.

In~\cite{Choi2007}, Choi \textit{et al}. use the average cyclic coefficient to describe the network they extract from a biographical dictionary of Greek and Roman myths (cf. Figure~\ref{fig:eigenvector}b). This allows them to confirm their assumption that this network is treelike due to its mainly genealogical nature. They also study the distribution of the cyclic coefficient and identify two classes of characters: very peripheral characters embedded in a locally treelike neighborhood, vs. characters whose narrative is more or less tightly intertwined with others', and belonging to clique-like structures.

\paragraph{Vertex Attributes.}
Certain authors extract character networks possessing vertex attributes, i.e. each vertex is described by certain individual fields such as gender, age, or social group. This information is leveraged in different way when studying the network. The simplest is to assess the prevalence of the attribute values in the graph. When studying their science-fiction corpus in~\cite{Rochat2017}, Rochat \& Triclot do so with the vertex attribute describing the main occupation of each character: scientist, politician, technician, religious figure, animal, etc. For instance, they observe that the most widespread occupation among the main characters is scientist.

If the network is large enough and sufficiently connected, it is also possible to study the induced subgraphs corresponding to sets of vertices possessing certain attribute values. On the \textit{Iliad}, Kydros \textit{et al}.~\cite{Kydros2015} consider four categories of vertices (\textit{Greeks}, \textit{Trojans}, \textit{Gods}, and \textit{Others}). They study them separately, by computing a collection of standard topological measures on the induced subgraphs. For instance, they find out that the Trojan subgraph is much denser than its Greek counterpart, revealing a much higher cohesion of these characters in the story. They also study how these four groups of vertices are interconnected. As an example, the number of relationships between gods and mortals are similar for either side, denoting a carefully balanced interventionism. 

Another approach is to compute the \textit{assortativity} (or \textit{homophily}) of the network for an attribute of interest. It is obtained by applying an association measure to the two series constituted by the attribute values of \textit{connected} pairs of vertices~\cite{Newman2003a}. In the case of numerical attributes, one generally uses Pearson's correlation coefficient, and for categorical ones it is Cohen's $\kappa$. Put differently, the assortativity measures whether the edges tend to appear between \textit{similar} vertices (positive value), in terms of the considered attribute, or \textit{dissimilar} ones (negative value). In the first case, the network is said to be \textit{assortative} (or \textit{homophilic}), and in the latter, it is \textit{disassortative} (or \textit{heterophilic}). Yose \textit{et al}. compute the assortativity of the network extracted from the Irish epic text \textit{Cogadh Gaedhel re Gallaibh}, which describes a war between Irishmen and Vikings. They leverage this measure to determine whether hostile interactions tend to be \textit{internal} (i.e. connect Irishmen), which would mean the text mainly describes a domestic conflict, or \textit{external} (i.e. connect Irishmen and Vikings), which would mean both sides are opposed. Bossaert \& Meidert~\cite{Bossaert2013} use a similar approach to check whether adolescent support is affected by school year, school house or gender in the \textit{Harry Potter} series of novels.

\subsection{Applications}
\label{sec:Applications}
In this section, we present a selection of methods and results aiming at solving a specific problem and relying on character networks for their purpose. We overlook the many articles from the field of Literary Studies, which focus on a single fiction work (or a few ones) and aim at describing it in great details~\cite{Rochat2014a, Rochat2014}: these are case studies, their goal is not to propose a generic tool that one could subsequently apply to other works. We also ignore articles only using character networks for visualization~\cite{Mutton2004, Venturini2016, Xanthos2016, Grener2017}, as they are somewhat limited, at least from this perspective.

\subsubsection{Assessment of Literary Theories}
\label{sec:AppAsstLitTh}
Some articles aim at assessing the validity of literary theories. Those are often evaluated qualitatively and/or on a small number of fiction works~\cite{Elson2010}. Automated approaches can help at testing them in a more quantifiable way, and/or on larger corpora.

In~\cite{Elson2010}, Elson \textit{et al}. focus on two literary assumptions related to the size of the community surrounding the protagonist in 19th century British novels. The first one is that it depends on the amount of dialogue occurring: it would not be possible to show many conversations when there are many characters to consider. This is formally translated as the presence of a strong correlation between the amount of dialogue and the number of characters in the novel. The second assumption is that it depends on the social setting of the novel: there would be more verbal interactions in rural than urban communities. This is handled by manually determining the setting of the considered novels.

Elson \textit{et al}. adopt conversational networks to model the novels. They use a number of features to assess both literary theories, some describing the network (numbers of $3$- and $4$-cliques, average degree, density) and others extracted from the detected verbal interactions (e.g. numbers of speaking and non-speaking characters, variance of the number of utterances by character). They apply their method to their corpus of sixty classic novels. They find a slightly positive correlation between the numbers of characters and utterances, and an even stronger one when counting only speaking characters: this invalidates the first literary theory. The same behavior is observed when considering the average degree and the density instead of the number of utterances, which the authors consider as a confirmation. Regarding the second theory, Elson \textit{et al}. do not observe any significant difference in the numbers of characters (speaking or not), or any tested feature, relative to the type of setting (rural vs. urban), which invalidates the second theory. Instead, the most influential factor seems to be the narrator's perspective: characters are much loosely connected when the story is told in the first person than in the third.

Jayannavar \textit{et al}.~\cite{Jayannavar2015} later revisit these results, based on SINNET, the tool developed by Agarwal \textit{et al}.~\cite{Agarwal2012, Agarwal2013, Agarwal2013a, Agarwal2014a}. It allows identifying two kinds of interactions: unilateral vs. bilateral interactions (cf. Section~\ref{sec:InterDetDirAct} for more details), the former supposedly subsuming conversations. The authors argue that identifying a wider range of interactions might change the results. Moreover, they observe that Elson \textit{et al}. misunderstood the second assumption: they confused the notions of literary \textit{setting} (the real-world environment) and \textit{space} (how it is portrayed in the novel, be it accurate or not). The theory expressed in the literature concerns space, and not setting. Some novels take place in a urban setting, but do not faithfully render the complexity of urban life.

Jayannavar \textit{et al}. propose a number of hypotheses derived from Elson \textit{et al}.'s and assess them using a comparable method. For all the tested types of networks, they find out that, as the number of characters increases, they tend to interact with more people, as Elson \textit{et al}., but less frequently, which supports the first original hypothesis. Moreover, the number of communities tend to increase too, while their average size gets smaller. Regarding the setting, they find out that it has no effect on any of the considered features, while underlining that the original second theory concerns the effect of space, and not setting.

Grayson \textit{et al}.~\cite{Grayson2016} study the co-occurrence network of Austen's \textit{Sense and Sensibility}. They remark that when considering the network built over the whole novel, the top characters in terms of betweenness and Eigenvector centralities are also the wealthiest. They interpret this as a confirmation of the literary theory regarding social exclusivity in the social systems described in Austen's novels. However, when considering networks focusing on chapters, this idea is questioned as poorer character hold a more important role (the main character still being the richest, though).

\begin{figure}[htb!]
	\centering
	\begin{tabular}{c c}
	    \includegraphics[width=0.45\textwidth]{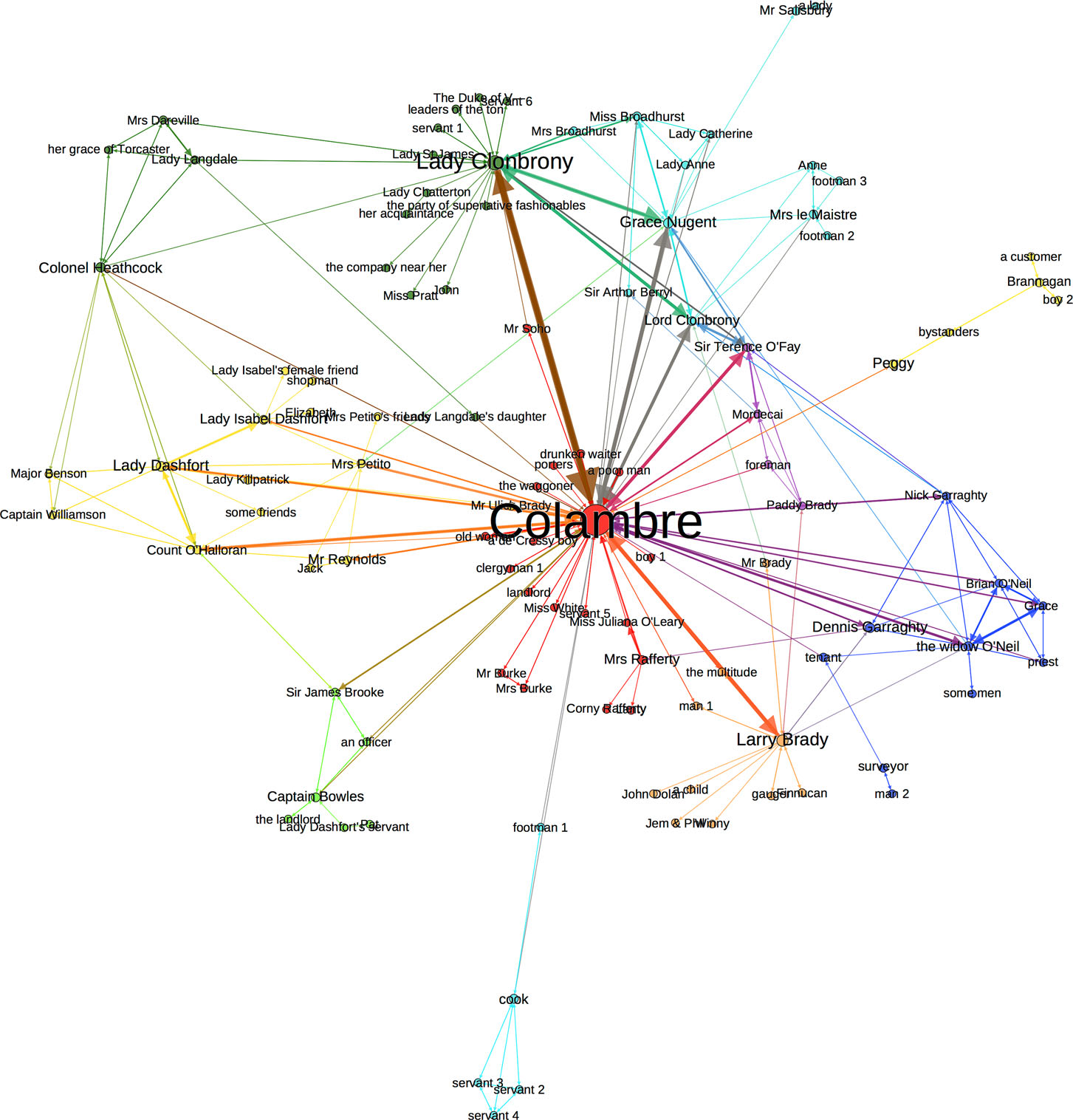} &
	    \includegraphics[width=0.45\textwidth]{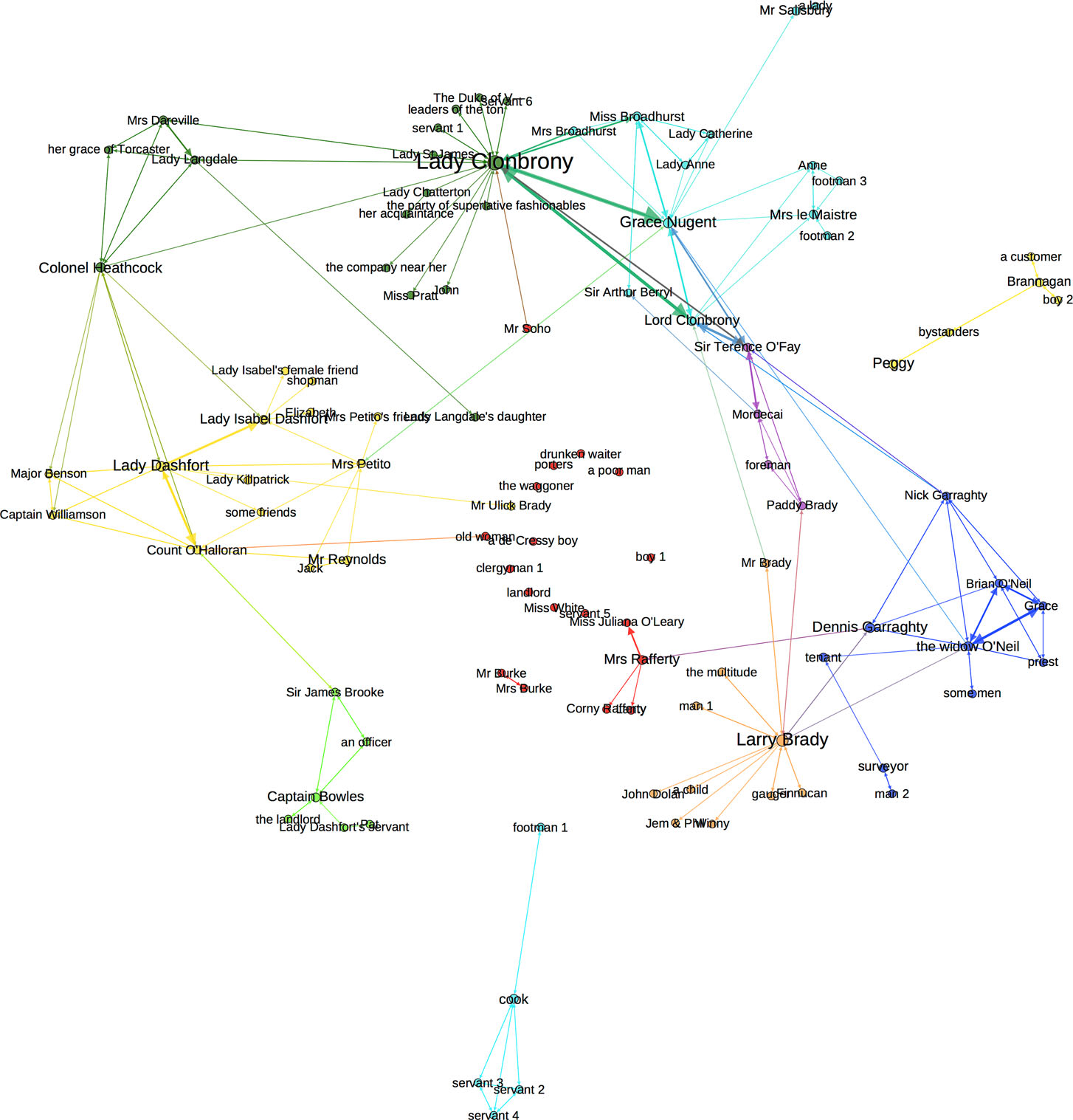} \\
	    a) & b) \\
    \end{tabular}
	\caption{Two versions of the character network of \textit{The Absentee}~\cite{Falk2016}: a) complete network; b) same network without \textsc{Colambre} (the main character).}
	\label{fig:absentee}
\end{figure}

Falk~\cite{Falk2016} studies the Bildungsroman, a category of novels dealing with the formative years of young characters. In the domain of literary analysis, some authors oppose two subclasses: \textit{domestic} vs. \textit{network} novels. In the former, the protagonist is a part of a social system and must evolve towards adulthood, whereas in the former he discovers a new social world to which he must adapt. Falk analyzes a conversational network representing Edgeworth's \textit{The Absentee} to show that both aspect can co-exist in the same novel (see Figure~\ref{fig:absentee}). He uses community detection to highlight the existence of two separate social systems connected by \textsc{Colambre} (the main protagonist), and centrality measures to show his central role among subgroups (major households) constituting his original social environment.

\subsubsection{Level of Realism and Historicity}
\label{sec:AppRealism}
A popular problem is to determine whether the network of social interactions extracted from a fictional narrative is realistic, in the sense that it displays topological properties similar to real-world social networks. Stiller \textit{et al}.~\cite{Stiller2003} study a corpus of ten plays by Shakespeare in an effort to determine whether the success of this playwright relies on his ability to mimic some basic properties of real social networks in his writing. Alberich \textit{et al}.~\cite{Alberich2002} and Gleiser~\cite{Gleiser2007} assess how realistic cooperation between \textit{Marvel} characters is, through the comparison of their co-occurrence network with real-world collaboration networks (e.g. IMDb, DBLP). In a series of articles studying myths, tales and other mythological stories, Mac Carron \textit{et al}.~\cite{MacCarron2012, MacCarron2013, MacCarron2014, MacCarron2014a, Kenna2016, Kenna2017, Yose2016, Yose2017} assume that such narratives convey both actual and fictional information. They assess their historicity by determining where they stand between completely real and purely fictional social networks.

\begin{table}[htb!]
	\centering
	\begin{tabular}{c c}
	    \includegraphics[height=3cm]{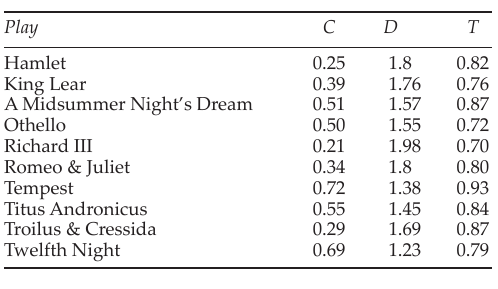} &
	    \includegraphics[height=3cm]{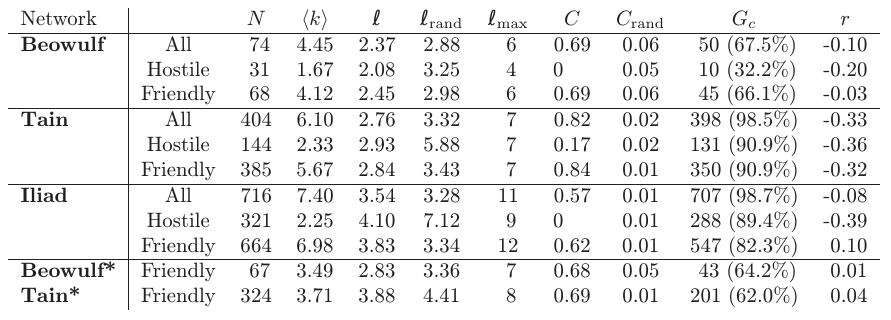} \\
	    a) & b) \\
    \end{tabular}
	\caption{Topological measures used to assess the realism of the character networks of: a)~Shakespeare's plays~\cite{Stiller2003}; b) mythological narratives~\cite{MacCarron2012}.}
	\label{tab:realism}
\end{table}

The approach adopted in these articles is globally the same: they describe their character networks using standard topological descriptors (e.g. degree distribution, average distance, transitivity, cf. Section~\ref{sec:DescriptiveTools}), as illustrated by Table~\ref{tab:realism}, and compare their counterparts obtained for real-world and/or random networks. They generally find both similarities and differences. For instance, the Marvel network possesses certain properties observed in real-world networks: presence of a giant component, small-world (small average distance and high transitivity), scale-free (power law-distributed degree). However, it also presents important differences: its transitivity is much lower, and its degree is much smaller, reflecting the fact that Marvel characters tend to collaborate more often with the same people than real-world agents do~\cite{Alberich2002}. This can be explained by legal matters related to the \textit{Comics Authority Code}~\cite{Gleiser2007}. In the case of myths, \textit{parts} of the networks are realistic whereas others are not. For instance, the Old English epic poem \textit{Beowulf} is modeled by a relatively realistic network, provided one ignores the eponymous character: this is likely caused by the concentration of fantastic plot elements on him, whereas the other characters are based on real people~\cite{MacCarron2012, MacCarron2014a, Kenna2016, Kenna2017}.

\subsubsection{Role Detection}
\label{sec:AppRoleDetection}
A number of authors leverage character networks for role detection in plays, novels and movies. i.e. assigning one among several predefined narrative roles to each character. A consensual definition would be that a character plays a role in the story in the sense that it follows ``characterization archetypes''~\cite{Rochat2014a}, such as being a protagonist or an antagonist. Once detected, roles can be used to solve higher-level problems such as plot classification (Section~\ref{sec:AppClassification}), storyline detection (Section~\ref{sec:AppStoryDecomposition}), and plot summarization (Section~\ref{sec:AppSummary}). Identifying the protagonists can help characterize the structure of the plot. For instance, in his series \textit{Les Rougon-Macquart}, Zola applies a technique called \textit{polyfocalisation} and consisting in switching the narrative focus between several characters, which are therefore all protagonists~\cite{Rochat2014a, Rochat2015}. On the contrary, Rousseau's \textit{Les confessions} is largely focused on a specific character.

\begin{figure}[htb!]
	\centering
	\begin{tabular}{c c}
	    \includegraphics[height=5cm]{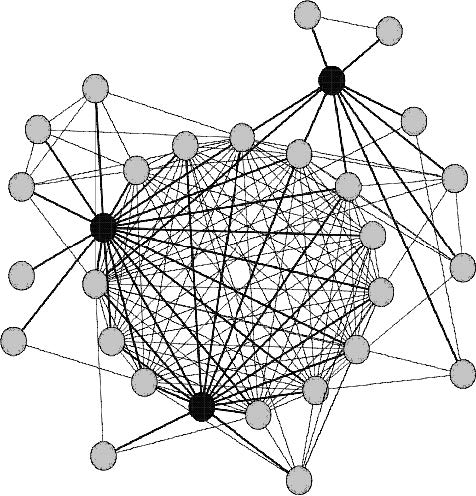} &
	    \includegraphics[height=5cm]{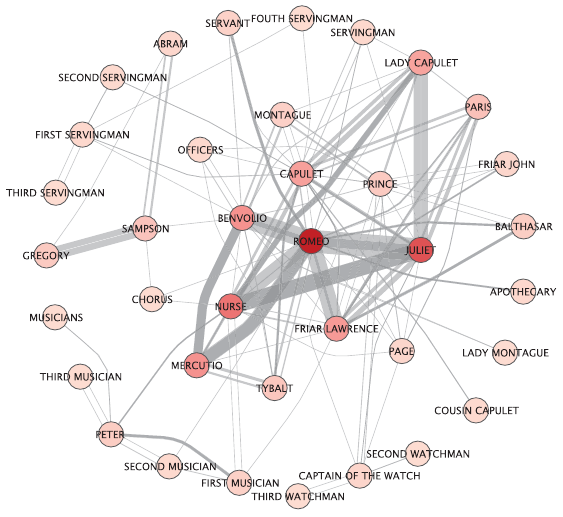} \\
	    a) & b) \\
    \end{tabular}
	\caption{Main characters in two plays by Shakespeare: a) \textit{Troilus and Cressida}~\cite{Stiller2005}; b) \textit{Romeo and Juliet}~\cite{Masias2017}.}
	\label{fig:protagonists}
\end{figure}

The different detected roles differ significantly from one author to the other, and some assign several roles to certain characters. In the simplest case, one wants to distinguish between \textit{major} and \textit{minor characters}~\cite{Stiller2005, Tran2015, Masias2017, Lee2018}, as illustrated by Figure~\ref{fig:protagonists}. Certain authors try to split the latter class into \textit{supporting characters} and \textit{extras}~\cite{Rochat2014, Rochat2014a, Park2011}. Dealing with adversarial stories, in which two sides are opposed, adds another dimension: character alignment, i.e. whether a character assists or opposes the hero. This leads to the detection of protagonists vs. antagonists~\cite{Kwon2017}, and can be combined with the major/minor distinction to get a finer typology~\cite{Jung2013a, Muhuri2018}. It is also possible to focus only on certain roles, for instance Agarwal \textit{et al}.~\cite{Agarwal2012} want to identify the main protagonist and the narrator (which is not necessarily a character). Finally, some roles are specific to the considered narrative, for instance Pope \textit{et al}.~\cite{Pope2016} distinguish cops, gangsters and informants in the TV series \textit{The Wire}.

All the methods appearing in the literature rely on nodal topological measures: degree or strength are generally considered as the most discriminative, then transitivity, betweenness, closeness and Eigenvector centrality. Most of them are generally distributed according to a power law (or at least, a strongly heterogeneous distribution), which fits the assumption that there are many more minor characters than primary ones~\cite{Moretti2011a}. The simplest approach is to focus on a single nodal topological measure, and use predefined or automatically estimated thresholds to distinguish between roles~\cite{Park2011, Jung2013a}. However, most authors prefer to use simultaneously several such measures, as these are deemed complementary. They adopt various approaches: multiple thresholding~\cite{Stiller2005, Agarwal2012, Kwon2017}, cluster analysis~\cite{Masias2017}, supervised or semi-supervised classification~\cite{Gorinski2015, Pope2016, Muhuri2018}, definition of a composite measure~\cite{Tran2015a, Tran2015a, Tran2017a, Lee2018}, PCA (Principal Components Analysis) completed by visual inspection~\cite{Rochat2014, Rochat2014a}. In addition to purely structural features, certain authors dealing with alignment-related or work-specific roles use the narrative content: part-of-speech associated to each role~\cite{Pope2016}, sentiment analysis applied to the dialogues~\cite{Jung2013a} or co-occurrence context~\cite{Gorinski2015}.

\subsubsection{Classification of Fiction Works}
\label{sec:AppClassification}
One can suppose that the social network of characters corresponding to a work of fiction depends on a number of factors. First, it can be affected by the way real-world people interact, be it at the time of release, or at the time the plot takes place. Second, it can be influenced by the genre of the work, which is likely to highlight specific aspects of the social interactions, or specific types of social interactions. Third, the writer or director himself can choose to stylize the way he presents social interactions, or can unconsciously introduce certain biases in the way these are described. 

Based on this assumption, several authors take advantage of the network structure to predict some of these traits. Most of them use a supervised approach: Suen \textit{et al}. train a classifier to distinguish scripts based on various criteria (e.g. cinema vs. theater, publication year)~\cite{Suen2013}; Hettinger \textit{et al}. classify German novels in terms of genre~\cite{Hettinger2015}; Li \textit{et al}. do the same for plays~\cite{Li2018h}; Holanda \textit{et al}. try to distinguish texts according to three classes (pure fictions, pure biographies, and legends --a mix of both)~\cite{Holanda2017}. A few authors adopt a non-supervised, more exploratory approach, through clustering: Ardanuy \& Sporleder group similar novels and study the uniformity of such groups in terms of genre and writer~\cite{Ardanuy2014, Ardanuy2015}; Waumans \textit{et al}. want to gather episodes of the same series. Rochat \& Triclot~\cite{Rochat2017} manually classify works in four plot classes (heroic-core, unicentered, acentric, polycentric). 

The general approach consists in extracting the character network of the considered fiction work, before computing a collection of topological measures used as features by the classifier: network size, density, transitivity, degree distribution, diameter, radius, centrality measures, and others. Li \textit{et al}. compare these to their weighted generalization of graph motifs~\cite{Li2018h}. Most authors additionally use meta-data (medium type, genre, writer, date, length/duration) and content-related features (point of view of the narrator, word frequency, topics). Ardanuy \& Sporleder~\cite{Ardanuy2014, Ardanuy2015} also leverage the plot dynamics: they split each narrative into a few predefined phases (exposition, rising action, climax...) and separately compute the features for each of them.

In the end, the nature of the most discriminative features largely depends on the considered data and classification task. For instance, Suen \textit{et al}.~\cite{Suen2013} observe that the best features to distinguish between plays and movies are degree-related, because plays generally have a single largely dominating character, whereas movies possess several of them. When discriminating between fictions and biographies, Holanda \textit{et al}.~\cite{Holanda2017} remark that the most discriminant feature is the number of \textit{hapax legomena} (single-occurrence characters), which increases with the level of realism of the book. They assume that novel writers tend to use characters several times once they have made the effort of introducing them, whereas character occurrences are dictated by historical events in biographies.

\subsubsection{Story Decomposition}
\label{sec:AppStoryDecomposition}
Character networks are used in the literature to break down a story into its constitutive parts. Solving this problem is relevant for numerous information retrieval tasks, such as plot summarization or comparison, and can be expressed under various forms. 

\textit{Storyline detection} consists in identifying the multiple subplots intertwined by the writer/director to constitute the plot of his narrative. The methods proposed in the literature are applied to movies~\cite{Weng2007, Park2011} and TV series~\cite{Weng2009}. They postulate that each storyline is organized around a leading character. They first use role detection to detect these main characters, and then community detection to determine their entourage. Each scene is then associated to one of the communities depending on the involved characters, and storylines are obtained by ordering them chronologically. Scenes involving characters from multiple communities are points of contact between the storylines: they generally correspond to key moments in the plot, and are therefore particularly interesting for higher-level tasks such as automatic summarization. Certain authors observe that main characters are likely to appear in multiple subplots, and therefore assume that they can belong to several communities~\cite{Weng2007a, Park2011}.

\begin{figure}[htb!]
	\centering
	\begin{tabular}{c@{} c@{} c}
	    \includegraphics[width=0.33\textwidth]{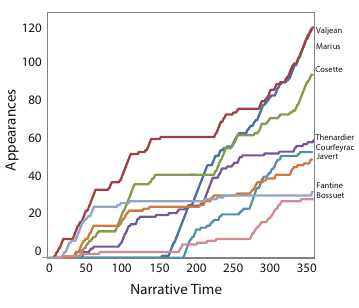} &
	    \includegraphics[width=0.33\textwidth]{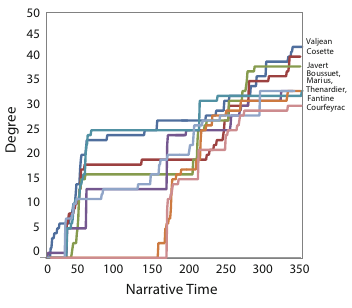} &
	    \includegraphics[width=0.33\textwidth]{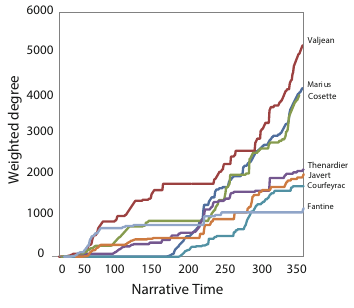} \\
	    a) & b) & c) \\
    \end{tabular}
	\caption{Evolution of the cumulative dynamic network of \textit{Les Misérables} used in~\cite{Min2016} to detect plot phases: a)~number of occurrences of the characters; b) degree; c) strength.}
	\label{fig:storyline}
\end{figure}

\textit{Story segmentation} constitutes another form of story decomposition. It consists in splitting the plot into consecutive and meaningful phases. When dealing with movies and TV series, this amounts to gather shots constituting scenes or substories (so-called \textit{sequences}, cf. Section~\ref{sec:InterDetCoocNarrUnit}). Traditional methods focus only on low-level audiovisual features~\cite{Weng2007a}, but recent approaches take advantage of the plot-related semantics encoded in character networks. Weng \textit{et al}.~\cite{Weng2009} define a character similarity measure based on neighborhood proximity in the character network, then use it to compute scene similarity based on the similarity of the involved characters. The plot is then split by identifying strongly dissimilar consecutive scenes. Min \& Park~\cite{Min2016} want to detect standard plot phases (exposition, rising action, climax, falling action, resolution) in novels. For this purpose, they use a dynamic cumulative network and detect significant increases of vertices and edges, as well as stable periods, to identify these phases (see Figure~\ref{fig:storyline}). They subsequently use sentiment analysis and topic detection to improve their tool by taking the content into account.

\subsubsection{Summarization}
\label{sec:AppSummary}
Several approaches use character networks to summarize audiovisual narratives, in place or in addition to the traditional low-level audiovisual features. This task consists in producing an extractive summary by combining important scenes picked from the original narrative.

Tran \textit{et al}.~\cite{Tran2015a, Tran2017a, Do2018} propose a method for the automatic summarization of movies. They first extract a co-occurrence network, and use a combination of standard topological measures to identify the main characters. They then assign a so-called \textit{social score} to each scene depending on the nature of the involved characters. They combine this score with video-related features such as inter-scene distance and duration, to identify the scenes of particular interest, which are finally used to generate the summary. Bost \textit{et al}.~\cite{Bost2018a, Bost2016c} propose an extractive method to summarize TV series seasons, from the perspective of a character of interest (by opposition to summarizing the whole plot). They extract a dynamic conversational network using narrative smoothing (cf. Section~\ref{sec:GraphExtrDynOthers}), in order to take into account the non-linear nature of the narratives (eg. parallel subplots). Their assumption is that important events occurring in the storyline of characters are associated to durable changes in their social environment. They use the vertex neighborhood as a proxy to such social environment, and detect these events by clustering the sequence representing the successive states of this neighborhood. Their algorithm combines this graph-based approach to a more traditional analysis of the filmmaking grammar to generate the character-oriented summaries.

\begin{figure}[htb!]
	\centering
	\begin{tabular}{c c}
	    \includegraphics[width=0.4\textwidth]{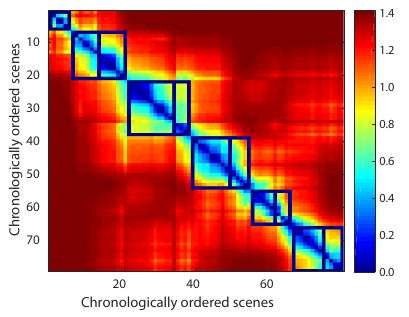} &
	    \includegraphics[width=0.4\textwidth]{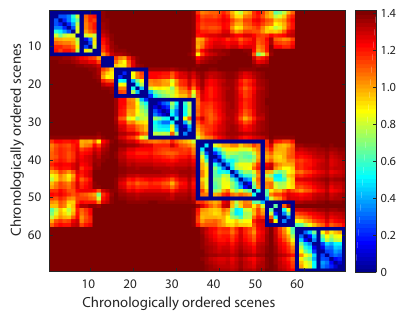} \\
	    a) & b) \\
    \end{tabular}
	\caption{Evolution of the vertex neighborhood, and segmentation of the narrative, for two characters from the \textit{Game of Thrones TV} series~\cite{Bost2016c}: a) \textsc{Jaime Lannister}; and b) \textsc{Arya Stark}.}
	\label{fig:summarization}
\end{figure}

Tsai \textit{et al}.~\cite{Tsai2013} use a network of character groups, instead of individual characters (cf. Section~\ref{sec:GraphExtrStatCount}): each vertex represents possibly overlapping subsets of characters, and edges between them model inclusion relations involving these subsets. The authors assume that a movie is typically built by first exposing the characters in scenes involving few of them, before developing the proper story by making more of them interact during more crowded scenes. They use an incoming degree-based centrality measure to identify the most important character groups in the plot and partition the network around them. The video summary is finally built by solving an optimization problem that consists in selecting the scenes involving the most central vertices (character groups), under some user-specified length constraints.

\subsubsection{Other Applications}
\label{sec:AppOtherApp}
In the literature, a number of less studied problems are also solved through the analysis of character networks: we mention them briefly in this section. 

\paragraph{Improvement of pre-processing.} Some authors leverage character networks to improve the execution of tasks which are considered as pre-processing in our extraction process. As a part of their method to perform scene segmentation in movies, Xu \textit{et al}.~\cite{Liang2009a} separately extract two co-occurrence networks: one based on the movie video, one based on its script. These networks are used to match faces from the video to character names in the script. 

Yeh \& Wu~\cite{Yeh2014} propose an iterative approach taking advantage of a co-occurrence networks to perform face clustering in movies. They start by applying a generic face detection method to obtain a raw estimation of the face clusters. They use their own method~\cite{Yeh2012} to extract a co-occurrence network, assuming each face cluster corresponds to a character. Similar clusters are merged based on features extracted from both the network and the video. The resulting clusters are used to extract a new co-occurrence network, and the same process is iteratively repeated to improve the quality of both face clusters and character network.

\begin{figure}[htb!]
	\centering
	\begin{tabular}{c c}
	    \includegraphics[width=0.4\textwidth]{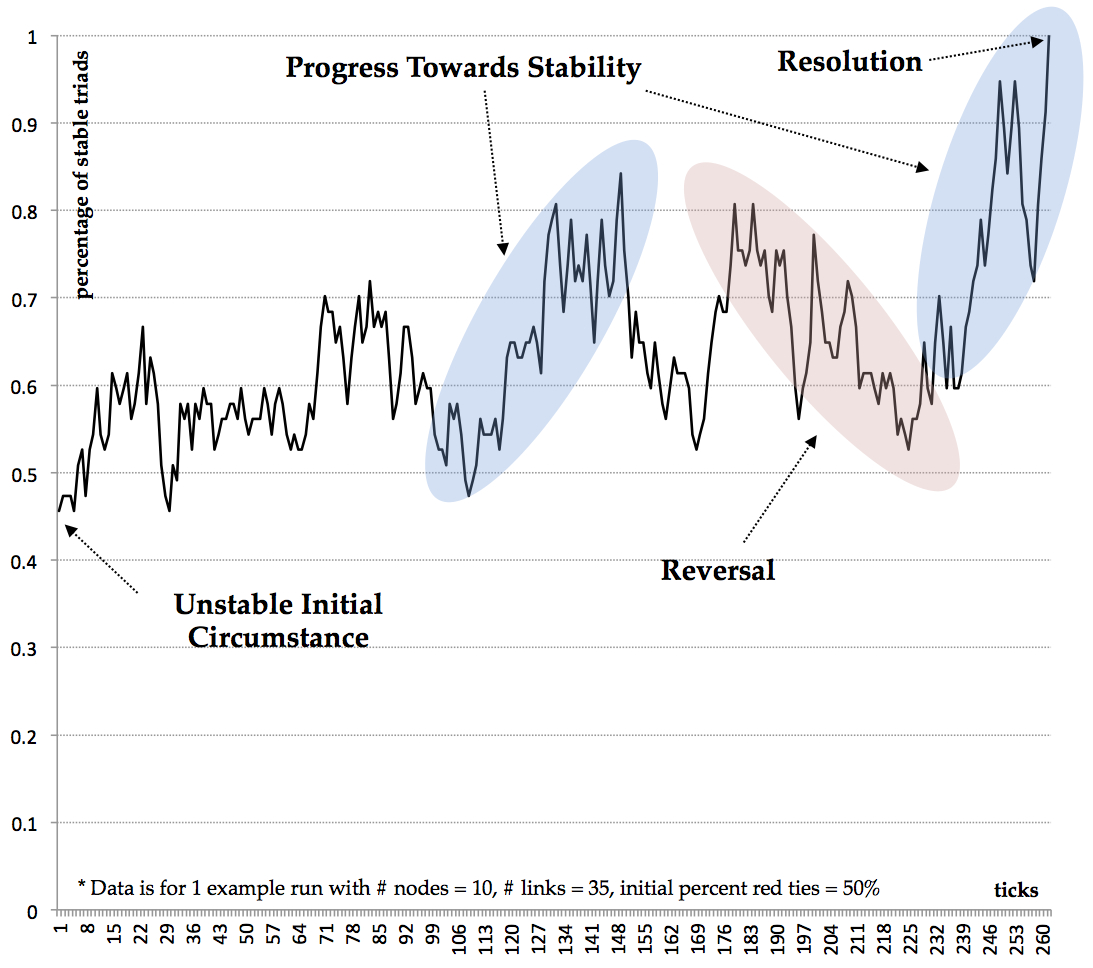} &
	    \includegraphics[width=0.4\textwidth]{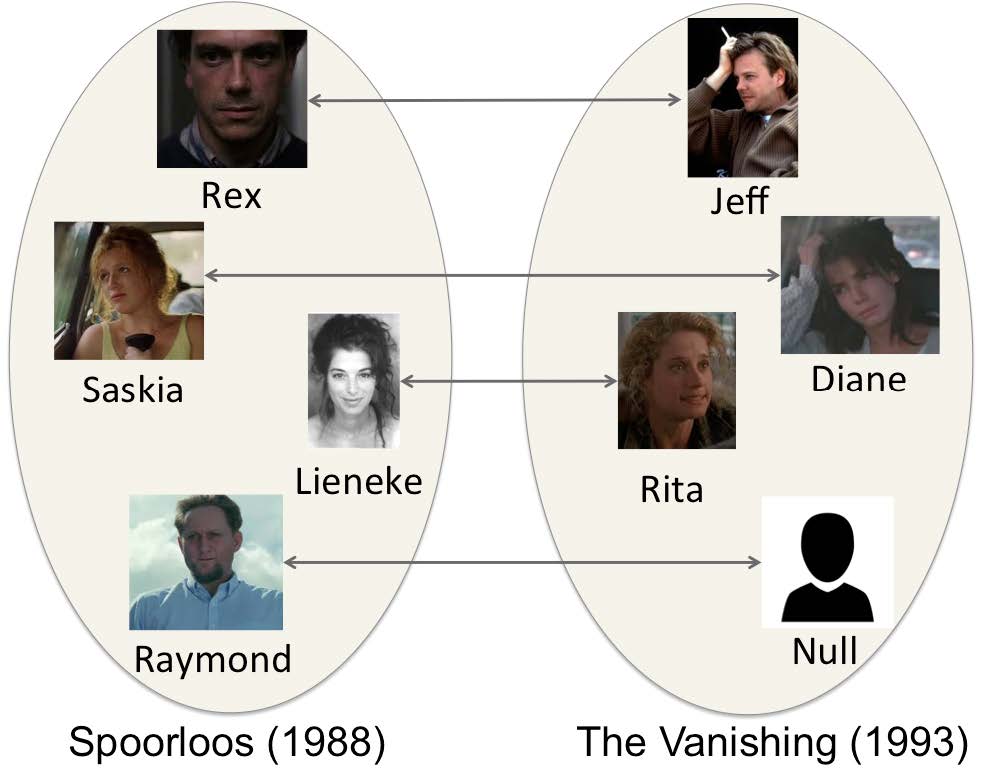} \\
	    a) & b) \\
    \end{tabular}
	\caption{Evolution of the proportion of stable triads, used in~\cite{Sack2013}; b) Alignment between the characters of two different movies (\textit{Spoorloos} and \textit{The Vanishing})~\cite{Chaturvedi2018}.}
	\label{fig:AppMisc}
\end{figure}

\paragraph{Generation and prediction of narratives.} Other authors use character networks to predict or generate stories. Sack~\cite{Sack2012, Sack2013, Sack2014} proposes a method to produce proto-narratives, i.e. sequences of events involving characters, useful to later build stories. Starting from a randomly generated unbalanced signed network, it iteratively switches one edge sign in a randomly picked unbalanced triangle, thus making it balanced (cf. Figure~\ref{fig:AppMisc}a). The whole network ultimately becomes balanced. Each sign switch is considered as an event in the proto-narrative: either befriending (negative edge becoming positive) or betraying (vice versa). The number of switches undergone by a vertex can be used \textit{a posteriori} to determine the type of behavior of the corresponding character. 

Nalisnick \& Baird have a similar approach, when they try to use structural balance to predict the evolution of Shakespeare's plays~\cite{Nalisnick2013a}. However, they use Marvel \textit{et al}.'s model~\cite{Marvel2011}, which consists in repeatedly squaring the signed adjacency matrix.

\paragraph{Recommendation systems.} Lee \& Jung~\cite{Lee2018} propose a recommendation system, allowing to propose a movie to a user depending on the ones he already likes. The authors first extract a so-called \textit{affective} dynamic co-occurrence network, which is basically a signed network representing the friendly vs. hostile relationships between characters. Based on the evolution of the centrality of the vertices and the relationships between them, they identify significant affective changes, which are assumed to correspond to major plot events. They then build a lattice-based representation of the plot using these events. These are finally used as proxies to compare plots.

\paragraph{Alignment of narratives.} In~\cite{Chaturvedi2018} Chaturvedi \textit{et al}. aim at automatically detecting movie remakes. They do not work with the videos themselves, but rather with textual summaries obtained from Wikipedia. They define two similarity measures between the narratives: one based on the plot and the other on the characters. The former is the cosine similarity applied to bag-of-words representations of the events and entities detected in the summaries. The latter is the average character alignment between the narratives. This alignment is computed by matching the characters of two distinct narratives (cf. Figure~\ref{fig:AppMisc}b) depending on their name, gender, prominence in the narrative, and relationships with other characters (the only character network-based aspect of the process). The authors combine both measures to train a classifier into predicting so-called \textit{remake clusters} (classes containing a movie and all its remakes).

Elsner~\cite{Elsner2012} also works with texts, but these are 19th century romantic novels. In order to compare their narratives, he defines a character similarity measure relying on both individual and relational aspects. The former include the characters' prominence as well as lexical properties of the text surrounding the characters' occurrences, represented as unigram distributions. Relationships are detected based on a variant of the co-occurrence approach described in Section~4.1 of the \textit{Main Article}, and Esner uses the resulting edge weights when comparing characters. The characters are first aligned based on the individual features. Then, the relational similarity between two characters belonging to two distinct narratives depends on how much the edge weights of their aligned neighbors match. In addition, Elsner takes the narrative dynamics into account by considering how all these aspects evolve over the narrative. He then trains a classifier to distinguish between actual novels and randomly generated ones.

\section{Discussion, Opportunities and Perspectives}
\label{sec:Perspectives}
Our review of the literature reveals that there are still a number of issues to solve and directions to explore, at each of the steps constituting the network extraction (Section~\ref{sec:PerspExtrac}) and exploitation (Section~\ref{sec:PerspAnalysis}) processes.

\subsection{Network Extraction}
\label{sec:PerspExtrac}
We identify three main types of open problems regarding the extraction of character networks: the improvement and development of tools to identify characters and interactions in the various types of media used to create narratives (Section~\ref{sec:PerspExtracIdentif});of approaches to construct dynamic character networks based on this information (Section~\ref{sec:PerspExtracDynNet}); and of evaluation methods to assess the performance of these identification and construction tools (Section~\ref{sec:PerspExtracEval}).

\subsubsection{Character and Interaction Identification}
\label{sec:PerspExtracIdentif}
It appears from this survey that all media are not equal regarding the character and interaction identification processes. This is because they rely on the resolution of a number of medium-specific lower-level problems, for which the state-of-the-art performance is not always as advanced in all domains. 

\paragraph{Textual Narratives} 
With textual narratives, the literature shows that automatic methods work relatively well when dealing with the simplest tasks: character occurrences can be identified through NER~\cite{Goyal2018}, and detecting their co-occurrences is quite straightforward. A number of authors extract character networks based on these tools only. However, there is still work to do in order to leverage more of the information conveyed by text. Character unification is not efficiently solved, as it requires detecting all forms of character aliases, and most of all performing anaphora resolution, which is a difficult NLP problem~\cite{Poesio2016}. There are also research opportunities with the identification of advanced types of interactions between characters, besides simple co-occurrences. This task is much harder, as it requires analyzing the semantics of the text, for instance by detecting action verbs and their respective subjects and objects. For all these open problems, as explained in the survey, the specific characteristics of fiction offer both additional difficulties and leverage compared to non-fiction content.

\paragraph{Audiovisual Narratives} 
By comparison with text, the results obtained with automatic tools on audiovisual narratives are much less satisfying, to the point where many authors perform certain tasks such as face recognition manually. This difference might explain why there are many more articles dealing with text than audiovisual narratives. The low-level tasks necessary to detect character occurrences (face detection and tracking, speaker segmentation) and unify them (face track clustering, speaker clustering) are still open problems when performed in such uncontrolled conditions. Regarding interaction detection, in practice authors basically focus on co-occurrences only, as identifying more precise interactions requires solving even lower-level problems. For instance, to get a conversational network, one has to perform lip motion detection or speaker identification to determine who is speaking. Furthermore, interlocutors may be tricky to detect when multiple characters are involved in the same scene. Obtaining an action-based network would require determining which character performs which action on which other character~\cite{Vrigkas2015}: there is no trace of such attempt in the literature.

It seems difficult to automate the whole character network extraction pipeline as long as all these basic processing steps cannot be efficiently performed. Moreover, unlike for textual narratives, here the specific properties of fiction result in significant  difficulties compared to non-fiction, and yield little additional leverage. On the bright side, multimodal approaches constitute a very promising perspective, as some authors try to combine information extracted from the video and audio streams, but also from text (scripts, transcriptions, external sources such as wiki pages).

\paragraph{Comics} 
This might be surprising, but comics appear to be even harder to process as a narrative than video content. This is because the information they convey can take a number of different forms (text, drawings, balloons, panels, onomatopoeia, effect lines, textures), each one needing some specific processing. Moreover, the medium itself is not very normalized, in the sense that distinct artists may express the same ideas in very different graphical ways. As a result, all the (few) authors extracting character networks from comics operate manually~\cite{Alberich2002, Gleiser2007, Rochat2017}.

Detecting and unifying character occurrences in comics require solving certain problems similar to those met for videos, such as face detection, but in an even more adverse context (non-anthropomorphic and/or highly deformed characters). Moreover, they also require performing specific lower-level tasks such as panel segmentation or bubble detection, which are still open problems. For interaction detection, like for audiovisual narratives, co-occurrences are easy to get once the characters have been identified (provided the panels have been properly segmented), but identifying more precise interactions is hard. Leveraging conversations would essentially depend on the extraction of text and locating of bubbles and captions, two difficult tasks which are not satisfactorily tackled by state-of-the-art methods. Identifying specific actions that characters perform on each other would require associating semantics to specific body postures, which is also an open problem. To finish with comics, it is worth noticing that we could not find any article dealing with \textit{photonovels}: on the one hand, their economical interest is questionable, as they felt out of fashion, but on the other hand they might be easier than comics, as they rely on photographs instead of drawings, while retaining certain conventions of comics.

\paragraph{General Remarks} 
To summarize, independently of the type of medium, there are many low-level open problems regarding the extraction of characters and interactions in fiction works. Recently, deep learning methods have proven to be efficient on classic NLP and multimedia tasks (e.g.~\cite{Ghannay2018}), and are therefore a promising perspective to solve most of these problems. Alternatively, a promising change of perspective would be to break with the pipeline approach, which is currently the norm for character network extraction, as illustrated by Figure~\ref{fig:process}. A pipeline is a chaining of separate tools, each one aiming at solving a specific subproblem. 

All approaches first attempt to detect characters before trying to capture interactions, possibly propagating errors made at the earlier stages of the process. Instead, both tasks could probably benefit from one another and be performed jointly, as demonstrated by Yeh \& Wu for face clustering~\cite{Yeh2014}. Another possible option to avoid or reduce the pipeline limitations is to leverage deep learning methods and adopt a supervised \textit{end-to-end} approach over parts, or even the whole extraction process. However, deep learning methods require to be trained on large corpora, and current fiction-related corpora are small, or do not even exist for certain tasks (especially for comics). Thus, before being able to leverage deep learning methods, it is likely that the community will have to constitute or extend such corpora, a tedious work often performed collectively. An alternative would be to take advantage of existing corpora to perform \textit{transfer learning}, based on existing models built on non-fiction data.

\subsubsection{Dynamic Networks}
\label{sec:PerspExtracDynNet}
Many authors using character networks to solve higher-level problems identify the dynamics of the story as a very important aspect to be taken into account. This is especially true in long-term narratives such as TV or novel series, in which relationships and characters are likely to change over time. Yet, the overwhelming majority of methods proposed in the literature rely on \textit{static} networks. This may be because extracting \textit{dynamic} networks requires making tough methodological choices such as choosing the size of the temporal window. Moreover, time can be modeled in several ways in a network~\cite{Holme2012, Aggarwal2014}. Finally, there are much fewer off-the-shelf tools to analyze dynamic networks, and these are more difficult to use. 

However, this issue deserves to be explored, as dynamic networks are likely to improve both our understanding of the structure of fictional character networks, and the performance of the methods taking advantage of this structure to solve problems such as those presented in Section~\ref{sec:Applications}. It is even likely that such networks are necessary to solve certain intrinsically dynamic problems such as narrative generation (see Section~\ref{sec:AppOtherApp}).

\subsubsection{Evaluation and Methodological Choices}
\label{sec:PerspExtracEval}
As described before, extracting a reliable character network requires solving efficiently a number of media-specific problems (ex. NER, face detection), most of them difficult. 

\paragraph{Evaluation Data and Methods} 
This raises several questions, the first one being that of \textit{performance assessment}. As mentioned before, all existing extraction methods rely on the pipeline approach. Consequently, each processing step constituting the pipeline must be assessed separately. Performing such a task requires an appropriate ground truth, designed for the considered problem. Such annotated corpora already exist for most problems, but they are generally based on non-fiction data, in which case they do not account for the specific characteristics of fiction described throughout this survey. To obtain a reliable performance evaluation, the test dataset must be built on fiction, which ties this point to the issue of machine learning training that we mentioned before: for a number of problems, such corpora are still to be constituted. But it is also important to estimate the relevance of the extracted character networks, i.e. the output of the pipeline. Up to now, only a few authors have tried to conduct this type of assessment on character networks~\cite{Lee2012f, Agarwal2013}. This task necessitates ground-truth networks to which one can compare the networks extracted using the pipeline, so here too there is a significant manual annotation work to do. Besides these data-related aspects, some methodological questions remain to be solved. There are many possibly ways to compare two graphs~\cite{Wills2019}, and it is not obvious which ones are the most appropriate to the case of character networks. In addition, this aspect could be dependent on the high-level use one wants to make of the network (ex. story decomposition, summarization).

\paragraph{Methodological Choices} 
Solving the data and methods issues described above is important not only to evaluate the performance of the processing steps constituting the extraction pipeline, but also to drive methodological choices. Indeed, there are a number of open questions regarding which approaches are the most appropriate to handle these steps, or if some steps should be handled at all, and assessing their effect on the overall pipeline performance is necessary to discriminate between them. For instance, in textual narratives, it is not clear whether detecting all forms of character mentions (including nominals) instead of just looking for proper nouns, and/or solving anaphoras instead of ignoring them, would significantly improve the quality of the extracted networks. The same goes when detecting interactions: we do not know if conversation-based networks are more informative than co-occurrence-based ones, yet they are much harder to extract. To generalize, we do not know whether solving such harder low-level problems would result in better character networks: this remains to be tested. Solving efficiently these problems require significant extra efforts, so this is a very important question, yet it is still open. Very few~\cite{Jayannavar2015, Edwards2018} of the reviewed articles try to tackle the problem, by assessing the quality of the extracted networks. Solving the evaluation-related issues would also allow assessing the relevance of using dynamic networks instead of static ones, and/or to identify which parameter values are optimal (e.g. size of the narrative unit when detecting co-occurrences, or of the sliding window when building dynamic networks). It would also allow comparing the different ways of computing edge weights, as recent articles on real-world conversational networks have shown their importance for classification tasks~\cite{Papegnies2019}. Finally, it is important to note that the relevance of the extracted network is likely to depend greatly on the high-level task for which one plans to leverage this network.

\subsection{Network Analysis}
\label{sec:PerspAnalysis}
We distinguish between the development of methods aiming at describing character networks (Section~\ref{sec:PerspAnalysisDescr}), and the exploitation of character networks to solve high-level problems (Section~\ref{sec:PerspAnalysisApps}).

\subsubsection{Descriptive Tools}
\label{sec:PerspAnalysisDescr}
\paragraph{Exploration/Adaptation of Existing Tools} 
The tools used in the literature to describe character networks are standard, very widespread complex network topological measures (see Section~\ref{sec:DescriptiveTools}). Yet, the field has produced many more tools which are not as known (e.g. motif-based measures, backbone detection, core-periphery structures, community structure-related vertex roles), but still likely to be relevant to study character networks (see~\cite{FontouraCosta2007} for a review). Therefore, there is a large number of opportunities to explore here. Some of these tools (e.g. motif-based measures) are designed for large networks, which means they would be relevant only for networks representing collections of narratives, like the comic networks mentioned in the survey~\cite{Alberich2002, Gleiser2007}. Another interesting possibility is to adapt existing measures to the specific case of character networks. For instance, certain authors identify one or two characters as extremely central in the studied narrative (protagonist vs. antagonist). Then when computing distance-based centrality measures such as the closeness, why not focusing only on the shortest path joining the character of interest to the protagonist, instead of considering all pairs of characters?

\paragraph{Additional Information} 
Another promising perspective is to follow the evolution currently taking place in the field of complex network analysis, and to integrate more information in the network, in order to apply tools able to leverage this additional information. A few authors mentioned in the survey have started extracting such networks, by including time (dynamic networks), relationship polarity (signed networks), or individual information (attributed networks) in their model. However, they focus more on the extraction process than on the way to efficiently use this information. For instance, several approaches exist to detect significant changes in dynamic networks~\cite{Aggarwal2014}, which could be applied to identify plot twists, or to segment the narrative. Regarding signed networks, there are alternatives to the concept of structural balance, for instance relaxations allowing to identify mediator groups in a situation of conflict: this could be of interest when analyzing so-called \textit{adversarial} movies~\cite{Ding2010}. Vertex attributes (see Section~\ref{sec:CharIdAddProcCharAttr}) can be the object of the description through measures such as assortativity, which allows quantifying how much homophilic the network is. But they can also help to interpret the output of other tools, for instance in the context of community detection, by associating certain attribute values to specific communities. To conclude, note that it is possible to combine all these aspects (time, attributes, polarity) in one network, which would require very specific descriptive tools.

\paragraph{Network Simplification} 
Another, more minor, open question, is that of network simplification. Most narratives contain a number of extras, i.e. very minor characters used only to strengthen the level of realism of the story, or set the atmosphere. They might be considered as noise though, as they are basically part of the set or background, and do not affect the story in any way. Certain authors try to remove such characters, for example by trimming leaves from the character network. This allows, in particular, easing the visual exploration of the network. However, the objective effect on the network topological properties and high-level problem at hand remains to be identified and measured, and some other methods might be more appropriate.

\subsubsection{Applications}
\label{sec:PerspAnalysisApps}
\paragraph{Methods Transposition} 
It is worth noting that the articles using character networks to solve high-level problems are quite segregated depending on the medium of the considered narratives (textual vs. audiovisual). This makes sense for the low-level part of the network extraction process, as it is medium-specific, but not for the exploitation of the extracted networks, as these do not vary much depending on the narrative medium. Consequently, it seems possible to transpose the methods designed to solve certain problems on a given medium to another one. For instance, text-based networks have been used to assess literary theories, but we could not find any article trying to do the same for cinematographic ones, e.g. the widespread use of the so-called \textit{Hollywood Formula} to design movie plots. On the contrary, character networks have been used to generate summaries of TV series: the same general principle could be applied to novels.

\paragraph{Narrative Classification} 
A number of high-level applications taking advantage of character networks consist in solving some classification problem, such as grouping narratives depending on their genre, time of publication, or creator. A related task is that of narrative recommendation (e.g. movie recommendation), as it can also be formulated as a prediction problem. In both cases, all authors feed their classifier with a few topological measures, generally the same very standard measures mentioned in Section~\ref{sec:PerspAnalysisDescr}. But there is no reason why these specific measures would be more appropriate for a classification task than the other measures defined in the literature. Also, the most relevant measures are not necessarily the same depending on the considered classification task. A first improvement would therefore be to adopt a more exhaustive approach, and consider a much larger number of topological measures, picking them so that they characterize the network at various scales and scopes (see for instance~\cite{Papegnies2019} on real-world conversational networks). Another improvement would be to use representation learning instead of manually selecting discriminant features. This could be performed by using \textit{graph embeddings}~\cite{Cui2019}, a recent transposition to graphs of the NLP concept of \textit{word embeddings}~\cite{CamachoCollados2018}. This approach allows directly learning the most appropriate numeric representation of the character network for the considered classification task. However, it requires access to a large amount of data.

\paragraph{Trans- and Cross-media}
Recently, cross- and trans-media narratives have become popular, and their automatic processing constitutes another promising perspective. Crossmedia means that the same narrative is repeated over several different media, e.g. a novel and its movie adaptation. Transmedia means that different narratives belonging to the same fictional universe are expressed using different media, e.g. the sequel of a movie taking the form of a novel. This raises the interesting problem of \textit{network alignment}, i.e. determining which character in one network corresponds to which character in the other (see also Section~\ref{sec:AppOtherApp}). In the case of transmedia, one could use character networks to look for plot intersections between the different narratives, and possibly detect divergences. They could also be used to help the user navigate among the sometimes numerous and complex narratives (e.g. superhero universes). In the case of crossmedia, adapting a narrative for a different medium generally involves changing the plot or even the story, and adding or removing characters. Identifying which changes a character network undergoes during such adaptation could help to understand and maybe automating this process. Moreover, having access to the same story under different forms could be leveraged to improve network extraction itself, by adopting and extending the multimodal approach already mentioned in Section~\ref{sec:PerspExtracIdentif}.

\paragraph{Narrative Generation} 
Finally, automatic plot generation constitutes another promising perspective. Performing this task is likely to require leveraging some additional information, as mentioned in Section~\ref{sec:PerspAnalysisDescr}, in particular: dynamic networks, in order to represent existing plots and study them to model their typical evolution; and signed networks to deal with antagonistic stories, which are very frequent. A few recent articles deal with plot generation (see Section~\ref{sec:AppOtherApp}), but there is still much work to do. Designing models to generate networks which would be realistic according to some criteria of interest constitutes an important part of the complex networks field\footnote{E.g. the famous Watts-Strogatz~\cite{Watts1998} and Barabási-Albert~\cite{Barabasi1999} models, which respectively produce small-world and scale-free networks}, so this task could benefit from the work already conducted on this issue. 

However, it is worth noting that if generating a sequence of events corresponding to a plot looks achievable in the short-term, automatically converting it into a proper narrative seems like a very difficult problem. It is possible to adopt an extractive approach, i.e. to build the narrative by selecting narrative bits among a predefined collection. The alternative is to generate the narrative outright, but this seems realistic only for certain media, in particular text, for which efficient language-based models exist.

\subsection*{Acknowledgments}
The authors would like to thank the anonymous reviewers for their work and feedback, which helped significantly improve this article. Part of this work was funded by \href{https://agorantic.univ-avignon.fr/en}{Agorantic FR 3621}.

\appendix
\section{Methods for the Extraction of Fictional Character Networks}
\begin{figure}[htb!]
	\centering
	\includegraphics[width=\textwidth]{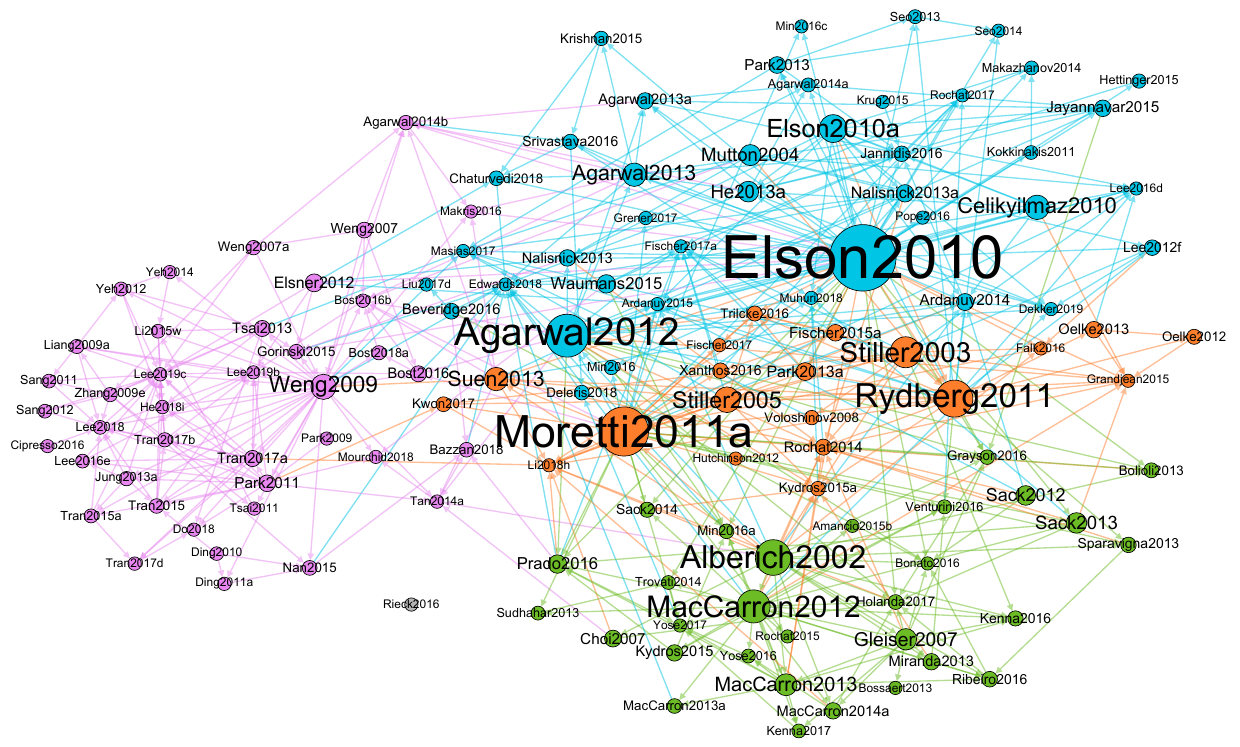}
	\caption{Citation network for the articles of Table~\ref{tab:ArticleList}. Figure and data available at \href{https://doi.org/10.6084/m9.figshare.7993040}{10.6084/m9.figshare.7993040} under CC-BY license. Note that the figure does not include the latest additions to Table~\ref{tab:ArticleList}.}
	\label{fig:CitationNet}
\end{figure}

\begin{figure}[htb!]
	\centering
	\includegraphics[width=\textwidth]{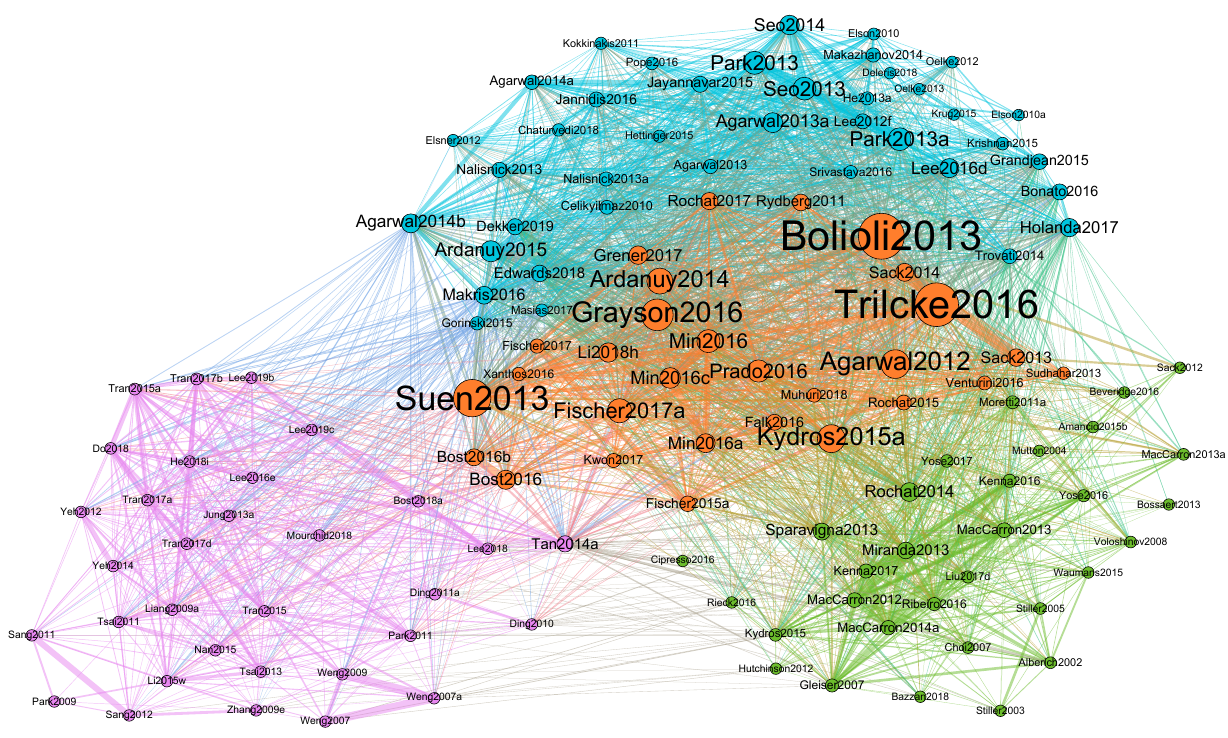}
	\caption{Co-citation network for the articles of Table~\ref{tab:ArticleList}. Figure and data available at \href{https://doi.org/10.6084/m9.figshare.7993040}{10.6084/m9.figshare.7993040} under CC-BY license. Note that the figure does not include the latest additions to Table~\ref{tab:ArticleList}.}
	\label{fig:CocitationNet}
\end{figure}

Table~\ref{tab:ArticleList} shows the list of articles describing methods to extract character networks from works of fiction. Although many other articles are mentioned through the survey when dealing with specific points of the process described in Figure~\ref{fig:process}, in this table we only focus on methods aiming at extracting proper networks. Figures~\ref{fig:CitationNet} and \ref{fig:CocitationNet} display how these articles relate. 

\paragraph{Resource Availability.} We made these resources publicly available online\footnote{\url{https://doi.org/10.6084/m9.figshare.7993040}}: the table as a spreadsheet file, the figures as separate files, the networks as Gephi and Graphml files, and the bibliographic entries as a BibTeX file. 

\textbf{Note:} The up-to-date list of academic articles is now available directly as a Web page\footnote{\url{https://compnet.github.io/CharNetReview/}}. This paper will not be updated anymore.

\afterpage{\begin{landscape}
\scriptsize
\begin{longtable}{>{\raggedright\arraybackslash}p{5.3cm} >{\raggedright\arraybackslash}p{2.0cm} l l l l l@{\hskip 0.8cm} l l l l >{\raggedright\arraybackslash}p{4.2cm}}
	\hline
	\textbf{Work of fiction} & \textbf{Ref.} & \multicolumn{5}{l}{\textbf{Relationships}} & \multicolumn{4}{l}{\textbf{Graph}} & \textbf{Application} \\
	 & & \textbf{Cc.} & \textbf{Cv.} & \textbf{M.} & \textbf{Ac.} & \textbf{Af.} & \textbf{W.} & \textbf{Di.} & \textbf{S.} & \textbf{Dy.} & \\
	\hline\endhead
	\hline\endfoot\endlastfoot
	
	Mozart's \textit{Cosi Fan Tutte} & \cite{Harary1963} & N & N & N & N & Y & N & N & Y & Y & Descriptive analysis \\
    Murdoch's \textit{A Severed Head} & \cite{Harary1966a} & N & N & N & N & Y & N & N & N & N & Descriptive analysis \\
    Shakespeare's \textit{A Midsummer Night's Dream} & \cite{Stanton1967} & N & N & N & N & Y & N & N & Y & Y & Descriptive analysis \\
    Wagner's \textit{Der Ring des Nibelungen} & \cite{Mayer1973} & N & N & N & N & Y & N & Y & Y & N & Descriptive analysis \\
    Shakespeare's \textit{Twelfth Night} & \cite{Harary1975} & N & N & N & N & Y & N & Y & Y & N & Descriptive analysis \\
    16 fairy tales & \cite{Auster1980} & N & N & N & N & Y & N & N & Y & Y & Descriptive analysis \\
    Sophocles' \textit{Oedipus Rex} & \cite{Harary1982} & N & N & N & N & Y & N & Y & N & N & Descriptive analysis \\
    Priestley's \textit{Dangerous Corner} & \cite{Harary1985} & N & N & N & N & Y & N & Y & Y & N & Comparative study \\
    5 classic novels & \cite{Knuth1993} & Y & N & N & N & N & N & N & N & N & Benchmark for graph processing tools \\
	\textit{Marvel} universe & \cite{Alberich2002} & Y & N & N & N & N & N & N & N & N & Level of realism \\
	10 Shakespeare's plays & \cite{Stiller2003, Stiller2005} & Y & N & N & N & N & N & N & N & N & Level of realism \\
	All of Shakespeare's plays & \cite{Mutton2004} & N & Y & N & N & N & Y & N & N & N/Y & Visualization \\
	Dictionary of Greek and Roman mythology & \cite{Choi2007} & N & N & Y & N & N & N & N/Y & N & N & Descriptive analysis \\
	\textit{Marvel} universe & \cite{Gleiser2007} & Y & N & N & N & N & Y & N & N & N & Level of realism \\
	Hollywood movies, and TV series & \cite{Weng2007, Weng2007a, Weng2009} & Y & N & N & N & N & Y & N & N & N & Storyline identification/segmentation \\
	9 classical plays & \cite{Voloshinov2008} & N & Y & N & N & N & N & N & N & N & Level of realism \\
	Movies and TV series & \cite{Park2009, Park2011} & N & Y & N & N & N & Y & Y & N & N & Role detection, story segmentation \\
	\textit{Friends} TV show & \cite{Yuan2009, Yuan2010} & Y & N & N & N & N & Y & N & N & N & Method assessment \\
	15 Hollywood movies & \cite{Zhang2009e, Liang2009a, Sang2011, Sang2012} & Y & N & N & N & N & Y & N & N & N & Face-name matching, Scene segmentation \\
	Austen's \textit{Pride \& prejudice}, and \textit{Emma} & \cite{Celikyilmaz2010} & N & Y & N & N & N & Y & N & N & N & Method assessment \\
	10 adversarial movies & \cite{Ding2010} & Y & N & N & N & N & Y & N & Y & N & Method assessment \\
	60 19th century British novels & \cite{Elson2010, Elson2010a, Elson2012} & N/Y & N/Y & N/Y & N & N & Y & N & N & N & Check literary theories \\
	20 Hollywood movies & \cite{Ding2011a} & Y & N & N & N & N & Y & N & Y & Y & Community detection \\
	300 19th century Swedish novels & \cite{Kokkinakis2011} & Y & N & N & Y & Y & N & N & N & N & Method assessment \\
    Toriyama's \textit{Dragon Ball} vol.32 & \cite{Murakami2011} & Y & N & N & N & N & Y & N & N & N & Method assessment \\
	Shakespeare's \textit{Hamlet}, \textit{Macbeth}, and \textit{King Lear} & \cite{Moretti2011a} & N & Y & N & N & N & N & N & N & N & Discussion about the characters \\
	Collection of Greek Tragedies & \cite{Rydberg2011} & N & Y & N & N & N & N & Y & N & N & Plot structure comparison \\
	8 Hollywood movies & \cite{Tsai2011} & Y & N & N & N & N & N & N & N & Y & Scene segmentation \\
	Carroll's \textit{Alice in Wonderland} & \cite{Agarwal2011, Agarwal2012, Agarwal2013, Agarwal2013a, Agarwal2014a} & N & N & N & Y & N & N & N/Y & N & N/Y & Role detection \\
	41 19th century novels & \cite{Elsner2012} & Y & N & N & N & N & Y & N & N & Y & Genre comparison \\
	Rowling's \textit{Harry Potter} novels & \cite{Hutchinson2012} & Y & N & N & N & N & Y & N & N & N & Method assessment \\
	The \textit{Old Testament} & \cite{Lee2012f} & N & Y & N & Y & Y & N & N & N & N & Method assessment \\
	4 European tales & \cite{MacCarron2012, MacCarron2013, MacCarron2013b, MacCarron2013a, MacCarron2014a, Kenna2016, Kenna2017} & Y & Y & Y & Y & Y & N/Y & N & Y & N & Level of historicity \\
	19th and 20th century Swedish novels & \cite{Oelke2012, Oelke2013} & Y & N & N & N & N & Y & N & N & N/Y & Visualization \\
	6 canonical European novels & \cite{Sack2012, Sack2013, Sack2014} & Y & N & N & N & N & Y & N & N & N & Proto-narrative generation \\
	Frankel's \textit{The Devil wears Prada} & \cite{Yeh2012, Yeh2014} & Y & N & N & N & N & Y & N & N & N & Face clustering \\
	Manzoni's \textit{I promessi sposi} & \cite{Bolioli2013} & N/Y & N/Y & N & N & N & Y & N & N & N & Method assessment \\
	Rowling's \textit{Harry Potter} novels & \cite{Bossaert2013} & N & N & N & Y & N & N & Y & N & N & Study of peer support \\ 
	3 classic European novels & \cite{He2013a, Makazhanov2014} & N & Y & N & N & Y & N & N & N & N & Method assessment \\
	10 Hollywood movie scripts & \cite{Jung2013a} & N & Y & N & N & N & Y & Y & N & N & Role detection \\
	Homer's \textit{Odyssey} & \cite{Miranda2013, Miranda2018} & Y & Y & Y & Y & Y & N & N & N & N & Level of historicity \\
	Shakespeare's plays & \cite{Nalisnick2013, Nalisnick2013a} & N & Y & N & N & N & Y & Y & Y & Y & Method assessment \\
	20 novels & \cite{Park2013, Park2013a, Seo2013, Seo2014} & Y & N & N & N & N & Y & N & N & N & Descriptive analysis \\
	Rowling's \textit{Harry Potter and the Philosopher's Stone} & \cite{Sparavigna2013} & N & Y & Y & Y & Y & N & N & N & N & Level of realism \\
	3 story books & \cite{Sudhahar2013} & N & N & N & Y & N & Y & Y & N & N & Role detection \\
	173 theater and 580 movie scripts & \cite{Suen2013} & N/Y & N/Y & N & N & N & Y & N & N & N & Classification of works \\
	12 Hollywood movies & \cite{Tsai2013} & Y & N & N & N & N & N & N & N & N & Video summarization \\
	674 movie scripts & \cite{Agarwal2014b, Agarwal2016} & N & Y & N & N & N & N & N & N & N & Performance evaluation \\
	238 novels & \cite{Ardanuy2014, Ardanuy2015} & Y & N & N & N & N & Y & N & N & N & Classification of works \\
	250,000 novels, myths and fary tales & \cite{Bodrova2014} & Y & N & N & Y & N & N & N & N & N & Method assessment \\
	1,800 19th century British and American novels & \cite{Condello2014} & Y & N & N & N & N & Y & N & N & N & Comparative study \\
	Tales and myths & \cite{MacCarron2014} & Y & Y & Y & Y & Y & N & N & Y & N & Level of historicity, Categorization \\
	\textit{Das Nibelungenlied}, and Shakespeare's \textit{Hamlet} & \cite{Marazzato2014, Sparavigna2015} & Y & N & N & N & N & Y & N & N & N & Method assessment \\
	Rousseau's \textit{Les confessions} & \cite{Rochat2014a, Rochat2014} & Y & N & N & N & N & N/Y & N & N & N/Y & Role detection \\
	Shakespeare's \textit{Hamlet} and \textit{Othello} & \cite{Sparavigna2014a} & N & Y & N & N & N & N & N & N & N & Descriptive analysis \\
    \textit{Star Trek} and \textit{Star Gate} movies and TV series scripts & \cite{Tan2014a, Tan2017, Tan2018a} & Y & N & N & N & N & Y & N & N & N & Comparative study \\
	Knowles' \textit{The Legends of King Arthur and his Knights} & \cite{Trovati2014} & N & N & N & Y & N & N & N & Y & N & Method assessment \\
	Rowling's \textit{Harry Potter} series & \cite{Zhang2014q} & N & N & N & N & Y & N & N & N & N & Descriptive analysis \\
    20 novels & \cite{Amancio2015b} & Y & N & N & N & N & Y & N & N & N & Descriptive analysis \\
	1,276 movie scripts & \cite{Gorinski2015} & Y & N & N & N & N & Y & N & N & N & Script summarization \\
	1,682 16--20th century German novels & \cite{Hettinger2015} & Y & N & N & N & N & N & N & N & N & Classification of works \\
	60 19th century British novels & \cite{Jayannavar2015} & N & Y & N & Y & N & N/Y & N/Y & N & N & Check literary theories \\
	200 17--18th century French plays & \cite{Karsdorp2015} & Y & N & N & N & N & Y & N & N & N & Character ranking, Visualization \\
    617 movie scripts & \cite{Krishnan2015} & N & Y & N & N & N & N & N & Y & N & Infer formality level of interactions \\
	58 German novels & \cite{Krug2015, Jannidis2016, Krug2020} & N/Y & N/Y & N & N & N & Y & N & N & N & Role detection \\
	Homer's \textit{Iliad} & \cite{Kydros2015} & Y & N & N & N & N & N & N & N & N & Descriptive analysis \\
	Embirikos' \textit{The Great Eastern} & \cite{Kydros2015a} & Y & Y & Y & Y & Y & Y & Y & N & N & Descriptive analysis \\
	Shakespeare's \textit{Julius Caesar} & \cite{Lotker2015} & N & Y & N & N & N & N & Y & N & N & Detect communities of characters \\
	6 Hollywood movies & \cite{Li2015w} & Y & N & N & N & N & N & N & N & Y & Video summarization \\
	183 \textit{Friends} episodes & \cite{Nan2015} & Y & N & N & N & N & Y & N & N & Y & Method assessment \\
	\textit{The Wire} script & \cite{Pope2016} & N & Y & N & N & N & Y & N & N & N & Role detection \\
	Zola's \textit{Les Rougon-Macquart} & \cite{Rochat2015} & Y & N & N & N & N & Y & N & N & N & Plot structure comparison \\
	First 6 \textit{Star Wars} movies & \cite{Tran2015} & Y & N & N & N & N & Y & N & N & N & Role detection \\
	Corpus of Hollywood movies & \cite{Tran2015a, Tran2017a, Do2018} & Y & N & N & N & N & Y & N & N & N & Movie summarization \\
	Large corpora of European plays & \cite{Fischer2015a, Trilcke2016, Fischer2016, Fischer2017, Fischer2017a} & Y & N & N & N & N & N & N & N & N & Check literary theories \\ 
	46 fantasy novels, and Hugo's \textit{Les misérables} & \cite{Waumans2015} & N & Y & N & N & N & N & N & N & Y & Classification of works \\
	Martin's \textit{A Storm of Swords} & \cite{Beveridge2016} & Y & N & N & N & N & Y & N & N & N & Descriptive analysis \\
	3 modern novels & \cite{Bonato2016} & Y & N & N & N & N & Y & N & N & N & Model fitting \\
	3 TV series & \cite{Bost2016, Bost2016b, Bost2016c, Bost2018a} & N/Y & N/Y & N & N & N & Y & N & N & N/Y & Video summarization \\
	Zhi Ning's \textit{Journey to the West Prequel} & \cite{Chen2016k} & Y & N & N & N & N & Y & N & N & N & Descriptive analysis \\
    Tarentino's \textit{Pulp Fiction} & \cite{Cipresso2016} & N & N & N & Y & N & Y & Y & N & N & Descriptive analysis \\
	Edgeworth's \textit{The Absentee} & \cite{Falk2016} & N & Y & N & N & N & N & N & N & N & Check literary theories \\
	19th century British novels & \cite{Grayson2016, Grayson2016a} & Y & N & N & N & N & Y & N & N & N/Y & Descriptive analysis \\
	4 novels & \cite{John2016, John2017, John2019} & Y & N & N & N & N & Y & N & N & N & Visualization \\
    Chinese Buddhist Canon & \cite{Lee2016d} & N & Y & N & N & N & Y & N & N & N & Descriptive analysis \\
	10 Hollywood movies & \cite{Lee2016e, Lee2018} & Y & N & N & N & N & Y & N & N & Y & Recommendation system \\
	Shakespeare's \textit{Julius Caesar}, \textit{Hamlet}, and \textit{Othello} & \cite{Lotker2016} & N & Y & N & N & N & N & Y & N & Y & Study of time flow in dynamic networks \\
    501 movie scripts & \cite{Makris2016} & Y & N & N & N & N & Y & N & N & N & Method assessment \\
	The \textit{Pentateuch} 
	    & \cite{Massey2016} & N & Y & Y & Y & Y & N & N & N & N/Y & Level of historicity \\
    Hugo's \textit{Les misérables} & \cite{Min2016, Min2016a, Min2016c} & Y & N & N & N & N & N & N & N & N/Y & Story segmentation \\
	\textit{La chanson de Roland}, and 
	    \textit{Alice in Wonderland} & \cite{Prado2016} & Y & Y & Y & Y & Y & N & N & N & N/Y & Descriptive analysis \\
	Tolkien's \textit{Middle-Earth} novels & \cite{Ribeiro2016} & Y & Y & Y & Y & N & N & N & N & N & Descriptive analysis \\
	All of Shakespeare's plays & \cite{Rieck2016} & Y & N & N & N & N & Y & N & N & N & Comparative study \\
	CMU Movie Summary corpus & \cite{Srivastava2016} & Y & N & N & Y & Y & N & N & Y & N & Method assessment \\
	Homer's \textit{Iliad} & \cite{Venturini2016} & N & N & N & Y & N & N & N & N & N & Visualization \\
	First 4 books of Martin's \textit{A Song of Ice and Fire} & \cite{Wohlgenannt2016} & Y & N & N & N & N & N & N & N & N & Method assessment \\
    Molière's \textit{L'école des femmes} & \cite{Xanthos2016} & Y & N & N & N & N & Y & N & N & Y & Visualization \\
	\textit{Ossian} corpus of Scottish epic poems & \cite{Yose2016} & Y & Y & Y & Y & Y & N & N & Y & N & Level of historicity \\
	3,568 English-language plays from 1550--1900 & \cite{Algee-Hewitt2017, Piper2017} & N & Y & N & N & N & Y & Y & N & N & Check literary theories \\
    817 movie scripts & \cite{Chen2017v} & N & Y & N & N & N & N/Y & N/Y & N & N & Identify generative model \\
    4 gospels of the \textit{New Testament} & \cite{Grandjean2015} & Y & N & N & N & N & N & N & N & N & Comparative study \\
	Staël's \textit{Corinne ou l'Italie} & \cite{Edmondson2017} & Y & N & N & N & N & Y & N & N & N & Visualisation \\
    British Victorian novels & \cite{Grener2017, Luczak-Roesch2018} & Y & N & N & N & N & N & N & N & N/Y & Visualization \\
	9 fictionary, legendary and biographical books & \cite{Holanda2017} & N & N & N & Y & N & N & N & N & N & Level of historicity \\
	\textit{Game of Thrones} TV series (Seasons 1–7) & \cite{Janosov2017, Janosov2017a, Janosov2021a} & Y & N & N & N & N & Y & N & N & N & Event prediction \\
    Shakespeare's \textit{Hamlet} & \cite{Kwon2017} & N & N/Y & N/Y & N & N & Y & Y & N/Y & N/Y & Descriptive analysis \\
	All of Shakespeare's plays & \cite{Lee2017d} & N & Y & N & N & N & Y & N & N & N & Visualization \\
    \textit{Game of Thrones} TV series (Seasons 1--6) & \cite{Liu2017d} & N & N & N & Y & N & N & N & Y & Y & Descriptive analysis \\
	Shakespeare's \textit{Romeo \& Juliet} & \cite{Masias2017} & N & Y & N & N & N & N/Y & N & N & N & Role detection \\
	\textit{Wertheriaden} corpus & \cite{Murr2017, Barth2018} & Y & N & N & N & N & Y & N & N & N & Comparative study, Method assessment \\
	Corpus of science-fiction works & \cite{Rochat2017} & Y & N & N & N & N & Y & N & N & N & Plot structure comparison \\
	Leo Tolstoy's \textit{War and Peace} & \cite{Skorinkin2017} & Y & N & N & N & N & Y & N & N & N & Check narrative theory \\
	17 Hollywood movies & \cite{Tran2017b, Tran2017d} & Y & N & N & N & N & Y & N & N & Y & Role detection \\
    \textit{Friends} TV series & \cite{Bazzan2018} & N & Y & N & N & Y & N & N & N & Y & Check narrative theory \\
	Scripts of \textit{Game of Thrones} seasons 1, 3 \& 5 & \cite{Beveridge2018} & Y & Y & Y & N & N & N & N & N & N & Descriptive analysis \\
    Cao \& Gao's \textit{The Story of the Stone} & \cite{Bi2018} & Y & N & N & N & N & Y & N & N & Y & Check literary theories \\
    956 movie scripts \& 1 comic & \cite{Chao2018} & Y & N & N & N & N & Y & Y & Y & Y & Dynamic patterns \\
    Summaries of 577 original movies and remakes & \cite{Chaturvedi2018} & N & N & N & N & Y & Y & N & N & N & Plot structure comparison \\
	First 2 seasons of \textit{Friends} & \cite{Deleris2018} & N & Y & Y & N & N & Y & Y & N & N & Method assessment \\
	\textit{Friends} TV series  & \cite{Edwards2018, Edwards2019} & N/Y & N/Y & N/Y & N & N/Y & Y & N & N & N & Comparison of extraction methods \\ 
	37 Shakespeare plays & \cite{Evalyn2018, Shukla2018, Shukla2018a} & N & Y & N & N & N & Y & Y & N & N & Classification of works \\
    \textit{The Big Bang Theory} scripts & \cite{FronzettiColladon2019, FronzettiColladon2020} & N & Y & N & N & N & Y & Y & N & N & Classification of works \\
    Hugo's \textit{Les misérables} & \cite{Ginsburg2018} & N & Y & N & N & Y & Y & N & N & Y & Literary analysis \\
    8 Hollywood movies & \cite{He2018i} & Y & N & N & N & N & Y & N & N & N & Role detection \\
	\textit{Wonder Woman}, \textit{Thor}, and \textit{Hunger Games} scripts & \cite{Jones2018a, Jones2020a} & N & Y & N & N & N & Y & Y & N & N & Role detection, Check theory \\
	65 classic plays & \cite{Li2018h} & N & Y & N & N & N & Y & Y & N & N & Classification of works \\
	7 Hollywood movies and their scripts & \cite{Mourchid2018, Mourchid2019a, Lafhel2020, Lafhel2021} & N & Y & N & N & N & N & N & N & N & Descriptive analysis \\
	Tagore's \textit{Raktakarabi} and \textit{Muktodhara} & \cite{Muhuri2018} & N & Y & N & N & N & Y & Y & N & N & Role detection \\
	4 popular mangas & \cite{Murakami2018, Murakami2020} & Y & N & N & Y & N & Y & Y & N & N & Method assessment \\
    1 TV show and 3 movies & \cite{Lv2018} & Y & Y & N & N & N & Y & N & N & N & Method assessment \\
	53 Slovene fables & \cite{Markovic2018} & Y & N & N & N & N & N & N & N & N & Descriptive analysis \\
	Medieval Irish text \textit{Cogadh Gaedhel re Gallaibh} & \cite{Yose2017} & Y & Y & Y & Y & Y & N & N & N/Y & N & Level of historicity \\
	51 movies & \cite{Vicol2018} & Y & Y & N & Y & Y & N & Y & N & Y & Annotated corpus presentation \\
	4 plays & \cite{Zawislak2018, Zawislak2019} & N & N & N & Y & N & N & N & N & N & Descriptive analysis \\
    2 Chinese and 1 American TV series & \cite{Zhang2018ah} & N & N & N & N & Y & N & N & N & N & Descriptive analysis \\
    5 Victorian and Modernist novels & \cite{Alexander2019} & N & Y & N & N & N & N & N & N & N & Check narrative theory \\
    6 Jane Austen novels & \cite{Bipasha2019} & Y & N & N & N & N & Y & N & N & N & Method assessment \\
    Cao's \textit{Dream of the Red Chamber} & \cite{Chen2019q} & Y & N & N & N & N & Y & N & N & N & Method assessment \\
    2 novels and their movie adaptations & \cite{Chowdhury2019} & N & Y & N & N & N & N/Y & N/Y & N & N & Plot structure comparison \\
	20 classic and 20 modern novels & \cite{Dekker2019} & Y & N & N & N & N & Y & N & N & N & Descriptive analysis \\
	Rowling's \textit{Harry Potter} novels & \cite{Everton2019} & Y & N & N & N & N & W & N & N & N & Descriptive analysis \\
    930 Polish 19th and 20th century novels & \cite{Kubis2019, Kubis2021} & N & Y & N & N & N & N & N & N & N & Comparative study \\
    12 Hollywood movies & \cite{Lee2019c} & Y & N & N & N & N & Y & N & N & Y & Method assessment \\
	142 movies & \cite{Lee2019b, Lee2019d, Lee2020, Lee2020a, Lee2020c} & Y & N & N & N & N & Y & N & N & Y & Story and character embedding \\
	\textit{The Lord of the Rings} scripts and \textit{Harry Potter} novels & \cite{Li2019o} & Y & N & N & N & N & N & N & N & N & Descriptive analysis \\
    The \textit{Gospels} and \textit{Acts of the Apostles} & \cite{Massey2019} & N & N & N & Y & Y & N & N & N & N & Descriptive analysis, Comparative study \\
    1 Chinese TV series episode and 1 Hollywood movie & \cite{Pan2019b} & Y & N & N & N & N & Y & N & N & Y & Visualization, Descriptive analysis \\
    4 Swedish and Finnish 19th century plays & \cite{Pikkanen2019} & N & Y & N & N & N & Y & Y & N & N & Descriptive analysis \\
	170 contemporary Dutch novels & \cite{Smeets2019, Volker2020, Smeets2021} & Y & N & N & N & N & Y & N & N & N & Check literary theories, Level of realism \\
    \textit{Game of Thrones} TV series & \cite{Stavanja2019} & N & N & N & N & Y & N & N & N & N & Link prediction \\
	Rowling's \textit{Harry Potter} novels & \cite{Vani2019} & Y & N & N & N & N & Y & N & N & N & Visualization \\
	40 movies & \cite{Zeng2019b} & N & Y & N & N & N & N & N & N & N & Quality prediction \\
	5,269 novels & \cite{Sims2020} & N & Y & N & N & N & Y & N & N & N & Information propagation \\
	4 TV comedy series & \cite{Bazzan2020} & N & Y & N & N & Y & N & N & N & N & Comparative study \\	
	2 mythological and 1 historical texts & \cite{Besnier2020} & Y & N & N & N & N & N & N & N & N/Y & Plot structure comparison \\
    Twain's \textit{Adventures of Huckleberry Finn} & \cite{Feild2020} & Y & N & N & N & N & N & N & N & N & Presentation of a tool \\
	Martin's \textit{A Song of Ice and Fire} & \cite{GesseyJones2020} & Y & Y & Y & Y & Y & N & N & N & N & Descriptive analysis \\
    894 movie scripts & \cite{Hopp2020} & N & Y & N & N & N & Y & N & N & N & Check narrative theory \\
    Jin Yong's 15 novels & \cite{Jia2020a, Jia2020c} & N & Y & N & N & N & N & N & N & N & Method assessment \\
    \textit{The Force Awakens} script & \cite{Jones2020, Jones2020a} & N & Y & N & N & N & N & Y & N & Y & Role detection, Check theory \\
	15,540 movie subtitles & \cite{Kagan2019} & Y & N & N & N & N & Y & N & N & N & Identification of gender biases \\
    Heterogeneous selection of 26 novels & \cite{Kim2019b, Kim2020a} & N & N & N & N & Y & N & N & N & N & Classification of works \\
    6 Hollywood movies & \cite{Kulshreshtha2020} & Y & N & N & N & N & Y & N & N & Y & Segment story, Role detection \\
    Korean webtoons & \cite{Lee2020f} & Y & N & N & N & N & Y & N & N & Y & Recommendation system \\
    Script of Phillips' \textit{Joker} & \cite{Lee2020h} & Y & N & N & N & N & Y & N & N & Y & Mise-en-scène identification \\
    No experimental assessment & \cite{Lee2020j} & Y & N & N & N & N & Y & N & N & Y & Movie summarization \\
    Rowling's \textit{Harry Potter} and Alcott's \textit{Little Women} & \cite{Mellace2020, Vani2020} & Y & Y & Y & Y & Y & N & N & N & N/Y & Visualization \\
    Collection of classical tragedies & \cite{Moretti2020} & N & Y & N & N & N & Y & N & N & N & Network generation \\
    Rebreanu's \textit{Ion} & \cite{Pojoga2020} & N & Y & N & N & N & Y & N & N & N & Visualization \\
    Milton's \textit{Paradise Lost}, Homer's \textit{Iliad} & \cite{Ruegg2020} & N & Y & N & N & N & Y & N & N & N/Y & Comparative study \\
    Summaries of 4 novels & \cite{Shahsavari2020} & N & N & N & N & Y & N & Y & N & N & Estimation from summaries \\
    6 contemporary US novels & \cite{Thomas2020} & Y & N & N & N & N & Y & N & N & N & Check literary theories \\
    Shakespeare's \textit{Macbeth} and \textit{Othello} & \cite{Yavuz2020} & N & Y & N & N & N & Y & N & N & N & Descriptive analysis \\
    7 Chinese novels & \cite{Zhao2020f} & Y & N & N & N & N & Y & N & N & Y & Detect communities of characters \\
    \textit{Bleak House}, \textit{Ulysses}, and \textit{The Wire} & \cite{Alexander2021} & N & Y & N & N & Y & N & N & N & N & Comparative study \\
    Lafayette's \textit{La Princesse de Clèves} & \cite{Bilis2021} & Y & Y & Y & N & Y & Y & N & N & N & Visualization \\
    Galsworthy's \textit{Strife} and Bhattacharya's \textit{Nabanna} & \cite{Chakraborty2021} & N & Y & N & N & N & Y & Y & N & Y & Community detection \\
    Bulgakov's \textit{The Master and Margarita} & \cite{Danilova2021} & N & N & N & N & Y & N & N & N & N & Descriptive analysis \\
    2 classic Chinese novels & \cite{Fan2021, Fan2021c} & Y & N & N & N & N & Y & N & N & N & Descriptive analysis \\
    144 Russian plays and Tolstoy's \textit{War \& Peace} & \cite{Fischer2021} & Y & N & N & N & N & Y & N & N & N/Y & Descriptive analysis \\
    Lowe's \textit{Long Time No See} & \cite{Huang2021b} & N & Y & N & N & Y & Y & N & N & N & Descriptive analysis, Visualization \\
    969 IMSDb scripts & \cite{Kounelis2021} & Y & N & N & N & N & Y & N & N & N & Recommendation system \\
    Scripts of 12 Hollywood movies & \cite{Lee2021} & Y & N & N & N & N & Y & N & N & Y & Subplot identification \\
    89 popular movies & \cite{Malik2021a} & Y & N & N & N & N & Y & N & N & N & Identification of racial biases \\
    10 Tagore's plays and novels & \cite{Mandal2021a} & N & Y & N & N & N & N & N & N & N & Presentation of a tool \\
    Shakespeare's \textit{Measure for Measure} & \cite{Marceniuk2021} & N & N & N & N & Y & N & N & N & N & Visualization \\
    500 fanfiction books & \cite{Rahul2021} & Y & N & N & N & N & Y & N & N & N & Genre prediction \\
    Grimm's \textit{Children's and Household Tales} & \cite{Schmidt2021} & N & N & N & Y & Y & N & N & N & N & Annotated corpus presentation \\
    Lynch's \textit{Twin Peaks} & \cite{Siljak2021} & Y & N & N & N & N & Y & N & N & N & Descriptive analysis \\
    Sophocles' \textit{Antigone} & \cite{Yavuz2021} & N & Y & N & N & N & Y & N & N & N & Speech sequence modeling \\
    Luo's \textit{Romance of the Three Kingdoms} & \cite{Zhang2021o} & N & Y & N & N & N & N & N & N & Y & Comparative study \\
    Fan-Fictions & \cite{Zadeh2022} & Y & N & N & N & N & N & N & N & N & Popularity prediction \\
    \hline
	 & & & & & & & & & & & \\
    \caption{List of methods designed to extract character networks from fictional works. The \textit{Relationships} column indicates whether the edges correspond to co-occurrences (Cc.), conversations (Cv.), mentions (M.), direct actions (Ac.), or affiliations (Af.). The \textit{Graph} column shows whether the extracted networks are weighted (W.), directed (Di.), signed (S.), and dynamic (Dy.). When the authors experiment with different methods, several values may appear in the same cell. Note that this table has been completed after the publication of the official ACM CS article. An up-to-date version of this table is now available directly as at \url{https://compnet.github.io/CharNetReview/}.}
	\label{tab:ArticleList}
\end{longtable}
\end{landscape}}

\paragraph{Citation Network.} Figure~\ref{fig:CitationNet} shows the citation network of the articles listed in Table~\ref{tab:ArticleList}. Note that these articles also cite many bibliographic references not included in this table: we do not represent them in the figure, in order to improve its readability. 

Each vertex corresponds to an article and an edge represents the fact that an article cites another article. As is usually done for this type of network, edges are directed from the cited article to the citing one, in order to show the flow of ideas. Each vertex is identified by its BibTeX key, which corresponds to the key used in the BibTeX file publicly available online. The colors of the vertices represent the communities detected by the Louvain algorithm~\cite{Blondel2008} (a widespread modularity optimization-based method). The size of the vertices are proportional to their Hub score~\cite{Kleinberg1999}, a generalization of the degree designed to take higher order influence into account, in directed networks. It focuses on the outgoing edges, i.e. on how influential an article is.

The community structure shows the existence of 4 groups of articles with dense internal edges but comparatively sparsely connected to the other groups. They can be interpreted as groups of authors working in a relatively independent way, and possibly unaware of the work conducted in other communities. The purple community located in the left-hand part contains almost only articles dealing with video narratives, and almost all of them. Only a few articles from this community cite articles handling textual narratives, which highlights the lack of communication between these fields. The remaining communities are built around a core of several articles written by the same team: MacCarron \textit{et al}. (green community), Elson \textit{et al}. + Agarwal \textit{et al}. (blue community), Rochat \textit{et al}. + Moretti \textit{et al}. \textit{et al}. (orange community). The nature of the extracted character network (co-occurrence, conversational, etc.) does not seem to be related to the community. Rather, they seem to correspond to the high-level problem tackled by the concerned articles. For instance, articles in the green group tend to study the historical relevance of certain texts, those in the orange one mainly originate from the field of literary analysis, and works from the blue group correspond to computer science methods aiming at improving the character extraction process. The article of Rieck \& Leitte is completely isolated, as it does not cite any other article from the table, and is cited by none of them.

The centrality allows identifying a few important articles. The 2002 article of Alberich \textit{et al}.~\cite{Alberich2002} is a forerunner of the use of complex network analysis tools to study character networks, here for comics. In~\cite{Moretti2011a}, Moretti makes the connection between graph theory and literary studies. Elson \textit{et al}. are among the first to automate the extraction of conversational networks from novels in their 2010 article~\cite{Elson2010}, and so are Agarwal \textit{et al}. for action-based networks~\cite{Agarwal2012}. The 2012 article of MacCarron \& Kenna~\cite{MacCarron2012} is the first of a long series of papers related to mythological works. In the video community, the 2009 paper of Weng \textit{et al}.~\cite{Weng2009} describes the \textit{RoleNet} method, which inspired many other approaches in the community.

\paragraph{Co-citation Network.} Figure~\ref{fig:CocitationNet} shows the co-citation network of the articles from Table~\ref{tab:ArticleList}. Again, articles not listed in Table~\ref{tab:ArticleList} are not represented. Each vertex corresponds to an article and the weighted edges represent how similar two articles are in terms of the articles they cite. To get this similarity value, we compute Jaccard's coefficient based on both sets of bibliographic entries contained in the two considered articles. Like in Figure~\ref{fig:CitationNet}, each vertex is identified by its BibTeX key, and its color represents its community as detected by Louvain. Unlike the previous network, the vertex size is proportional to its eigencentrality~\cite{Bonacich1987}, as this network is not directed.

The community structure is constituted of 4 groups of articles relatively similar to those from Figure~\ref{fig:CitationNet}, as highlighted by the matching colors. Each community can be considered as a group of articles tending to cite the same set of papers, a set different from those of the other communities. The purple community corresponds to a group of articles dealing with video narratives, a bit like in Figure~\ref{fig:CitationNet}, except some of these articles are actually placed in other communities (e.g. Bost \textit{et al}. our own previous work, in the orange group). Still, all 3 remaining communities mainly focus on textual narratives. The green one contains all the articles of MacCarron \textit{et al}.'s team, which rely on manually extracted networks. But it seems its main characteristic is that these articles, as well as others in this community such as Miranda \textit{et al}.'s 2013 article~\cite{Miranda2013}, study the level of historicity of narratives. Although the article of Rieck \& Leitte is not directly connected to any other work from Table~\ref{tab:ArticleList} in Figure~\ref{fig:CocitationNet}, here it is placed in the green group, which means the article it mentions are mostly similar to the bibliography usually cited by those papers.

\bibliographystyle{ACM-Reference-Format}
\bibliography{charnets_long}

\end{document}